\documentclass[a4paper,11pt]{article}
\pdfoutput=1 

\usepackage{jheppub} 

\usepackage[T1]{fontenc} 

\usepackage{amssymb}
\usepackage{enumerate} 
\usepackage{multirow} 
\usepackage{booktabs} 
\usepackage[english]{babel} 
\usepackage[normalem]{ulem}
\usepackage{caption} 
\usepackage{comment}
\usepackage{makecell}
\title{\boldmath ${\cal CP}$ structure of the top-quark Yukawa 
interaction: NLO QCD corrections and off-shell effects}

\author{Jonathan Hermann,}
\author{Daniel Stremmer}
\author{and Malgorzata  Worek}
\affiliation{ Institute for Theoretical Particle Physics
and Cosmology, RWTH Aachen University, \\D-52056 Aachen, Germany}

\emailAdd{jonathan.hermann@rwth-aachen.de}
\emailAdd{daniel.stremmer@rwth-aachen.de}
\emailAdd{worek@physik.rwth-aachen.de}

\abstract{ 
Since its discovery at the Large Hadron Collider in 2012 the Higgs
boson has arguably become the most famous of the Standard Model
particles and many measurements have been performed in order to assess
its properties. Among others, these include measurements of the Higgs
boson's ${\cal CP}$ state which is predicted to be ${\cal CP}$-even.
Even though a pure ${\cal CP}$-odd state has been ruled out, a
possible admixture of a ${\cal CP}$-odd Higgs state has yet to be
excluded. In this work we present predictions for the associated
production of a leptonically decaying top quark pair and a stable
Higgs boson $p p \to e^+ \nu_e\, \mu^- \bar{\nu}_\mu\,b\bar{b}\,H$
with possible mixing between ${\cal CP}$-even and ${\cal CP}$-odd
states at NLO in QCD for the LHC with $\sqrt{s} = 13$ TeV. Finite
top-quark and gauge boson width effects as well as all double-,
single- and non-resonant Feynman diagrams including their interference
effects are taken into account. We compare the behaviour of the ${\cal
CP}$-even, -odd and -mixed scenarios for the integrated fiducial
cross-sections as well as several key differential distributions. In
addition, we show that both NLO corrections and off-shell effects play
an important role even at the level of integrated fiducial
cross-sections and that these are further enhanced in differential
distributions. Even though we focus here on the Standard Model Higgs
boson, the calculations could be straightforwardly applied to models
that have an extended Higgs-boson sector and predict the existence of
${\cal CP}$-odd Higgs-like particles, such as the two-Higgs-doublet
model. }

\dedicated{\rm TTK-22-19, P3H-22-054}

\keywords{Higher Order Perturbative Calculations, Top Quark, Higgs
  Production, Higgs Properties}

\begin{document} 
\maketitle
\flushbottom

%
\section{Introduction}
\label{sec:introduction}
%

The discovery of the Higgs boson with a mass of $125$ GeV by the ATLAS
and CMS collaborations in 2012 \cite {ATLAS:2012yve, CMS:2012jmk} has
launched an extensive research program to establish the true origin of
this particle and probe all its properties. Determination of its
couplings to the Standard Model (SM) particles and verification of its
quantum numbers are among the most important tasks for current and
future runs of the LHC and, in fact, for any future collider. Accurate
measurements of the Higgs boson's properties are critical for
validating the SM or potentially discovering new physical effects by
detecting deviations from the SM predictions. The ${\cal CP}$ nature
of the Higgs boson has been extensively studied at the LHC. As a
matter of fact, LHC measurements have already ruled out the hypothesis
of the pure ${\cal CP}$-odd Higgs boson
\cite{CMS:2014nkk,ATLAS:2015zhl}. However, only rather weak constrains
exist on a possible admixture between a ${\cal CP}$-even and ${\cal
CP}$-odd component. Moreover, most existing experimental analysis
concentrate on the coupling of the Higgs boson to two massive gauge
bosons $HVV$, where $V=Z, W$. In particular, both ATLAS and CMS have
investigated Higgs boson production via vector-boson fusion as well as
the following decays $H\to ZZ^*\to 4\ell$ and $H\to WW^* \to 2\ell 2
\nu$, see e.g.  Refs.
\cite{ATLAS:2016ifi,CMS:2017len,CMS:2019ekd,CMS:2019jdw,ATLAS:2020evk,
CMS:2021nnc}. It seems that so far the properties of the discovered
Higgs boson agree with the predictions of the SM
\cite{ATLAS:2013dos,CMS:2014fzn,ATLAS:2016neq}. It remains an
intriguing possibility, however, that the observed Higgs boson might
just be one member of an extended Higgs boson sector.  A very good
motivation for such a beyond the SM Higgs boson sector is the fact
that it allows for a new source of ${\cal CP}$ violation.  The latter
is required to explain the matter–antimatter asymmetry problem in the
observed universe \cite{Sakharov_1991}.  Several competing hypotheses
exist to explain the imbalance of matter and antimatter that resulted
in baryogenesis. As of yet, there is no consensus theory explaining
this phenomenon.  However, it is clear that ${\cal CP}$ violation in
the SM is insufficient for this purpose \cite{Kuzmin:1985mm}. Thus, we
are forced to look elsewhere and, in particular, beyond the well-known
SM. Among the simplest Higgs boson extensions are the
Two-Higgs-Doublet models (2HDMs) where the SM is extended by one extra
Higgs boson doublet with the same quantum numbers as the SM one
\cite{Gunion:1989we,Gunion:2002zf,Branco:2011iw,Grzadkowski:2014ada}.
These 2HDMs, both in their ${\cal CP}$-conserving and ${\cal
CP}$-violating versions, are usually used as benchmark models to
search for new scalars at the LHC. The 2HDM has four extra degrees of
freedom with two extra neutral scalars and two charged scalars. In the
${\cal CP}$-conserving version of a generic 2HDM the two neutral
states are ${\cal CP}$-even (often denoted as $h, H$) and one is
${\cal CP}$-odd (often denoted as $A$). Moreover, in this case there
are no tree-level couplings between the pseudoscalar Higgs boson and
vector bosons as the bosonic sectors of most 2HDM extensions conserve
parity. Thus, $AZZ$ and $AWW$ couplings must be induced trough fermion
loops. The branching ratios for $A\to ZZ$ and $A\to WW$ decays are
usually expected to be small, however, Higgs-fermion couplings can be
enhanced by large fermion masses and other model parameters. On the
other hand, in the ${\cal CP}$-violating version of a generic 2HDM,
the three neutral states are a mixture of ${\cal CP}$-even and ${\cal
CP}$-odd states. In addition, one of the three states is identified as
the $125$ GeV Higgs boson and all of them have non-zero $HZZ$ and
$HWW$ type interactions. Consequently, one must also concentrate on
investigating the couplings of the observed Higgs boson to the SM
fermions. Such studies are of the utmost importance to settle the
question of the true nature of this recently discovered Higgs boson.
Of all the SM fermions, the top quark plays a special role due to its
large mass of $m_t=172.5$ GeV. The top-quark Yukawa coupling $(Y_t)$
is the only coupling with the magnitude of the order of one in the
SM. Any hint of the ${\cal CP}$-odd or ${\cal CP}$-mixed component
would first manifest itself in the connection with the top quarks, for
example in the $t\bar{t}H$ production process.  Such a modified
coupling would be a direct indication of the presence of new
physics. Furthermore, it would impact not only integrated
cross-sections but also various differential cross-section
distributions, see e.g.
\cite{Ellis:2013yxa,Boudjema:2015nda,Buckley:2015vsa,
AmorDosSantos:2017ayi}.

Indirect measurements of the Yukawa coupling between the Higgs boson
and the top quark have been carried out at the LHC using the gluon
fusion process, very often with $H \to \gamma \gamma$ decays
\cite{CMS:2017dib,CMS:2018piu,ATLAS:2018doi,CMS:2018uag}.  In such a
case, the production and, depending on the final state, the decay
occur via virtual loops
\cite{CMS:2017dib,CMS:2018piu,ATLAS:2018doi,CMS:2018uag}. Therefore,
these measurements have been largely based on the assumption that no
unknown particles are allowed in these loops. A direct test of the
$Y_t$ coupling can be performed through the production of the Higgs
boson in association with a top quark pair.  In 2018 this process was
observed by the ATLAS and CMS collaborations
\cite{ATLAS:2018mme,CMS:2018uxb}. Recently, ATLAS and CMS have
reported first experimental Higgs-top ${\cal CP}$ studies, exploring
the $t\bar{t}H$ process with various Higgs boson decay channels
\cite{ATLAS:2020ior,CMS:2020cga,CMS:2022def}. Even though the
measurements favor the SM Higgs-top coupling, the ${\cal
CP}$-violating coupling has not yet been excluded. Using the Higgs
boson decays to two photons, the ATLAS collaboration has been able to
exclude the purely ${\cal CP}$-odd hypothesis with $3.9$ standard
deviations and establish a $95\%$ C.L. exclusion limit for the mixing
angle of $43^{\circ}$. On the other hand, the CMS collaboration has
performed the first measurement of the ${\cal CP}$ mixing angle for
the $\tau$ lepton Yukawa coupling \cite{CMS-PAS-HIG-20-006} which has
been found to be $4^{\circ} \pm 17^{\circ}$. This has allowed them to
set an exclusion limit on the mixing angle of $36^{\circ}$. Of course
for current and future comparisons with increasingly accurate
experimental measurements by ATLAS and CMS, equally precise
theoretical predictions must also be provided.

The effective Lagrangian for the spin ${\cal J}=0$ state can be
written down using the so-called Higgs characterisation model
\cite{Artoisenet:2013puc}.  In this parametrisation, the SM case,
i.e. the ${\cal CP}$-even state that is invariant under charge-parity
(${\cal CP}$) inversion and described by ${\cal J}^{\cal CP}=0^{++}$,
can easily be recovered. Furthermore, a ${\cal CP}$-odd state as
described by ${\cal J}^{\cal CP}=0^{+-}$, which is typical for a
generic 2HDM, is included.  In addition, ${\cal CP}$ mixing between
the ${\cal CP}$-even and ${\cal CP}$-odd state is allowed and it is
parameterised in terms of the mixing angle $\alpha_{CP}$. Finally,
within this framework the ${\cal CP}$-odd, ${\cal CP}$-even and ${\cal
CP}$-mixed states can all couple to SM particles.  In practice, the
effective Lagrangian that describes the top Yukawa coupling can be
written as a superposition of a ${\cal CP}$-even and a ${\cal CP}$-odd
phase. Any deviation from the SM value for the coupling would indicate
${\cal CP}$ violation in the top-Higgs sector and could be immediately
translated into new physics effects beyond the SM.

The Higgs characterisation model has been studied extensively in the
literature.  These studies include, among other processes, the Higgs
boson production in association with top quarks, see e.g. Refs.
\cite{Hou:2018uvr,Bahl:2020wee,Azevedo:2020fdl,Bortolato:2020zcg,
Martini:2021uey,Bahl:2021dnc}.  The impact of higher-order corrections
on ${\cal CP}$-sensitive observables has been investigated e.g.  in
Refs. \cite{Frederix:2011zi,Garzelli:2011vp,Artoisenet:2012st,Biswas:2014hwa,
  Demartin:2014fia,Hartanto:2015uka,Demartin:2015uha,Demartin:2016axk}.
Specifically, $pp\to t\bar{t}H$ production, the combined $t$-channel
$tH+\bar{t}H$ production process and $tWH$ have been studied including
parton shower effects.  For all three processes next-to-leading-order
(NLO) QCD corrections have only been taken into account for the
production stage while top-quark decays have been treated using parton
shower approximations.  In some cases spin-correlated LO top-quark
decays have been included as well.  In practise, however, such
simulations are always based on the on-shell top-quark approximation.
Moreover, for the computation of NLO QCD corrections to the $tWH$
process the isolation of the $t\bar{t}H$ process and elimination of
its double resonant contributions is required. In order to achieve
this, the authors have used subtractions that are fully local in the
phase space. Specifically, they have applied the two main schemes that
are present in the literature and are known under the name of diagram
removal and diagram subtraction \cite{Frixione:2008yi}. 
However, the only way to perform theoretically consistent simulations
that comprise both $t\bar{t}H$ and $tW(b)H$ is to compute the $pp \to
W^+b\, W^-\bar{b}\, H$ process using a complex top-quark mass. We
point out here that the $tWH$ contribution in the final state with one
$b$-jet can be obtained with massive bottom quarks using the
four-flavour scheme (see e.g. \cite{Jezo:2016ujg} for a simpler case of $t\bar{t}$
and $tW$). Instead, for $tWbH$ in the final state with two
$b$-jets, massless bottom quarks and the five flavour-scheme can be
employed. A simulation of the $pp \to W^+b\, W^-\bar{b}\, H$
process would include all contributions that are, on the one hand,
gauge invariant and, on the other hand, include interference and
finite top-quark width effects.  It 
would also contain contributions from the amplitude without any
resonant top-quark propagator and, in turn, interference effects
between single-top and single-antitop contributions. The latter are
absent in $tWbH$ simulations. A particular final state that we are
interested in can be enhanced and probed by means of kinematic cuts,
closely following what is being done in the ATLAS and CMS
experiments. We note that such full NLO QCD predictions for the
$t\bar{t}H$ process with leptonic top-quark decays exist but only for
the ${\cal CP}$-even case
\cite{Denner:2015yca,Stremmer:2021bnk}. Later also electroweak
corrections have been calculated for this process
\cite{Denner:2016wet}. Very recently even various Higgs boson decays
have been added albeit in the narrow-width-approximation (NWA)
\cite{Stremmer:2021bnk}. Taking into account how crucial the
measurements of the top-quark Yukawa coupling are, it is of utmost
importance to provide the full NLO QCD calculation for the $t\bar{t}H$
process allowing beyond the SM top-Higgs couplings and taking into
account all effects whilst consistently avoiding any approximations.

The purpose of the article is, therefore, manifold. Firstly, we
provide state-of-the-art NLO QCD predictions for the $pp\to t\bar{t}H$
process in the di-lepton top-quark decay channel with an extended
top-Higgs Yukawa coupling within the Higgs characterisation
framework. Specifically, we calculate NLO QCD corrections to the
$e^+\nu_e \, \mu^-\bar{\nu}_\mu\, b\bar{b}\,H$ final state,
consistently taking into account double-, single and non-resonant
top-quark contributions together with all interference
effects. Moreover, in the computation off-shell, top quarks are
treated in the complex mass scheme with a physical top-quark width
$\Gamma_t$. Similarly, non-resonant and off-shell effects due to the
finite $W$ boson width are also incorporated in the
calculation. Schematically, our calculations for the $e^+\nu_e \,
\mu^-\bar{\nu}_\mu\, b\bar{b}\,H$ matrix element squared can be
written down as $|{\cal M}_{t\bar{t}H}|^2 = |{\cal M}_{2tH}+{\cal
M}_{1tH} +{\cal M}_{1\bar{t}H}+{\cal M}_{0tH}|^2$.  Secondly, we
investigate the sensitivity of the $t\bar{t}H$ process to beyond the
SM physics in the Higgs boson sector and identify observables that are
particularly useful in distinguishing the three different Higgs boson
${\cal CP}$ states.  Thirdly, we assess the impact of full off-shell
effects on the $t\bar{t}H$ process with the extended top-Higgs
coupling both at the integrated and differential fiducial
cross-section level.  To perform such a study, a second computation
based on the NWA is carried out for the $pp\to t\bar{t}H$ process. In
this case the $e^+\nu_e \,\mu^-\bar{\nu}_\mu\, b\bar{b}\,H$ final
state is generated in the $pp \to t\bar{t}H\to W^+b\, W^-\bar{b} \, H
\to e^+\nu_e \, \mu^-\bar{\nu}_\mu\, b\bar{b}\,H$ decay chain where
the top quarks and $W$ gauge bosons are always kept on-shell.  On the
one hand, we compare our results from the full off-shell calculation
to the predictions based on the full NWA. In that case NLO QCD
corrections are calculated both to the production stage and the
top-quark decays.  On the other hand, we also provide NLO QCD
predictions in the NWA but with leading order (LO) top-quark decays
(abbreviated as ${\rm NWA}_{\rm LOdec}$). By employing the ${\rm NWA}$
and ${\rm NWA}_{\rm LOdec}$ results, we are able to estimate the
effects of NLO QCD corrections to the top-quark decays.

The paper is organised as follows. In Section \ref{sec:description} we
summarise the framework of our calculation. The theoretical setup for
LO and NLO QCD results is given in Section \ref{sec:setup}. Results
for the integrated fiducial cross-sections for the SM (${\cal
CP}$-even), ${\cal CP}$-mixed and ${\cal CP}$-odd Higgs boson are
presented in Section \ref{sec:tth-int}.  In Section \ref{sec:tth-diff}
predictions for differential fiducial cross-sections are given and we
investigate how the different ${\cal CP}$ configurations are affected
by the higher-order effects in various phase-space
regions. Additionally, we present observables that can be used to
distinguish the ${\cal CP}$ states. Finally, in Section
\ref{sec:tth-diff} we study the impact of off-shell effects on
differential fiducial cross sections and examine the size of NLO QCD
corrections to top-quark decays at the differential level. We
summarise the results and outline our conclusions in Section
\ref{sec:sum}.

%
\section{Description of the calculation and Higgs parametrisation}
\label{sec:description}
%

\begin{figure}[t!]
	\begin{center}
		\includegraphics[width=\textwidth]{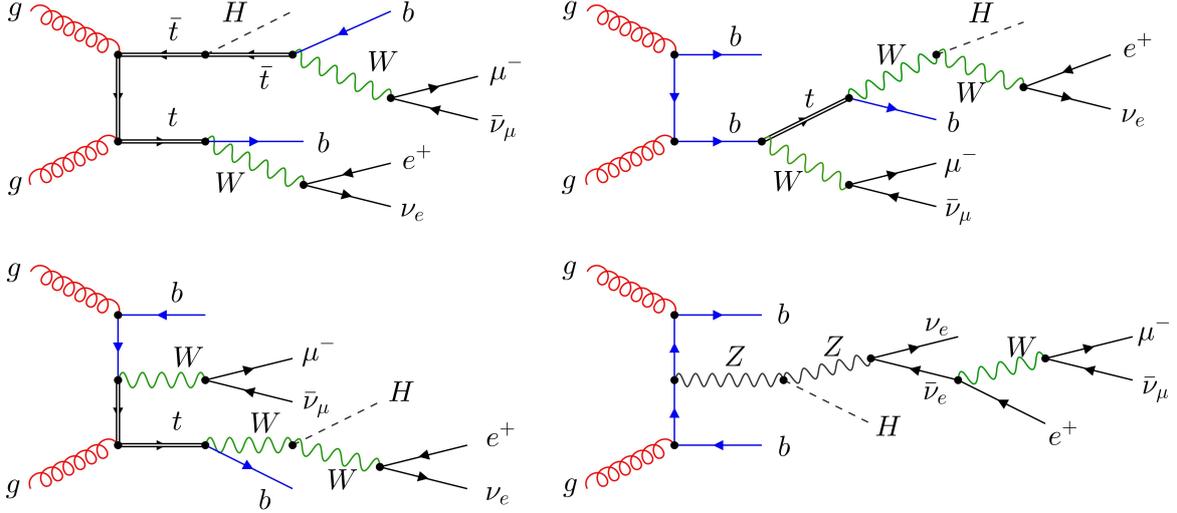}
	\end{center}
	\caption{\label{fig:FD} \it 
	Representative LO Feynman diagrams for the $p p \to e^+ \nu_e
\,\mu^- \bar{\nu}_\mu \, b \bar{b} \,H $ process involving two (top
left), only one (top right and bottom left) or no (bottom right)
top-quark resonances. The Feynman diagrams were created with the help
of the \textsc{FeynGame} \cite{Harlander:2020cyh} program.}
\end{figure}

In this work we consider the production of a stable ${\cal CP}$-even,
${\cal CP}$-odd and ${\cal CP}$-mixed Higgs boson in association with
a leptonically decaying top-quark pair at order $\mathcal{O}\left(
\alpha_s^3 \alpha^5 \right)$ in perturbation theory.  Specifically, we
study the process $p p \to e^+ \nu_e\, \mu^- \bar{\nu}_\mu \,b \bar{b}
\, H + X$, i.e. we calculate NLO QCD corrections to the LO process $p
p \to e^+ \nu_e \, \mu^- \bar{\nu}_\mu \, b \bar{b} \, H $ which is of
order $\mathcal{O}\left( \alpha_s^2 \alpha^5 \right)$.  Following the
Higgs characterisation framework \cite{Artoisenet:2013puc}, we allow
for a possible admixture of a ${\cal CP}$-odd Higgs state and extend
the top Yukawa interaction according to
\begin{equation} \label{eq:L_Yukawa}
    \mathcal{L}_{t\bar{t}H} = - \bar{\psi}_t  \frac{Y_t}{\sqrt{2}}
    \left( \kappa_{Ht\bar{t}}\, \cos(\alpha_{CP}) + i
      \kappa_{At\bar{t}} \, \sin(\alpha_{CP}) \gamma_5 \right)
    \psi_t\, H,
\end{equation}
where $\psi_t$ and $H$ are the top and Higgs fields, respectively,
whereas $\alpha_{CP}$ is the ${\cal CP}$-mixing angle.  Furthermore,
$Y_t = \sqrt{2} m_t /v $ is the top quark Yukawa coupling with the
Higgs vacuum expectation value $v \approx 246$ GeV and the top quark
mass $m_t$. In this parametrisation:
\begin{itemize}
\item $\alpha_{CP} = 0$ corresponds to the ${\cal CP}$-even case,
\item $\alpha_{CP} = \pi/2$ corresponds 
to the ${\cal CP}$-odd case,
\item $\alpha_{CP} = \pi/4$ corresponds 
to the ${\cal CP}$-mixed case.
\end{itemize}
\noindent
In practice, a ${\cal CP}$-mixed scenario could be realised by
choosing any value of $\alpha_{CP}$ from the range $\alpha_{CP} \in
\left(0,\pi/2 \right)$ but we focus here on the $\alpha_{CP} = \pi/4$
case for which $\cos(\alpha_{CP}) = \sin(\alpha_{CP})$.  Different
choices for the real-valued couplings $\kappa_{Ht\bar{t}}$ and
$\kappa_{At\bar{t}}$ will be discussed in Section \ref{sec:tth-int}.
In addition to changing the Yukawa interaction, we also introduce an
additional parameter $\kappa_{HVV}$ in the $HVV$ ($V=W^{\pm},Z$)
interaction term
\begin{equation} \label{eq:L_HVV}
    \mathcal{L}_{HVV} = \kappa_{HVV} \left( \frac{1}{2} \,g_{HZZ}
      Z_\mu Z^\mu + g_{HWW} W^+_\mu W^{- \mu}  \right) H,
\end{equation}
where $g_{HZZ} = 2 m_Z^2 / v$ and $g_{HWW} = 2 m_W^2 / v$ are the
$HVV$ couplings with $m_{Z,W}$ being the masses of the electroweak
(EW) gauge bosons.  We do not take into account higher dimensional
$HVV$ couplings and neglect any contribution from loop-induced
couplings of the Higgs boson to gluons or photons.  The above
described $HVV$ couplings are only of importance in full off-shell
calculations such as the one presented here since on-shell top quarks
are not heavy enough to produce both a Higgs and a $W$ boson.

In order to incorporate full off-shell effects, we consistently take
into account all Feynman diagrams with the $e^+\nu_e \, \mu^-
\bar{\nu}_\mu \,b \bar{b}\, H$ final state at order $\mathcal{O}\left(
\alpha_s^2 \alpha^5 \right)$ (for the LO computation) and interference
effects, irrespective of whether a $t\bar{t}$ pair actually occurs as
an intermediate state.  The same applies to the $W$ bosons appearing
in the calculation.  Examples for such diagrams with two, only one or
no top-quark resonances, to which we will be referring as double-,
single- and non-resonant contributions in the following, are depicted
in Figure \ref{fig:FD}.  Assuming a diagonal Cabibbo-Kobayashi-Maskawa
(CKM) matrix and using the $5$-flavor scheme, i.e. setting both the
mass and the Yukawa coupling of the bottom quark to zero, we obtain
$174$ diagrams for the $g g$ and $76$ for the $q \bar{q}$ initial
state at LO where $q\in \left\{ u,d,c,s \right\}$.  On the other hand,
$b\bar{b}$ initial states are neglected as they are at the per-mille
level \cite{Stremmer:2021bnk} and thus well within NLO scale
uncertainties.  The numbers of diagrams are independent of the ${\cal
CP}$-structure of the Higgs boson as long as
$\kappa_{HVV}\,,\kappa_{At\bar{t}}\,,\kappa_{Ht\bar{t}} \neq 0$.  To
include Breit-Wigner propagators for the top quark and gauge bosons in
a gauge invariant way, we use the complex-mass scheme
\cite{Denner:1999gp,Denner:2005fg,Bevilacqua:2010qb,Denner:2012yc}.
In addition to the full off-shell case we also provide results in the
NWA, i.e. the limit in which the unstable intermediate top quarks and
$W$ gauge bosons are put on-shell.  As only double-resonant diagrams
(see e.g. top-left diagram in Figure \ref{fig:FD}) are included when
employing the NWA, the numbers of considered diagrams for the $gg$ and
$q\bar{q}$ initial states are reduced to $8$ and $2$, respectively, at
LO.

All of the results presented in this work were obtained using the
\textsc{Helac-NLO} \cite{Bevilacqua:2011xh, Bevilacqua:2019quz} MC
program which consists of \textsc{Helac-Dipoles} \cite{Czakon:2009ss}
and \textsc{Helac-1Loop} \cite{vanHameren:2009dr}.  Amplitudes are
constructed in a recursive manner based on recursive Dyson-Schwinger
equations
\cite{Draggiotis:1998gr,Draggiotis:2002hm,Papadopoulos:2005ky} and the
phase-space integration is performed with the help of \textsc{Parni}
\cite{vanHameren:2007pt} and \textsc{Kaleu} \cite{vanHameren:2010gg}.
NLO corrections to the Born level process can be divided into virtual
and real corrections which are computed in \textsc{Helac-1Loop} and
\textsc{Helac-Dipoles}, respectively.  Note that from the point of
view of QCD corrections, this process is essentially identical to
$t\bar{t}\gamma$, $t\bar{t}W^\pm$ or $t\bar{t}Z$ production as
presented in Refs. \cite{Bevilacqua:2018woc, Bevilacqua:2019cvp,
Bevilacqua:2020pzy}.  For the virtual corrections we use the
\textsc{CutTools} \cite{Ossola:2007ax} implementation of the OPP
reduction \cite{Ossola:2006us} in order to reduce the occurring one
loop amplitudes to scalar integrals at the integrand level.  The
resulting scalar integrals are then evaluated using \textsc{OneLOop}
\cite{vanHameren:2010cp}.  The real corrections receive contributions
from $2\to 8$ partonic processes. In our case we need to consider the
following subprocesses:
\begin{align}
    \begin{split}
        g g &\to e^+ \nu_e \, \mu^-  \bar{\nu}_\mu \,b \bar{b} \,H g \\
        q \bar{q} 
        &\to e^+ \nu_e \, \mu^- \bar{\nu}_\mu \,b \bar{b} \,H g \\
        g q &\to e^+ \nu_e \, \mu^-  \bar{\nu}_\mu \, b \bar{b} \, H q \\
        g \bar{q} &\to e^+\nu_e \,  \mu^-  \bar{\nu}_\mu \, b \bar{b} \, H \bar{q} 
    \end{split}
\end{align}
Their respective contributions are computed using
\textsc{Helac-Dipoles}.  It implements two independent subtraction
schemes, the Catani-Seymour dipole formalism
\cite{Catani:1996vz,Catani:2002hc} and the Nagy-Soper subtraction
scheme \cite{Bevilacqua:2013iha}.  We use the Nagy-Soper scheme for
the full off-shell calculations while those done in the NWA are
performed in the Catani-Seymour formalism and its extension to
top-quark decays \cite{Campbell:2004ch,Bevilacqua:2019quz}.

Our full off-shell results are stored as (partially) unweighted events
\cite{Bevilacqua:2016jfk} in (modified) Les Houches Event Files
(LHEFs) \cite{Alwall:2006yp} and \textsc{Root} Ntuples
\cite{Antcheva:2009zz}.  This allows us to re-weight our results in
order to efficiently accommodate different parton distribution
functions (PDFs) as well as renormalisation and factorisation scale
settings.

The implementation of the modified $HVV$ and $Ht\bar{t}$ interactions
in \textsc{Helac-NLO} according to equations (\ref{eq:L_Yukawa}) and
(\ref{eq:L_HVV}) have been cross-checked with
\textsc{MadGraph\_aMC@NLO} \cite{Alwall:2014hca} and \textsc{MadLoop}
\cite{Hirschi:2011pa}. Furthermore, various tests have been performed
within the \textsc{Helac-NLO} code.  Specifically, we have compared
the LO full off-shell squared amplitudes for $g g \to e^+ \nu_e \,
\mu^- \bar{\nu}_\mu \, b \bar{b} \, H$ and $u \bar{u} \to e^+ \nu_e \,
\mu^- \bar{\nu}_\mu \, b \bar{b} \, H$ with \textsc{MadGraph\_aMC@NLO}
for a few random phase-space points. The latter have been generated
with the help of the \textsc{Rambo} \cite{Kleiss:1985gy} MC
phase-space generator.  In addition, we have checked the finite parts
along with the coefficients of the poles in $\epsilon$,
i.e. $1/\epsilon$ and $1/\epsilon^2$, of the virtual amplitudes for
stable $t\bar{t}H$ production for $gg$ and $u\bar{u}$ initial states
for all three ${\cal CP}$ states.  For the real corrections we have
ensured that the poles in $\epsilon$ occurring in the
$\mathbf{I}$-operator and those appearing in the virtual part cancel.
Finally, we have confirmed that the real emission part is independent
from any phase-space restriction imposed by the unphysical
$\alpha_{max}$ parameter
\cite{Nagy:1998bb,Nagy:2003tz,Bevilacqua:2009zn,Czakon:2015cla}.

%
\section{LHC setup}
\label{sec:setup}
%

We consider the ${\cal CP}$-even, ${\cal CP}$-odd and ${\cal
CP}$-mixed Higgs production in association with top quarks at the LHC
with $\sqrt{s}=13$ TeV and calculate higher-order corrections to the
$e^+\nu_e\, \mu^-\bar{\nu}_\mu \, b\bar{b}\,H$ final state at ${\cal
O}(\alpha_s^3\alpha^5)$. We work in the 5-flavour scheme ($N_f=5$) and
neglect the contribution from bottom quarks in the initial state. The
LHAPDF interface \cite{Buckley:2014ana} is used to provide an access
to parton density functions and we employ the
\texttt{NNPDF31-nlo-as-0118} PDF set \cite{Ball:2017nwa} with
$\alpha_s(m_Z)=0.118$ at LO and NLO in QCD. The running of the strong
coupling constant is performed with two-loop accuracy. The
$G_\mu-$scheme is used for the electroweak input parameters and we
adopt the same values as in Ref. \cite{Bevilacqua:2020pzy}. Thus, the
electromagnetic coupling constant $\alpha$ is given by
\begin{equation}
	\alpha =\frac{\sqrt{2}}{\pi} \,G_\mu \,
        m_W^2\,\left(1-\frac{m_W^2}{m_Z^2}\right)\,,
	~~~~~~~~~~~~~~~~~~~~~
	G_{ \mu}=1.166378 \cdot 10^{-5} \textrm{ GeV}^{-2}\,.
\end{equation}
We use the following masses and widths
\begin{equation}
\begin{array}{lll}
m_{t}=172.5 ~{\rm GeV} \,, &\quad \quad \quad
&m_{H}=125 ~{\rm GeV} \,, 
\vspace{0.2cm}\\
 m_{W}= 80.385  ~{\rm GeV} \,, &
&\Gamma_{W} =  2.09767 ~{\rm GeV}\,, 
\vspace{0.2cm}\\
  m_{Z}=91.1876   ~{\rm GeV} \,, &
&\Gamma_{Z} = 2.50775~{\rm GeV}\,,
\end{array}
\end{equation}
and neglect the masses of all other particles. From these input
parameters we calculate the top-quark width according to the formulae
derived in Refs. \cite{Denner:2012yc,Jezabek:1988iv}. For the values
provided above, the LO and NLO top-quark widths are given by
\begin{equation}
\Gamma_{t}^{\rm LO} =  1.45759 ~{\rm GeV}\,, \quad \quad \quad
\Gamma_{t}^{\rm NLO} =  1.33247   ~{\rm GeV}\,
\end{equation}
in the full off-shell case and by
\begin{equation}
\Gamma_{t, { \rm NWA}}^{\rm LO} =  1.48063 ~{\rm GeV}\,, \quad \quad \quad
\Gamma_{t, { \rm NWA}}^{\rm NLO} =  1.35355   ~{\rm GeV}\,
\end{equation}
in the NWA. The top-quark width is treated as a fixed parameter and we
use $\alpha_s(\mu_R=m_t)$ when calculating $\Gamma^{\rm NLO}_t$. Since
the Higgs boson is stable, its width is set to $\Gamma_H =
0$. IR-safety is ensured by the {\it anti}-$k_T$ jet algorithm
\cite{Cacciari:2008gp} which is used to cluster final state partons
with pseudo-rapidity $|\eta|<5$ into jets with the jet-resolution
parameter $R=0.4$. We require exactly two charged leptons, two
$b$-jets and one stable Higgs boson. All final states have to fulfil
the following experimental cuts
\begin{equation}
\begin{array}{ll l l  }
 p_{T,\, \ell} >25 ~{\rm GeV}\,,    &
\quad\quad \quad\quad\quad &
 p_{T,\, b} >25 ~{\rm GeV}\,, 
\vspace{0.2cm}\\
 |y_{\ell}|<2.5\,,&& |y_b|<2.5 \,,
\end{array}
\end{equation}
where $\ell=\mu^-,e^+$. No restrictions on the extra light jet and
missing transverse momentum, denoted as $p_{T,\, miss}=|\vec{p}_{T,\,
\nu_e} + \vec{p}_{T,\,\bar{\nu}_\mu}|$, are applied. We use a
dynamical scale setting $\mu_0=\mu_R=\mu_F=H_T/2$ as our central
scale, where $H_T$ is defined according to
\begin{equation} \label{eq:ht}
	H_T=p_{T,\, b_1}+p_{T, \, b_2}+p_{T,\, e^+}+p_{T, \,
          \mu^-}+p_{T, \, miss}+p_{T, \, H}\,,
\end{equation}
where $b_1$ and $b_2$ stand for the first and second hardest $b$-jet
in $p_T$. A detailed discussion and comparison with other scale
choices is given for the SM case ($\alpha_{CP}=0$) in
Ref. \cite{Stremmer:2021bnk}. In order to estimate the theoretical
uncertainties arising from neglected higher-order terms in the
perturbative expansion, we use the $7$-point scale variation in which
the factorisation and renormalisation scales are varied independently
in the range
\begin{equation}
	\frac{1}{2} \, \mu_0  \le \mu_R\,,\mu_F \le  2 \,  \mu_0\,, \quad
	\quad
	\quad \quad 
	\quad \quad \quad \quad \quad \quad \frac{1}{2}  \le
	\frac{\mu_R}{\mu_F} \le  2 \,.
\end{equation}
This leads to the following seven pairs
\begin{equation}
	\label{scan}
	\left(\frac{\mu_R}{\mu_0}\,,\frac{\mu_F}{\mu_0}\right) = \Big\{
	\left(2,1\right),\left(0.5,1  
	\right),\left(1,2\right), (1,1), (1,0.5), (2,2),(0.5,0.5)
	\Big\} \,,
\end{equation}
which are taken into account for the error estimation. We do not
include PDF uncertainties into the theoretical error estimation, since
these are significantly smaller than the scale uncertainties
\cite{Stremmer:2021bnk}, especially for the chosen PDF set.

%
\section{Integrated fiducial cross-sections}
\label{sec:tth-int}
%

\begin{figure}[t!]
	\begin{center}
		\includegraphics[width=0.7\textwidth]{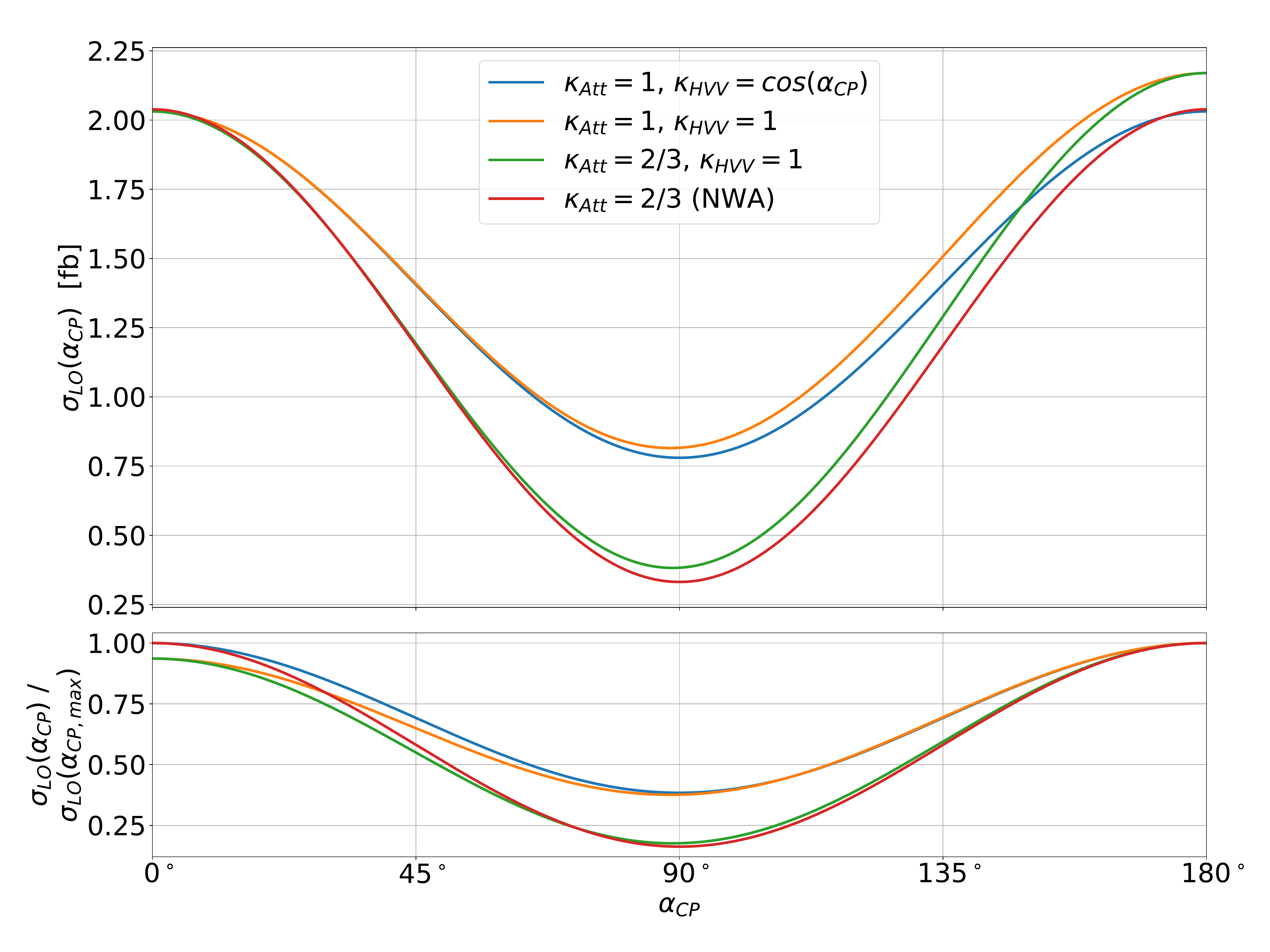}
	\end{center}
	\caption{\label{fig:integrated} \it LO integrated fiducial
cross-section depending on $\alpha_{CP}$ for the $pp\to e^+\nu_e\,
\mu^-\bar{\nu}_{\mu}\, b\bar{b}\,H$ process at the LHC with
$\sqrt{s}=13\textrm{ TeV}$. The red curve indicates the results
computed in the NWA for $\kappa_{At\bar{t}} = 2/3$ whereas the blue,
orange and green curves correspond to full off-shell results for
different choices of the coupling constants. In all cases we have
$\kappa_{Ht\bar{t}}=1$. In the lower panel the ratio to
${\alpha_{CP,\, max}}$ is displayed for each plotted case. }
\end{figure}

In this first part of our analysis we will discuss the results for
integrated fiducial cross-sections.  The two coupling parameters
$\kappa_{Ht\bar{t}}$ and $\kappa_{At\bar{t}}$ appearing in
Eq. (\ref{eq:L_Yukawa}) are chosen as follows. We take
$\kappa_{Ht\bar{t}} = 1$ everywhere to recover the SM results for
$\alpha_{CP} = 0$. For $\kappa_{At\bar{t}}$ we consider two different
case, $\kappa_{At\bar{t}} = 1$ and $\kappa_{At\bar{t}} = 2/3$.  The
first choice is simply designed to have the same coupling for both the
scalar and pseudoscalar Higgs boson.  The second one is motivated by
the measurements of Higgs boson production in gluon-fusion (GF)
process which indicate good agreement with the SM predictions
\cite{ATLAS-CONF-2021-014,CMS-PAS-HIG-18-029}.
Assuming that the top-loop induced terms dominate the GF we find
\begin{equation}
    \label{eq:GF}
    \frac{\sigma_{gg \to H}}{\sigma_{gg\to H}^{SM}} = \cos^2 \left(
      \alpha_{CP} \right) \kappa_{Ht\bar{t}}^2 + \sin^2 \left(
      \alpha_{CP} \right) \kappa_{At\bar{t}}^2\,
    \frac{g_{Agg}^2}{g_{Hgg}^2}
\end{equation}
for the ratio between the GF in our model and the SM
\cite{Demartin:2014fia,Demartin:2015uha}. Here $g_{Agg} = \alpha_s /
\left(2 \pi v\right)$ and $g_{Hgg} = - \alpha_s / \left(3 \pi
v\right)$ are the effective loop-induced gluon-Higgs couplings for a
${\cal CP}$-even and ${\cal CP}$-odd Higgs boson, respectively.
Hence, to recover the SM results for any value of $\alpha_{CP}$ one
must set $\kappa_{At\bar{t}} = 2/3$ and $\kappa_{Ht\bar{t}} = 1$.

For the remaining free parameter $\kappa_{HVV}$ we also analyse two
different cases, specifically $\kappa_{HVV} = \cos \left( \alpha_{CP}
\right)$ and $\kappa_{HVV} = 1$.  Note that in the first case
$\kappa_{HVV}$ actually becomes dependent on the mixing angle
$\alpha_{CP}$.  This choice is relevant for SM extensions such as the
${\cal CP}$-conserving $2$HDM where only ${\cal CP}$-even Higgs-like
particles couple to the EW gauge bosons, see
e.g. Refs. \cite{Aoki:2009ha, Branco:2011iw}.  Thus, the corresponding
couplings receive the same scaling as the scalar Higgs-top coupling.
However, when considering a possible non-zero $\alpha_{CP}$ for the SM
Higgs boson it is more sensible to choose $\kappa_{HVV} = 1$ in order
to recover results from Higgs boson production through vector boson
fusion (VBF) which, just as the GF measurements, indicate agreement
with the SM \cite{CMS:2018uag}.  Let us mention that in the NWA, the
choice of $\kappa_{HVV}$ is irrelevant since in this case the Higgs
boson can only be radiated off top quarks in the production stage.

In Figure \ref{fig:integrated} we present the LO integrated fiducial
cross-sections depending on the mixing angle $\alpha_{CP}$ for several
different combinations of the parameter choices outlined above.  The
blue, orange and green curves indicate results for the full off-shell
case with different choices for the couplings and the red one
describes the NWA with $\kappa_{At\bar{t}} = 2/3$.  To create these
curves one can make use of the fact that at LO the cross-section can
be split into six parts according to
\begin{align}
\label{eq:xSec_alpha_dependence}
    \begin{split}
    \sigma \left(\alpha_{CP}\right) &= \cos^2 \left(\alpha_{CP}\right)
    \kappa_{Ht\bar{t}}^2 \,\,\sigma_1 + \sin^2
    \left(\alpha_{CP}\right) \kappa_{At\bar{t}}^2 \,\,\sigma_2 + \cos
    \left(\alpha_{CP}\right) \sin \left(\alpha_{CP}\right)
    \kappa_{Ht\bar{t}}\, \kappa_{At\bar{t}} \,\,\sigma_3 \\
    & + \cos \left(\alpha_{CP}\right) \kappa_{Ht\bar{t}}\,
    \kappa_{HVV} \left(\alpha_{CP}\right) \,\sigma_4 + \sin
    \left(\alpha_{CP}\right) \kappa_{At\bar{t}} \,\kappa_{HVV}
    \left(\alpha_{CP}\right) \,\sigma_5 + \kappa_{HVV}^2
    \left(\alpha_{CP}\right) \,\sigma_6 \,,
    \end{split}
\end{align}
where the $\sigma_i$ coefficients are independent of the mixing angle,
$\kappa_{Ht\bar{t}}$, $\kappa_{At\bar{t}}$ and $\kappa_{HVV}$.  Thus,
it is enough to compute six integrated cross-sections with different
sets of input parameters $\{ \alpha_{CP}, \kappa_{Ht\bar{t}},
\kappa_{At\bar{t}}, \kappa_{HVV}\}$. From these, the cross-section
contributions $\sigma_i$ in Eq. \eqref{eq:xSec_alpha_dependence} can
then be calculated by solving a system of equations. The final results
for $\sigma(\alpha_{CP})$ with all required sets of parameters $\{
\alpha_{CP}, \kappa_{Ht\bar{t}}, \kappa_{At\bar{t}}, \kappa_{HVV}\}$
can then easily be assembled from
Eq. \eqref{eq:xSec_alpha_dependence}.  In case of the NWA it is even
sufficient to only evaluate the first three contributions as no $HVV$
couplings can occur. Therefore, we have $\sigma_{4,5,6}^{\rm NWA}=0$
both at LO and NLO in QCD. The numerical values for the cross-section
coefficients $\sigma_i$ are listed in Table
\ref{table:sigma_i}. Although there is a non-zero
  contribution from interference effects between the ${\cal
    CP}$-odd coupling and the SM contribution at the
matrix element level, any hadron collider observable is averaged over
charge conjugate processes. As a result, interference effects vanish in typical
hadronic (differential) cross-sections when integrated over a ${\cal
CP}$-symmetric part of the phase space. This observation still holds for
 fiducial cross-sections $(\sigma_3,\sigma_5)$ when  ${\cal
   CP}$-symmetric selection criteria are applied to all final states,
 see e.g. Ref. \cite{Plehn:2001nj}.
%
\begin{table*}[t!]
    \caption{\it Cross-section coefficients, $\sigma_i$, for the
splitting of the LO integrated fiducial cross-sections according to
Eq. (\ref{eq:xSec_alpha_dependence}) for the full off-shell and
narrow-width treatments.  For the latter the final three coefficients
do not contribute as no diagrams with couplings between the Higgs
boson and the EW gauge bosons occur in the NWA. The values in brackets
indicate the respective Monte Carlo integration error on the last
digit. }
    \label{table:sigma_i}
    \centering
    \renewcommand{\arraystretch}{1.5}
    \begin{tabular}{l@{\hskip 20mm}ll@{\hskip 10mm}}
        \hline\noalign{\smallskip}
          & ~~Off-shell & ~~~~~NWA \\
        \noalign{\smallskip}\midrule[0.5mm]\noalign{\smallskip}
        ~~~~~~$\sigma_1$ [fb]  &   $ ~~2.0643(4) $      & $ ~~2.0388(2) $ \\
        ~~~~~~$\sigma_2$ [fb]  &   $ ~~0.7800(1)$       & $ ~~0.74583(7) $ \\
        ~~~~~~$\sigma_3$ [fb]  &   $ -0.0002(8)$       &  $-0.0001(3)$ \\
        ~~~~~~$\sigma_4$ [fb]  &   $ -0.0693(8) $     & $ ~~~~~~~- $ \\
        ~~~~~~$\sigma_5$ [fb]  &   $ -0.0001(9) $     & $ ~~~~~~~- $ \\
        ~~~~~~$\sigma_6$ [fb]  &   $ ~~0.0363(9) $      & $ ~~~~~~~- $ \\
        \noalign{\smallskip}\hline\noalign{\smallskip}
    \end{tabular}
  \end{table*}

We note that Eq. \eqref{eq:xSec_alpha_dependence} is also instructive
in understanding and interpreting the different behaviours of the
curves depicted in Figure \ref{fig:integrated}.  First, one should
note that the production of pseudoscalar particles in association with
top quarks is suppressed compared to scalar ones for masses below
$\sim 200$ GeV if the two couplings $\kappa_{Ht\bar{t}}$ and
$\kappa_{At\bar{t}}$ are equal \cite{Haisch:2016gry}.  This difference
can be understood when looking at the $t \to t+H/A$ fragmentation
functions \cite{Dittmaier:2000tc, Dawson:1997im}
\begin{align} \label{eq:fragmentation}
    \begin{split}
        f_{t\to t+H}(x) &= \frac{\kappa^2_{Ht\bar{t}}}{(4 \pi)^2}
        \left[ \frac{4(1-x)}{x} + x \ln \left( \frac{s}{m_t^2} \right)
        \right]\,, \\      
        f_{t\to t+A}(x) &= \frac{\kappa^2_{At\bar{t}}}{(4 \pi)^2}
        \left[ x \ln \left( \frac{s}{m_t^2} \right) \right]\,,       
    \end{split}
\end{align}
where $x$ is the momentum fraction that the Higgs boson carries.  The
scalar fragmentation function has an additional $1/x$ term which
enhances the production of soft scalar particles compared to
pseudoscalar particles and thus results in a larger integrated
fiducial cross-section for a scalar Higgs boson.  This can clearly be
seen in Figure \ref{fig:integrated} as all depicted curves have a
minimum around $\alpha_{CP} = \pi/2$ which corresponds to the
production of a ${\cal CP}$-odd Higgs boson.  Reducing
$\kappa_{At\bar{t}}$ from $1$ to $2/3$ only enhances this effect which
is why around its minimum the orange curve $(\kappa_{At\bar{t}}=1)$ is
about twice as large as the green one $(\kappa_{At\bar{t}}=2/3)$.  For
the latter choice of $\kappa_{At\bar{t}}$, the production
cross-section of $2.03$ fb for a ${\cal CP}$-even Higgs boson is more
than five times as large as the $0.38$ fb for the ${\cal CP}$-odd
case. The mixed case falls almost exactly in the middle with $1.19$
fb.

One can also observe that the cross-sections computed in the NWA (red
curve) are symmetric with respect to $\alpha_{CP} \rightarrow \pi -
\alpha_{CP}$ which is equivalent to changing the sign of the top-quark
Yukawa coupling $Y_t$. As we have $\sigma_{4,5,6}^{\rm NWA}=0$ in the
NWA, we only need to consider the terms in the first line of
Eq. \eqref{eq:xSec_alpha_dependence} in order to understand this
phenomenon.  For the first two terms the symmetry is inherent in the
explicit $\alpha_{CP}$ dependence since $\cos^2 \left(\pi -
\alpha_{CP} \right) = \cos^2 \left(\alpha_{CP} \right)$ and $\sin^2
\left(\pi - \alpha_{CP} \right) = \sin^2 \left(\alpha_{CP} \right)$.
In contrast, for the third term we have $\cos \left(\pi - \alpha_{CP}
\right) \sin \left(\pi - \alpha_{CP} \right) = - \cos
\left(\alpha_{CP} \right) \sin \left(\alpha_{CP}\right)$ which means
that $\sigma_3$ must vanish in order to explain the observed behaviour
which is in line with our numerical results in Table
\ref{table:sigma_i}.  From this we can conclude that there is no
mixing between the production of ${\cal CP}$-even and ${\cal CP}$-odd
Higgs bosons which is known for stable top quarks, see
e.g. Ref. \cite{Bahl:2020wee}.

In the full off-shell case we only observe the same symmetry if we set
$\kappa_{HVV} \left(\alpha_{CP}\right) = \cos
\left(\alpha_{CP}\right)$.  With this choice the fourth and sixth term
in (\ref{eq:xSec_alpha_dependence}) have the same $\cos^2
\left(\alpha_{CP} \right)$ dependence as the first one and the fifth
term becomes proportional to $\cos \left(\alpha_{CP} \right) \sin
\left(\alpha_{CP}\right)$.  As the symmetry of the cross-section under
the replacement $\alpha_{CP} \rightarrow \pi - \alpha_{CP}$ is clearly
visible in the blue curve of Figure \ref{fig:integrated}, we conclude
that both $\sigma_3$ and $\sigma_5$ must either vanish or cancel,
which is again in line with our numerical results. This is broken if
we take the other choice for our $HVV$ coupling, $\kappa_{HVV}=1$,
which corresponds to the green and orange curves in Figure
\ref{fig:integrated}.  In this case the fourth term becomes
proportional to just $\cos \left(\alpha_{CP} \right)$.  As the
corresponding coefficient is negative, we find that the cross-section
for $\alpha_{CP} = \pi$ is actually larger than the one for
$\alpha_{CP} = 0$ by about $0.14 $ fb which translates to an increase
of $7 \%$.
%
\begin{table*}[t!]
    \caption{\it Comparison of LO and NLO QCD integrated fiducial
        cross-sections as calculated  in the NWA, NWA with LO
        top-quark decays and full off-shell approach for $\alpha_{CP}
        = 0, \pi/4$ and $\pi/2$. Also shown are their respective scale
        uncertainties and Monte Carlo integration errors. All values
        are given for the $pp\to e^+\nu_e\, \mu^-\bar{\nu}_{\mu}\,
        b\bar{b}\, H$ process at the LHC with $\sqrt{s}=13\textrm{
          TeV}$.} 
    \label{table:integrated}
    
    \centering
    \renewcommand{\arraystretch}{1.5}
    \begin{tabular}{l@{\hskip 10mm}l@{\hskip 10mm}ll@{\hskip 10mm}l}
        \hline\noalign{\smallskip}
        $\alpha_{CP}$ &   & Off-shell & NWA & \makecell{Off-shell \\ effects } \\
        \noalign{\smallskip}\midrule[0.5mm]\noalign{\smallskip} \multirow{3}{*}{$0$ (SM)}
        & $\sigma_{\text{LO}}$ [fb] & $2.0313(2)^{+0.6275\,(31\%)}_{-0.4471\,(22\%)} $
        & $ 2.0388(2)^{+0.6290 \, (31\%)}_{-0.4483 \, (22\%)}$ & $ -0.37\%$\\
        & $\sigma_{\text{NLO}}$ [fb] & $ 2.466(2)^{+0.027 \, (1.1\%)}_{-0.112 \, (4.5\%)} $
        & $ 2.475(1)^{+0.027 \, (1.1\%)}_{-0.113 \, (4.6\%)} $ & $ -0.36\%$\\
        & $\sigma_{\text{NLO}_{\text{LOdec}}}$ [fb] & $ - $
        & $ 2.592(1)^{+0.161 \, (6.2\%)}_{-0.242 \, (9.3\%)} $ & \\
        \noalign{\smallskip}\hline\noalign{\smallskip}
        &  ${\cal K} = \sigma_{\text{NLO}} / \sigma_{\text{LO}}$    & $ 1.21 $
        & $ 1.21 $ (LOdec: $ 1.27$) & \\    
        \noalign{\smallskip}\midrule[0.5mm]\noalign{\smallskip}
      \multirow{3}{*}{$\pi / 4$}   & $\sigma_{\text{LO}}$ [fb]
        & $ 1.1930(2)^{+0.3742 \, (31\%)}_{-0.2656 \, (22\%)} $
        & $ 1.1851(1)^{+0.3707 \, (31\%)}_{-0.2633 \, (22\%)} $ & $ 0.66\%$\\
        &  $\sigma_{\text{NLO}}$ [fb] & $ 1.465(2)^{+0.016 \, (1.1\%)}_{-0.071 \, (4.8\%)} $
        & $ 1.452(1)^{+0.015 \, (1.0\%)}_{-0.069 \, (4.8\%)} $ & $ 0.89 \%$\\
        &  $\sigma_{\text{NLO}_{\text{LOdec}}}$ [fb] & $ - $
        & $  1.517(1)^{+0.097 \, (6.4\%)}_{-0.144 \, (9.5\%)} $ & \\
        \noalign{\smallskip}\hline\noalign{\smallskip}
        &  ${\cal K} = \sigma_{\text{NLO}} / \sigma_{\text{LO}}$
                          & $ 1.23 $ & $ 1.23 $ (LOdec: $ 1.28$) & \\      
        \noalign{\smallskip}\midrule[0.5mm]\noalign{\smallskip}
      \multirow{3}{*}{$\pi / 2$}   & $\sigma_{\text{LO}}$ [fb]
        & $ 0.38277(6)^{+0.13123\, (34\%)}_{-0.09121\, (24\%)} $
        & $ 0.33148(3)^{+0.11240\, (34\%)}_{-0.07835 \, (24\%)} $ & $ 13.4\%$\\
        &   $\sigma_{\text{NLO}}$ [fb]
        & $ 0.5018(3)^{+0.0083 \,(1.2\%)}_{-0.0337 \, (6.7\%)}$
        & $ 0.4301(2)^{+0.0035 \, (0.8\%)}_{-0.0264 \, (6.1\%)} $ & $ 14.3 \%$\\
        &   $\sigma_{\text{NLO}_{\text{LOdec}}}$ [fb] & $ - $
        & $ 0.4433(2)^{+0.0323 \, (7.3\%)}_{-0.0470 \, (11\%)} $ & \\
        \noalign{\smallskip}\hline\noalign{\smallskip}
        &  ${\cal K} = \sigma_{\text{NLO}} / \sigma_{\text{LO}}$
        & $ 1.31 $ & $ 1.30 $ (LOdec: $ 1.34$) & \\
        \noalign{\smallskip}\hline\noalign{\smallskip}
    \end{tabular}
\end{table*}
%

Let us also mention that the final term in equation
(\ref{eq:xSec_alpha_dependence}) becomes independent of $\alpha_{CP}$
when choosing $\kappa_{HVV}=1$.  This means that in this case, the
breaking of the symmetry around $\alpha_{CP} = \pi/2$ in the full
off-shell case is only a result of the fourth term. This contribution
corresponds to the interference between diagrams where the Higgs boson
is radiated off a $W$ or $Z$ boson ( single- and non-resonant
top-quark contributions) with diagrams where the Higgs boson is
produced from one of the top quarks (double- and single-resonant
top-quark contributions), see Figure \ref{fig:FD}.  As single- and
non-resonant Feynman diagrams are absent in the NWA, the dependence on
$\alpha_{CP}$ is symmetric in this case.  The very different behaviors
in the full off-shell treatment and the NWA lead us to conclude that
the inclusion of off-shell effects is vital, in particular when
considering large mixing angles.  Since the discrepancy between the
two approaches is mostly driven by the $\sigma_4$ term, their
difference increases with increasing mixing angle.

This observation also holds at NLO, as can be seen from the integrated
fiducial cross-sections listed in Table \ref{table:integrated}.  Here,
we compare full off-shell and NWA as well as LO and NLO cross-sections
for $\alpha_{CP} = 0, \pi/4$ and $\pi/2$.  We do this comparison only
for a few values of $\alpha_{CP}$ and fixed $\kappa_{HVV}$ and
$\kappa_{At\bar{t}}$ since equation (\ref{eq:xSec_alpha_dependence})
does not hold at one-loop level for the full off-shell calculation,
making a simple interpolation impossible at NLO.  Following the
argumentation of Ref. \cite{Demartin:2016axk}, we will henceforth use
$\kappa_{HVV} = 1$ and $\kappa_{At\bar{t}} = 2/3$ in order to be
consistent with ATLAS and CMS measurements since we focus here on a
potential non-zero mixing angle $\alpha_{CP}$ for the SM Higgs boson.

Just like at LO, the off-shell effects at NLO increase for larger
values of $\alpha_{CP}$.  For the SM case we find that the full
off-shell results at LO and NLO in QCD are slightly smaller than those
obtained using the NWA.  Specifically, the full top-quark and $W$
boson off-shell effects change the integrated LO and NLO fiducial
cross-section by about $0.4\%$. The finding is consistent with the
expected uncertainty of the NWA \cite{Fadin:1993kt}, which for
sufficiently inclusive observables is of the order of ${\cal
O}(\Gamma_t/m_t) \sim 0.8\%$.  It is the other way around for ${\cal
CP}$-mixed and ${\cal CP}$-odd Higgs boson production. In these two
cases the NWA results are smaller than the full off-shell
ones. Moreover, the full off-shell effects are of the order of
$0.7\%-0.9\%$ ($13\%-14\%$ ) for the ${\cal CP}$-mixed (${\cal
CP}$-odd) case.  Even at the level of integrated fiducial
cross-sections, off-shell effects for the pure ${\cal CP}$-odd case
are substantial and therefore easily distinguishable from the other
two cases. A similar pattern will be visible at the differential
cross-section level.

Concerning the NLO QCD corrections we find that these are positive for
all three values of $\alpha_{CP}$ and consistent between the NWA and
full off-shell results.  Similarly to the off-shell effects, they
increase with larger mixing angle, albeit not as drastically.  For the
${\cal CP}$-even and -mixed scenarios we find corrections of $21 \% -
23 \%$ while they are at the level of $30 \%-31\%$ for the ${\cal
CP}$-odd case.  Scale uncertainties for the ${\cal CP}$-even and
-mixed cases, taken as the maximum of the lower and upper bounds,
amount to $31 \%$ at LO.  After the inclusion of NLO QCD corrections,
they are reduced substantially to $5\%$.  For the ${\cal CP}$-odd case
theoretical uncertainties arising from the scale dependence are of the
order of $34 \%$ at LO and $6 \% - 7 \%$ at NLO.  Thus, by including
higher-order effects in $\alpha_s$ we have reduced the theoretical
error by a factor of $6$.

In Table \ref{table:integrated} we additionally provide results for
the NWA$_{\text{LOdec}}$, i.e. for the NWA with NLO QCD corrections to
the production stage and with LO top-quark decays (denoted as
$\sigma_{\text{NLO}_{\text{LOdec}}}$ in Table
\ref{table:integrated}). In line with previous findings for top-quark
pair associated production with $\gamma$, $H$, $W^\pm$ and $Z$ bosons
\cite{Hermann:2021xvs,Stremmer:2021bnk,Bevilacqua:2019quz,
Bevilacqua:2020pzy,Bevilacqua:2022nrm}
we observe that NWA$_{\text{LOdec}}$ predictions are higher compared
to the NLO QCD results in the full NWA. Specifically, for the process
at hand we have an increase of $3 \% - 5\%$. The difference is largest
for the $\cal CP$-even and smallest for the $\cal CP$-odd Higgs boson.
The corresponding $\cal{K}$-factors, defined as ${\cal K}=
\sigma_{\text{NLO}_{\text{LOdec}}} / \sigma_{\text{LO}}$, are included
in Table \ref{table:integrated} in parenthesis. We observe that not
only the size of higher-order corrections is amplified in this case,
but also the associated scale uncertainties are larger.  Due to the
lack of higher-order effects in top-quark decays the latter amount to
$9\% - 11\%$.

Finally, let us also mention that, as expected from $t\bar{t}H$
production with stable top quarks, we find that the mixing term
$\sigma_3$ in Eq. (\ref{eq:xSec_alpha_dependence}) between the ${\cal
CP}$-even and ${\cal CP}$-odd states vanishes also for the NLO QCD
predictions in the full NWA as well as for the case with LO top-quark
decays. Indeed, we can write
\begin{equation}
    \sigma(\alpha_{CP}=\pi/4) = \frac{1}{2} \left\{
      \sigma(\alpha_{CP}=0) + \sigma(\alpha_{CP}=\pi/2)\right\}\,,
\end{equation}
which is fulfilled within statistical uncertainties  for both cases
(see Table \ref{table:integrated}).

%
\section{Differential fiducial cross-sections}
\label{sec:tth-diff}
%

In this section we perform the comparison between the ${\cal CP}$-even
(SM), ${\cal CP}$-mixed and ${\cal CP}$-odd Higgs boson at the
differential level. In the first step, we analyse the impact of NLO
QCD corrections on various differential cross-section distributions
for the three ${\cal CP}$ cases. Subsequently, the size of off-shell
and higher-order effects in top-quark decays is discussed.
%
\begin{figure}[t!]
	\begin{center}
		\includegraphics[width=0.45\textwidth]{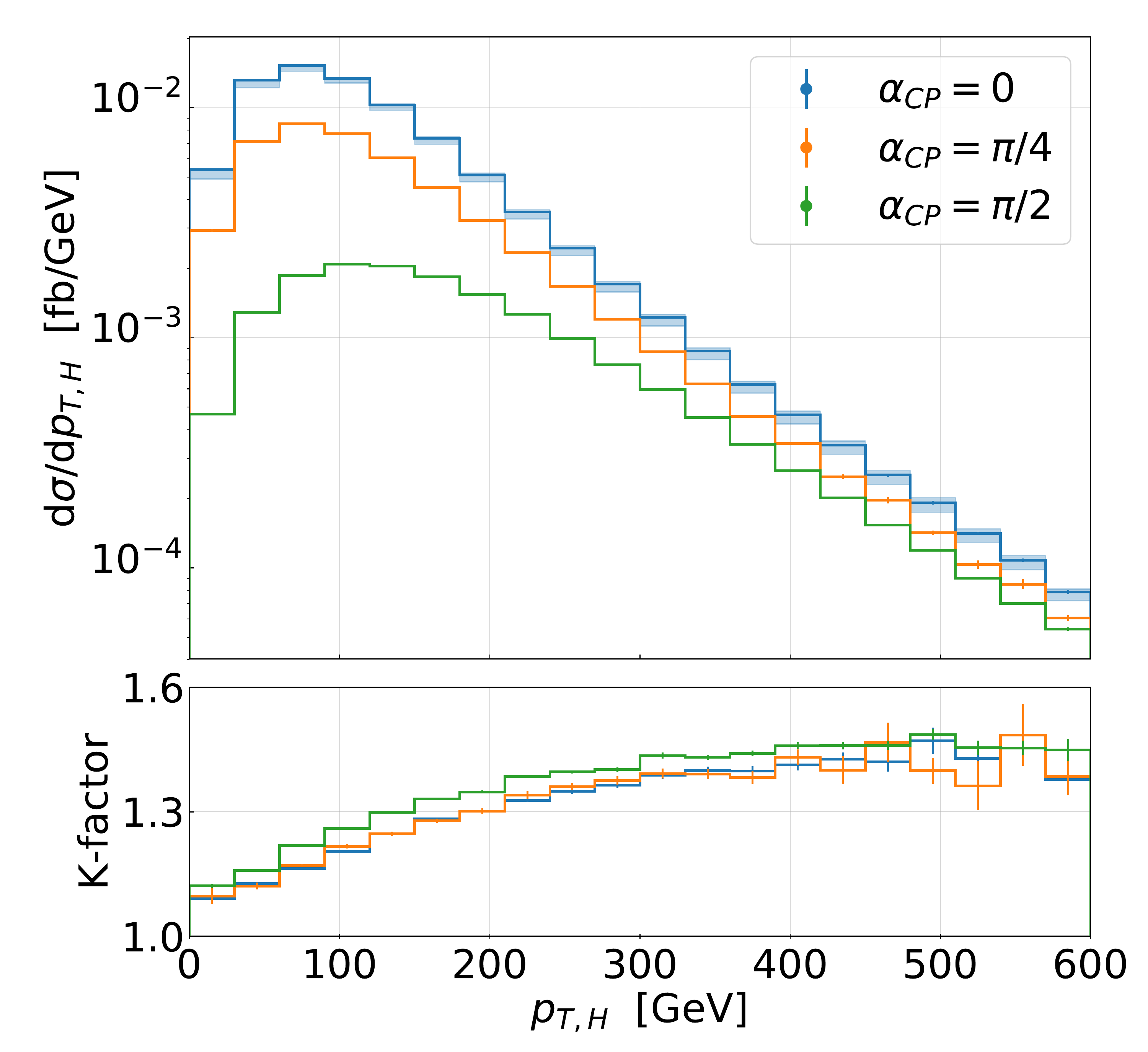}
		\includegraphics[width=0.45\textwidth]{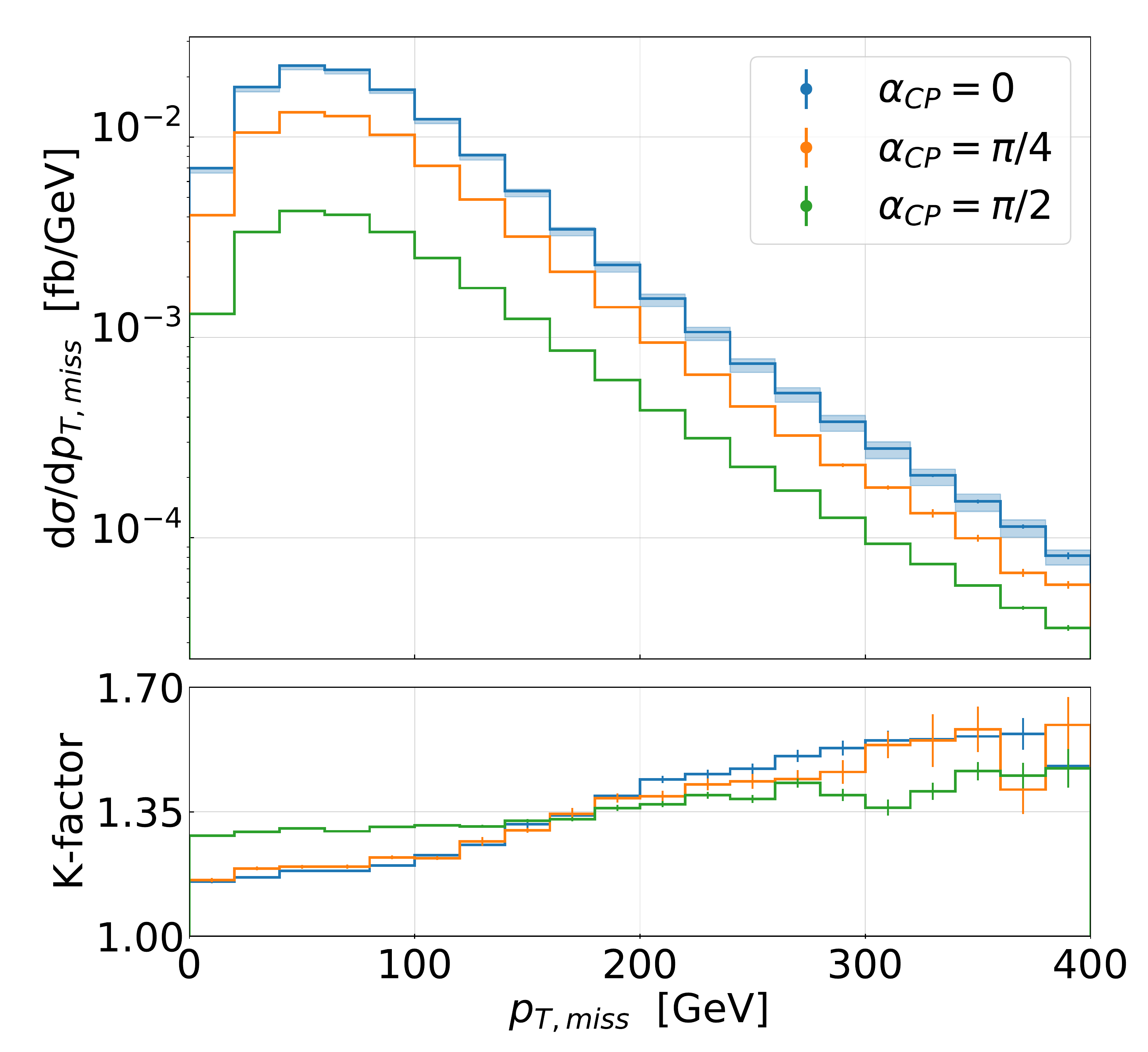}
		\includegraphics[width=0.45\textwidth]{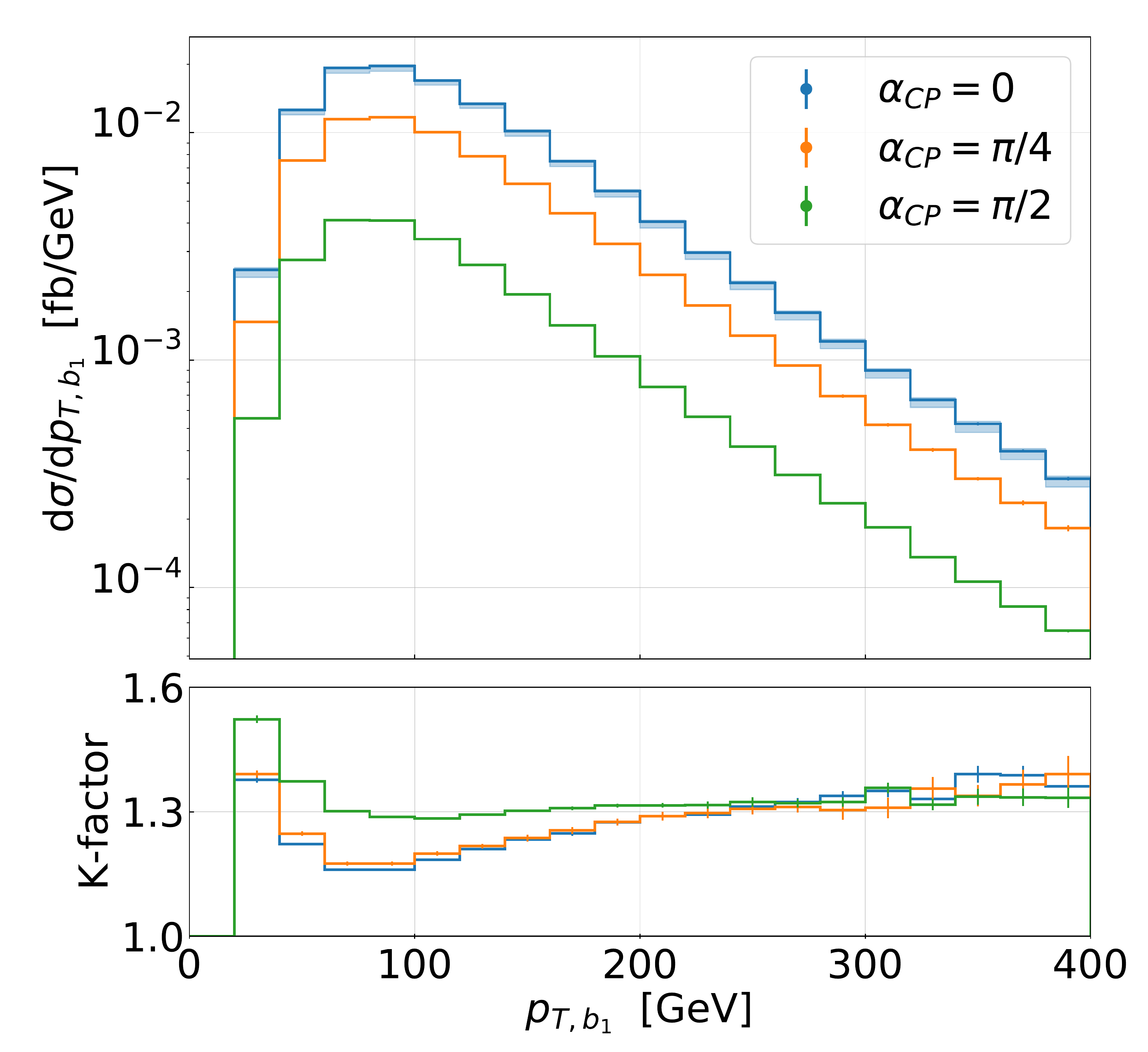}
		\includegraphics[width=0.45\textwidth]{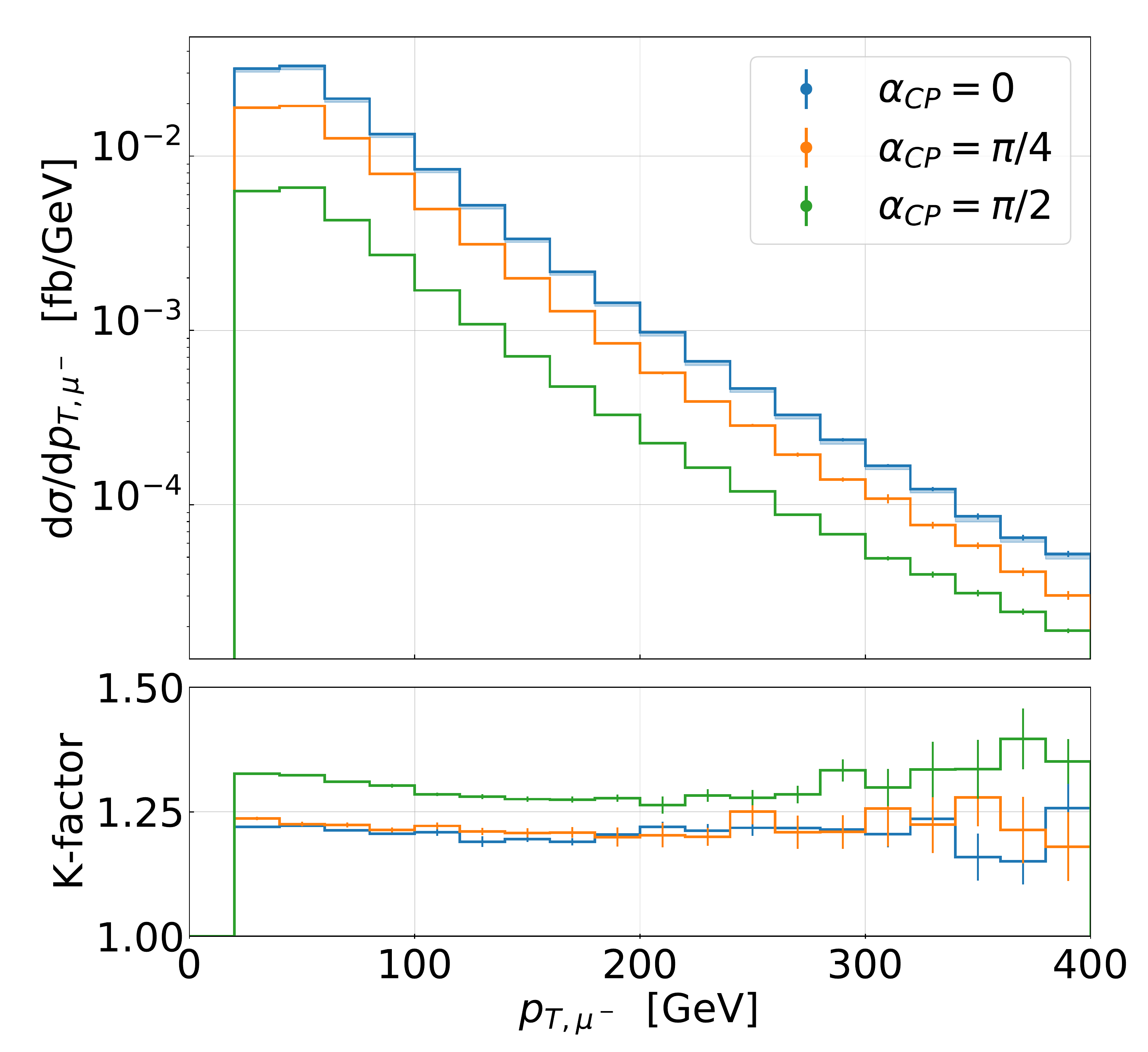}
	\end{center}
	\caption{\label{fig:kfac1a} \it
		Differential distributions at NLO in QCD for
                $\alpha_{CP}=0,\pi/4,\pi/2$ for the observables
                $p_{T,\,H}$, $p_{T,\,miss}$, $p_{T,\, b_1}$ and
                $p_{T,\,\mu^-}$ for the $pp\to
                e^+\nu_e\,\mu^-\bar{\nu}_{\mu}\,b\bar{b}\,H$ process
                at the LHC with $\sqrt{s}=13\textrm{ TeV}$. The lower
                panels show the differential
                $\mathcal{K}\textrm{-factor}$. Scale uncertainties are
                given for $\alpha_{CP}=0$ in the upper panel while
                Monte Carlo integration errors are displayed in both
                panels.}
\end{figure}
\begin{figure}[t!]
	\begin{center}
		\includegraphics[width=0.45\textwidth]{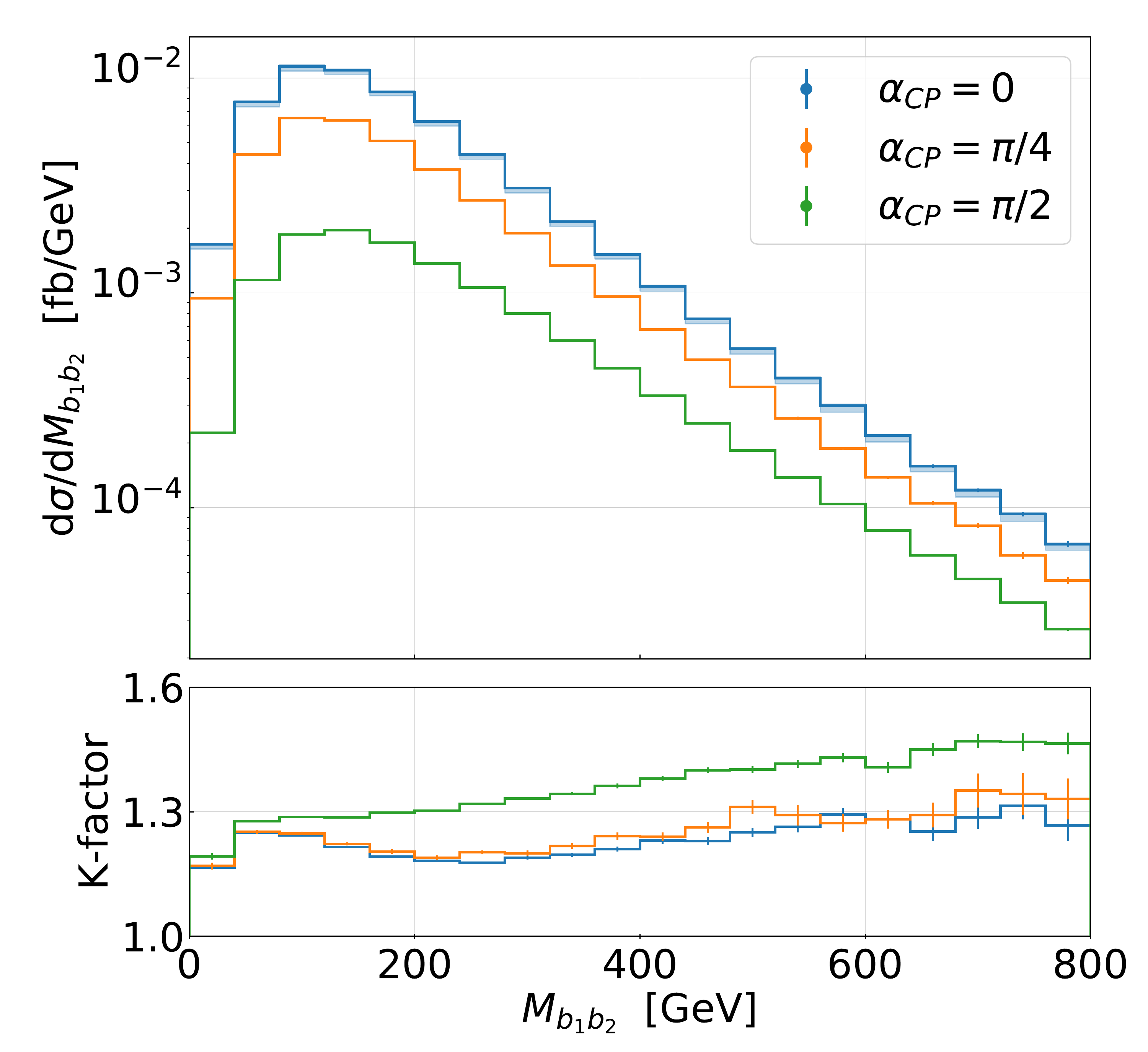}
		\includegraphics[width=0.45\textwidth]{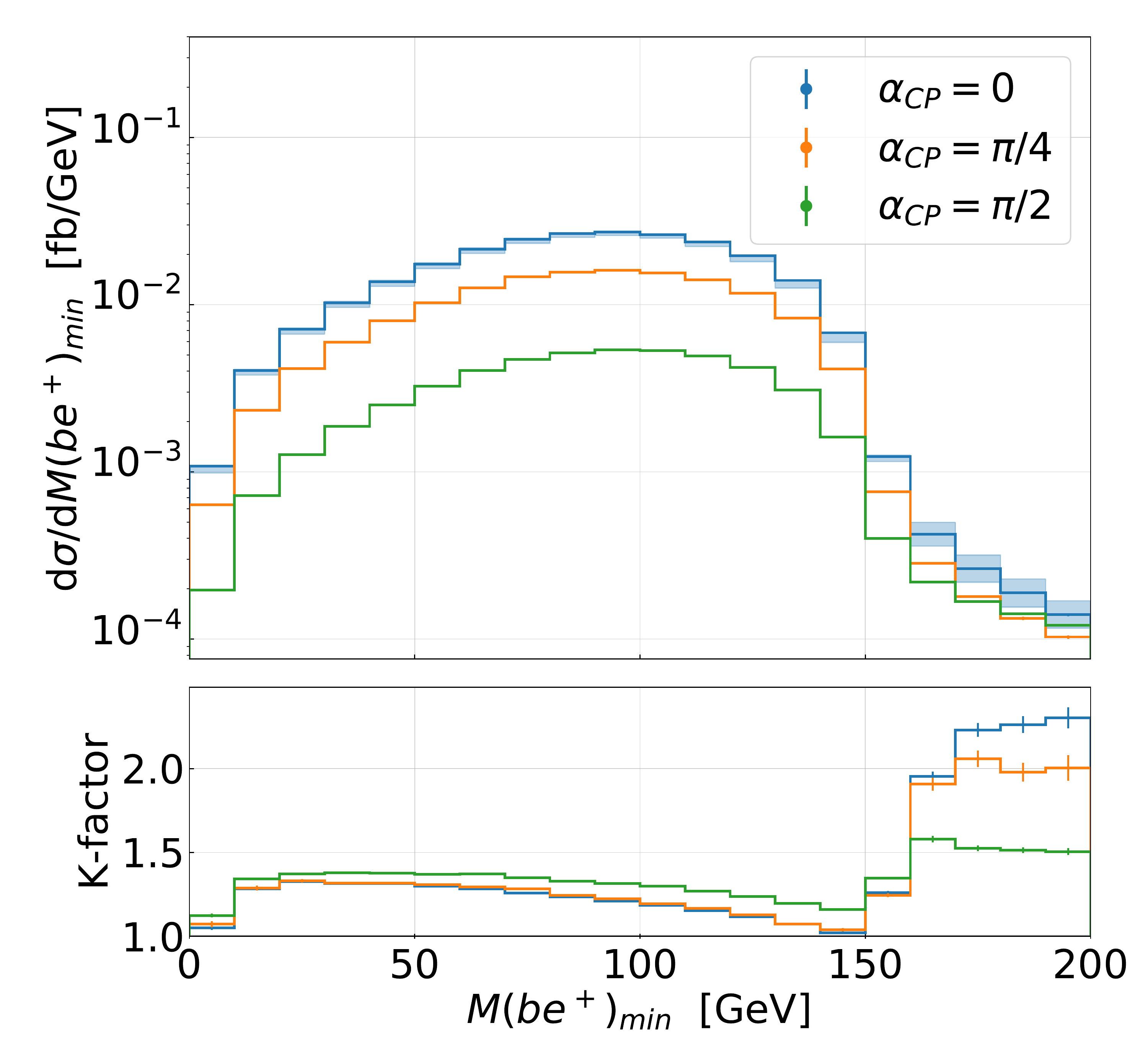}
	\end{center}
	\caption{\label{fig:kfac1b} \it
		Differential distributions at NLO in QCD for
                $\alpha_{CP}=0,\pi/4,\pi/2$  for the observables
                $M_{b_1b_2}$ and $M(be^+)_{min}$ for the $pp\to
                e^+\nu_e\,\mu^-\bar{\nu}_{\mu}\,b\bar{b}\,H$ process
                at the LHC with $\sqrt{s}=13\textrm{ TeV}$. The lower
                panels show the differential
                $\mathcal{K}\textrm{-factor}$. Scale uncertainties are
                given for $\alpha_{CP}=0$ in the upper panel while
                Monte Carlo integration errors are displayed in both
                panels.}
\end{figure}
%

In Figures \ref{fig:kfac1a} and \ref{fig:kfac1b} we present
differential distributions for dimensionful observables, specifically
the transverse momentum of the Higgs boson ($p_{T, \,H}$), the total
missing transverse momentum ($p_{T,\, miss}$), the transverse momentum
of the hardest $b$-jet ($p_{T, \,b_1}$), the transverse momentum of
the muon ($p_{T, \,\mu^-}$), the invariant mass of the two $b$-jets
($M_{b_1b_2}$) and the minimum mass of the positron and a $b$-jet
($M(be^+)_{min}$). In the upper panels of each plot the differential
cross-section distributions are given for the SM, ${\cal CP}$-mixed
and ${\cal CP}$-odd Higgs boson. Moreover, as reference, we display
the theoretical uncertainties obtained by scale variation for the SM
Higgs boson. The lower panels show the differential
$\mathcal{K}\textrm{-factor}$ for all three $\cal{CP}$ states. For
$p_{T,\,H}$ we observe that the relative QCD corrections are very
similar for the SM, ${\cal CP}$-mixed and ${\cal CP}$-odd Higgs
boson. They are about $9\%-12\%$ at the beginning of the spectrum,
whereas towards the tails they increase up to $45\%$. In contrast, the
QCD corrections for $p_{T,\, miss}$ behave quiet differently for the
${\cal CP}$-odd Higgs boson and increase from $28\%$ to $45\%$ for
increasing transverse momenta. For the other two ${\cal CP}$
configurations, the QCD corrections rise more drastically from $15\%$
at the beginning up to $60\%$ in the tail. Generally, in the case of
$t\bar{t}$ production, $p_T$ observables constructed from the decay
products of both top quarks, like for example $p_{T\, miss}$, $p_{T\,
b_1b_2}$ and $p_{T\, e^+\mu^-}$, are known to have huge ${\cal
K}$-factors in the tails, see
e.g. \cite{Denner:2012yc,Czakon:2020qbd}. At LO these observables
exhibit a strong suppression of the $t\bar{t}$ cross-section above
$150$ GeV. In the case of the dominant double-resonant top-quark
contributions, top-quark decay products are boosted via their parents,
which have opposite momenta. Even though the transverse momentum of
the particular decay products can be substantial, the $p_T$ of the
$bb$, $e^+\mu^-$ and $\nu_e \bar{\nu}_\mu$ systems can only acquire
rather small values. At NLO the $t\bar{t}$ system can attain large
transverse momentum by recoiling against extra jet radiation. Thus,
the kinematical constraint is partially lifted resulting in an
enhancement of the NLO cross-section and a huge ${\cal K}$-factor of
$2-4$. For the $t\bar{t}H$ production, where radiation of the Higgs
boson off top quarks is already taking place at LO, smaller NLO
corrections are foreseen. Indeed, the latter generate ${\cal
K}$-factors in the range of $1.6-1.8$
\cite{Denner:2015yca,Stremmer:2021bnk}. Additionally, the harder Higgs
boson spectrum in the ${\cal CP}$-odd case leads to a relaxation of
the kinematical suppression already at LO for $p_{T,\,miss}$. Similar
higher-order effects are observed for $p_{T,\, b_1b_2}$ and $p_{T,\,
e^+\mu^-}$. For $p_{T,\,b_1}$ we again find an enhancement of the NLO
QCD corrections at the beginning of the spectrum for the ${\cal
CP}$-odd Higgs boson leading to effects up to $52\%$ compared to
$40\%$ for the other ${\cal CP}$ configurations. These differences
decrease towards the high $p_T$-region and we observe
$\mathcal{O}(\alpha_s)$ corrections up to $40\%$ for all three
cases. For $p_{T,\,\mu^-}$ we find NLO QCD corrections of about
$20\%-25\%$ for the SM and ${\cal CP}$-mixed Higgs boson while they
increase up to $25\%-35\%$ for the ${\cal CP}$-odd one. For all three
${\cal CP}$ configurations, the higher-order effects mainly impact the
normalisation and we do not observe significant shape distortions.
The QCD corrections for the differential distribution $M_{b_1b_2}$ are
very similar for all three ${\cal CP}$ configurations at the beginning
of the spectrum and amount to $25\%-28\%$. Towards the tail they
increase up to $45\%$ for the ${\cal CP}$-odd Higgs boson and up to
$30\%-35\%$ for the other two ${\cal CP}$ configurations.

In the case of on-shell top quarks and $W$ gauge bosons, the
observable $M(be^+)_{min}$ is characterised by a sharp upper bound,
$\sqrt{m_t^2-m_W^2}\approx 153$ GeV, which renders it very sensitive
to the top-quark mass.  Due to additional radiation, this kinematical
edge is smeared at NLO in QCD for the NWA. For the off-shell
prediction, small but non-negligible single- and non-resonant
contributions that elude the kinematical bound around $153$ GeV are
already present at LO. Thus, this observable is particularly
interesting for investigating the top-quark modeling. At NLO this
feature becomes more pronounced because of QCD radiation that enters
the $b$-jet without being emitted from its parent $b$ quark. As a
result, we find very significant higher-order effects both below and
above this kinematical bound. Additionally, we observe that the QCD
corrections are very similar for small invariant masses while towards
the kinematical edge around $153$ GeV the ${\cal CP}$-odd Higgs boson
leads to $12\%-15\%$ larger QCD corrections. The QCD corrections
increase significantly for all three ${\cal CP}$ configurations for
invariant masses above this edge but their size differs
substantially. We find the smallest QCD corrections in this region for
the ${\cal CP}$-odd Higgs boson at abound $50 \%$ while they increase
even further to $100\%$ for the ${\cal CP}$-mixed case and to $130\%$
for the SM Higgs boson which means that the enhancement due to real
radiation above this kinematical edge is significantly reduced in the
${\cal CP}$-odd case. As we have already seen at the level of
integrated cross-sections (Table \ref{table:integrated}), the relative
size of full off-shell effects and, in turn, the contribution from
$HVV$ couplings, are significantly larger than for the other two
${\cal CP}$ states. Also here the single- and non-resonant
contributions play a crucial role since they are less affected by this
kinematical edge and thus typically receive smaller QCD corrections
than the double-resonant part in this phase-space region.
%
\begin{figure}[t!]
	\begin{center}
		\includegraphics[width=0.43\textwidth]{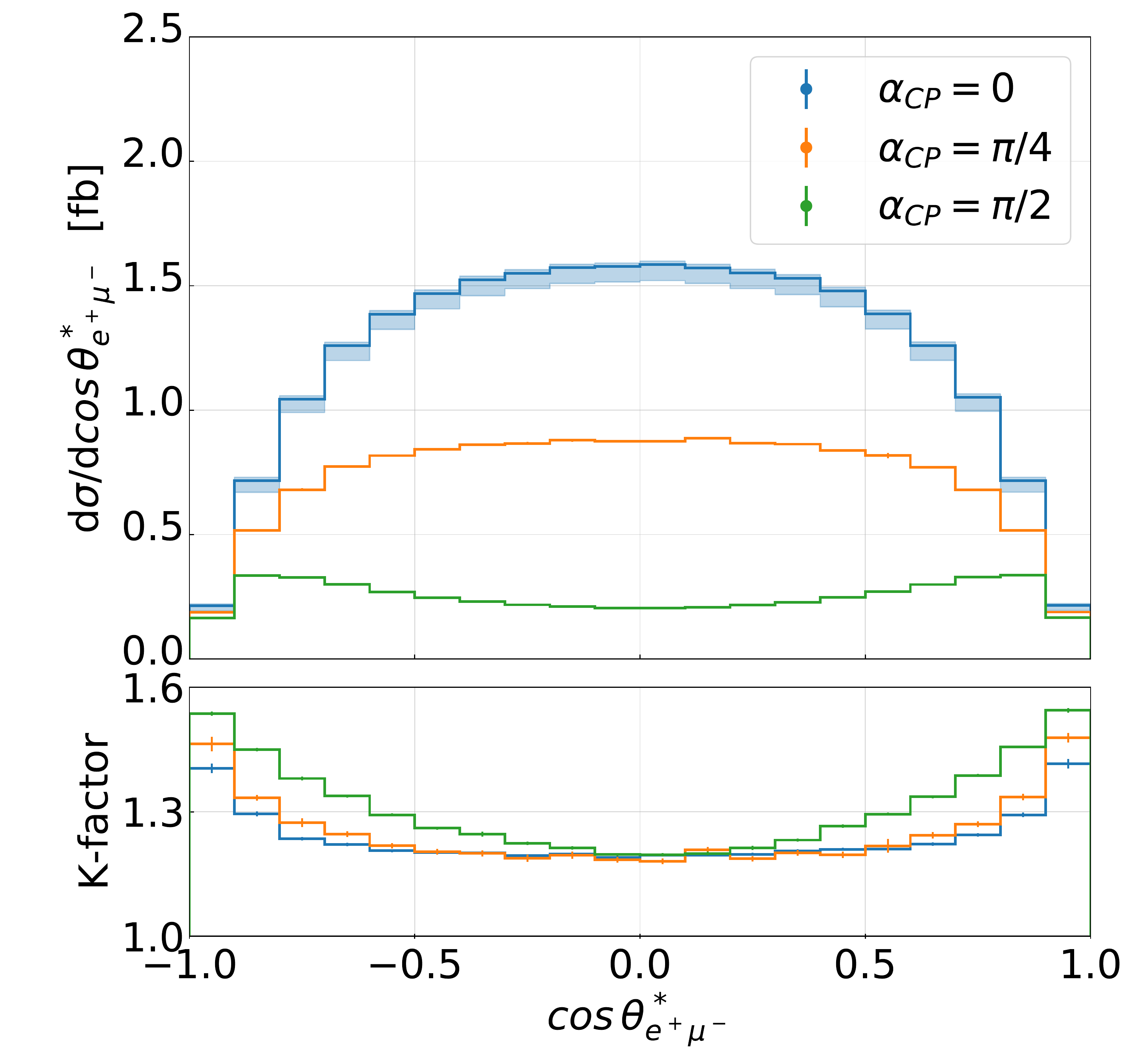}
		\includegraphics[width=0.43\textwidth]{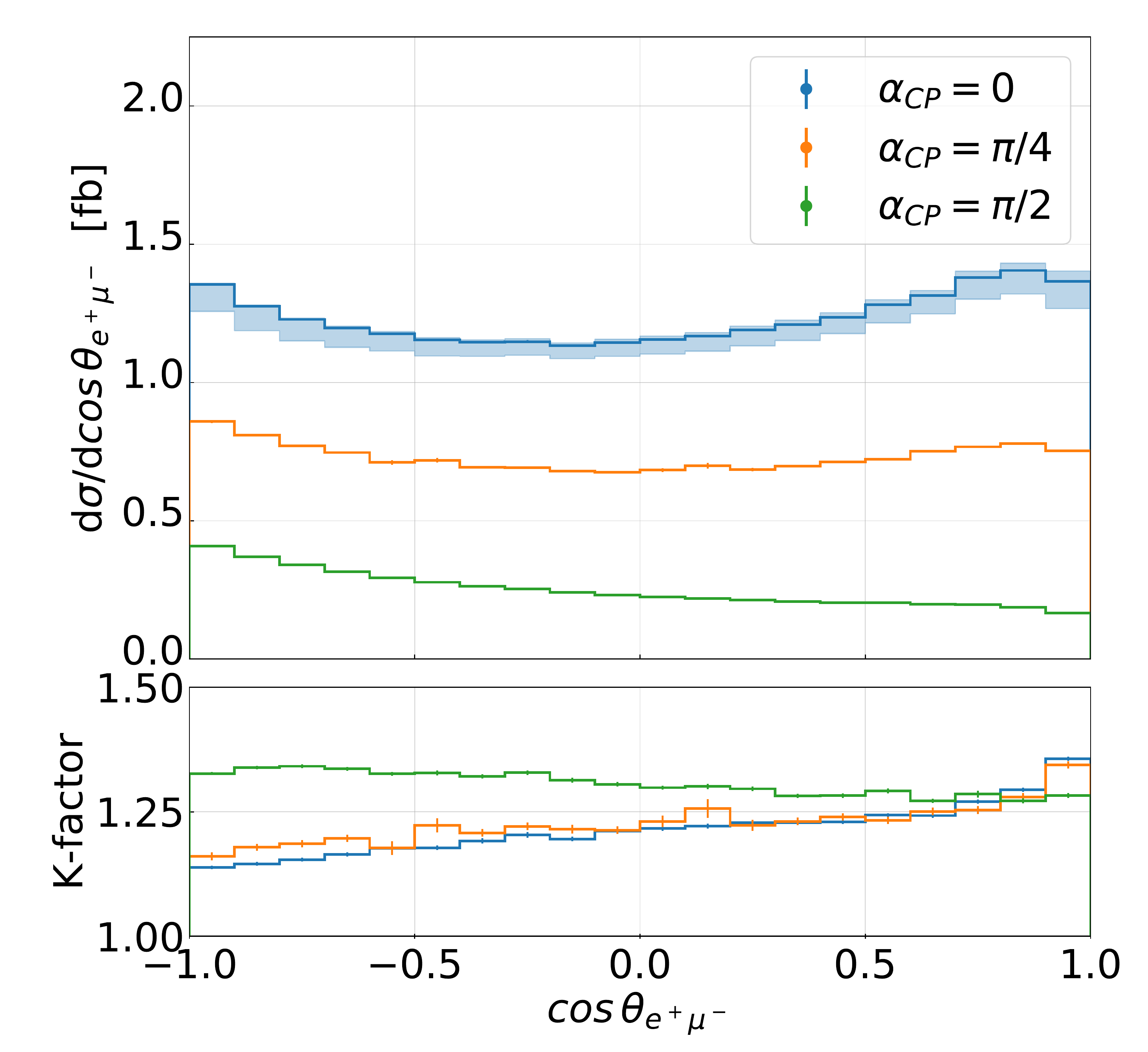}
		\includegraphics[width=0.43\textwidth]{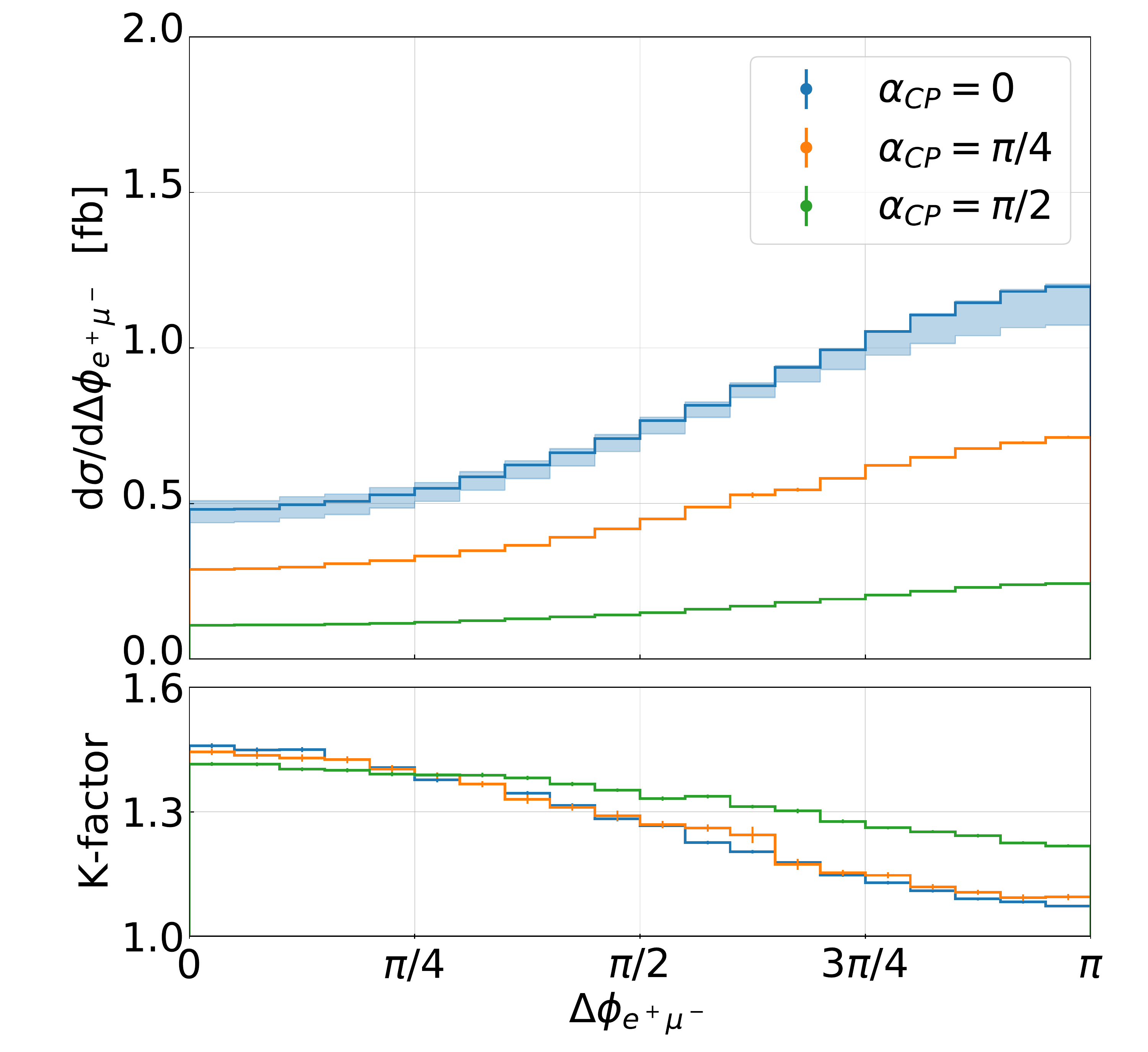}
		\includegraphics[width=0.43\textwidth]{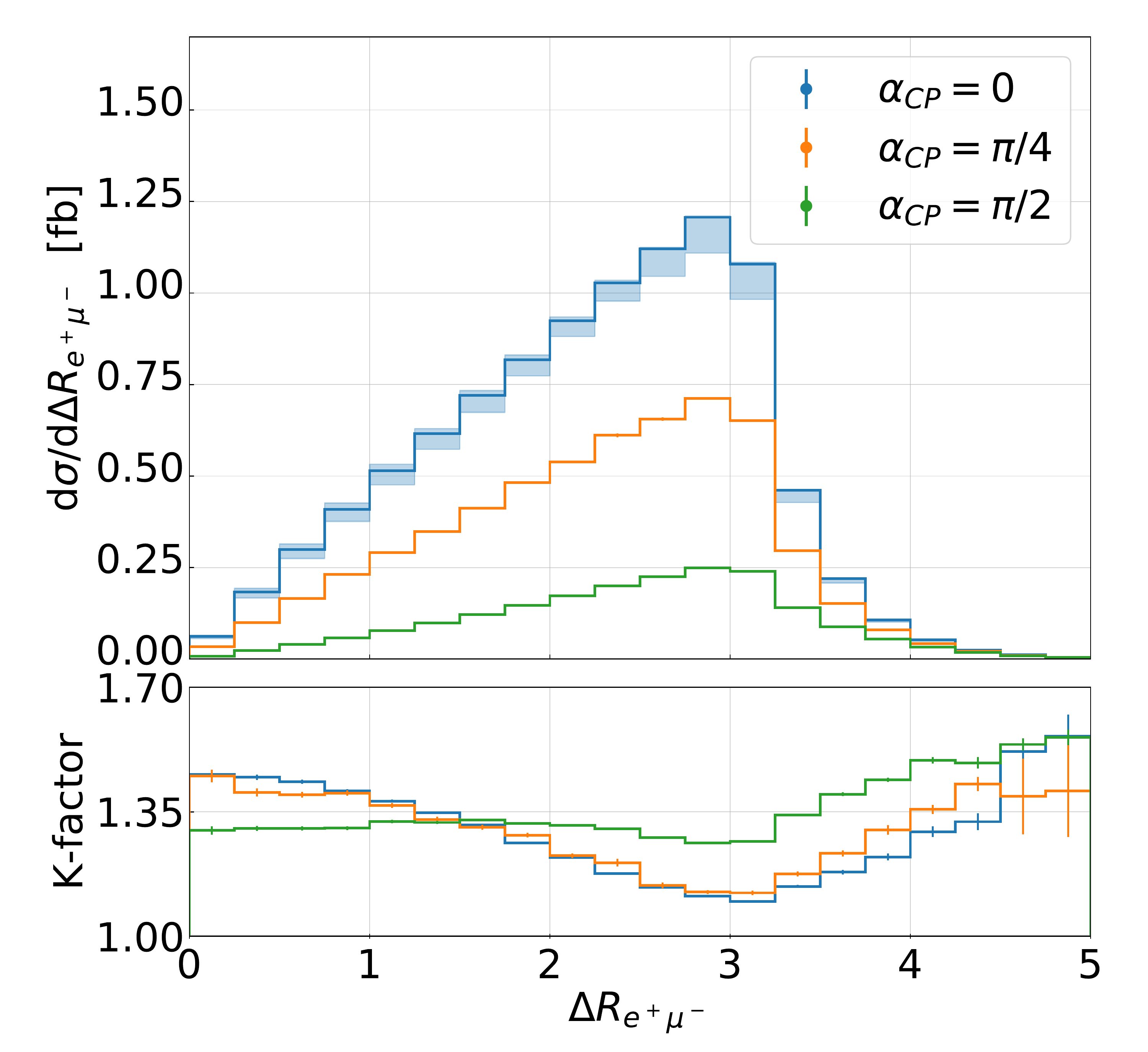}
	\end{center}
	\caption{\label{fig:kfac2a} \it
		Differential distributions at NLO in QCD for
                $\alpha_{CP}=0,{\pi}/{4}, {\pi}/{2}$ for the
                observables $\cos\theta^*_{e^+\mu^-}$,
                $\cos\theta_{e^+\mu^-}$, $\Delta\phi_{e^+\mu^-}$ and
                $\Delta R_{e^+\mu^-}$ for the $pp\to e^+\nu_e\,
                \mu^-\bar{\nu}_{\mu}\, b\bar{b}\, H$ process at the
                LHC with $\sqrt{s}=13\textrm{ TeV}$. The lower panels
                show the differential
                $\mathcal{K}\textrm{-factor}$. Scale uncertainties are
                given for $\alpha_{CP}=0$ in the upper panel while
                Monte Carlo integration errors are displayed in both
                panels.}
\end{figure}
\begin{figure}[t!]
	\begin{center}
		\includegraphics[width=0.43\textwidth]{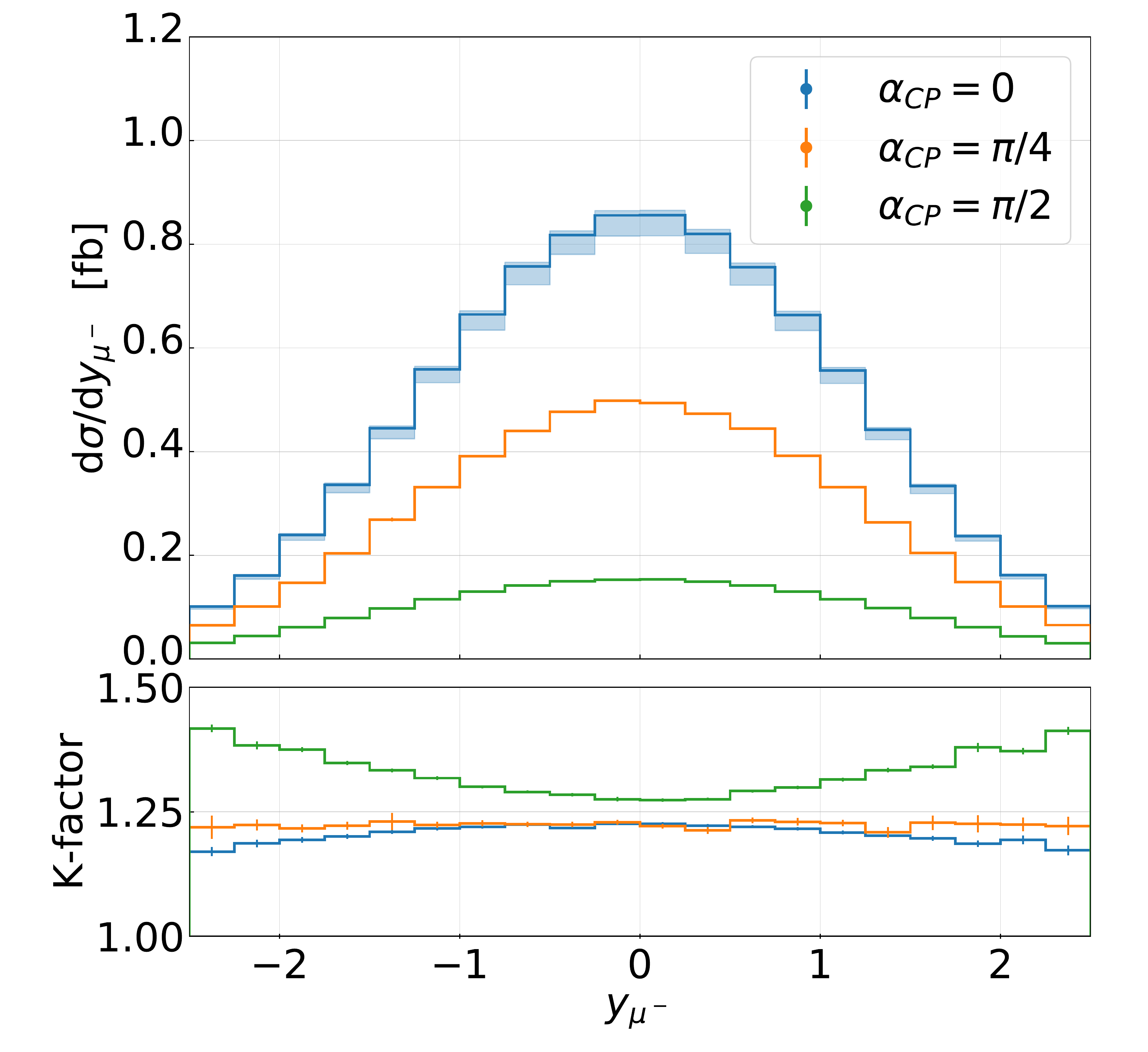}
		\includegraphics[width=0.43\textwidth]{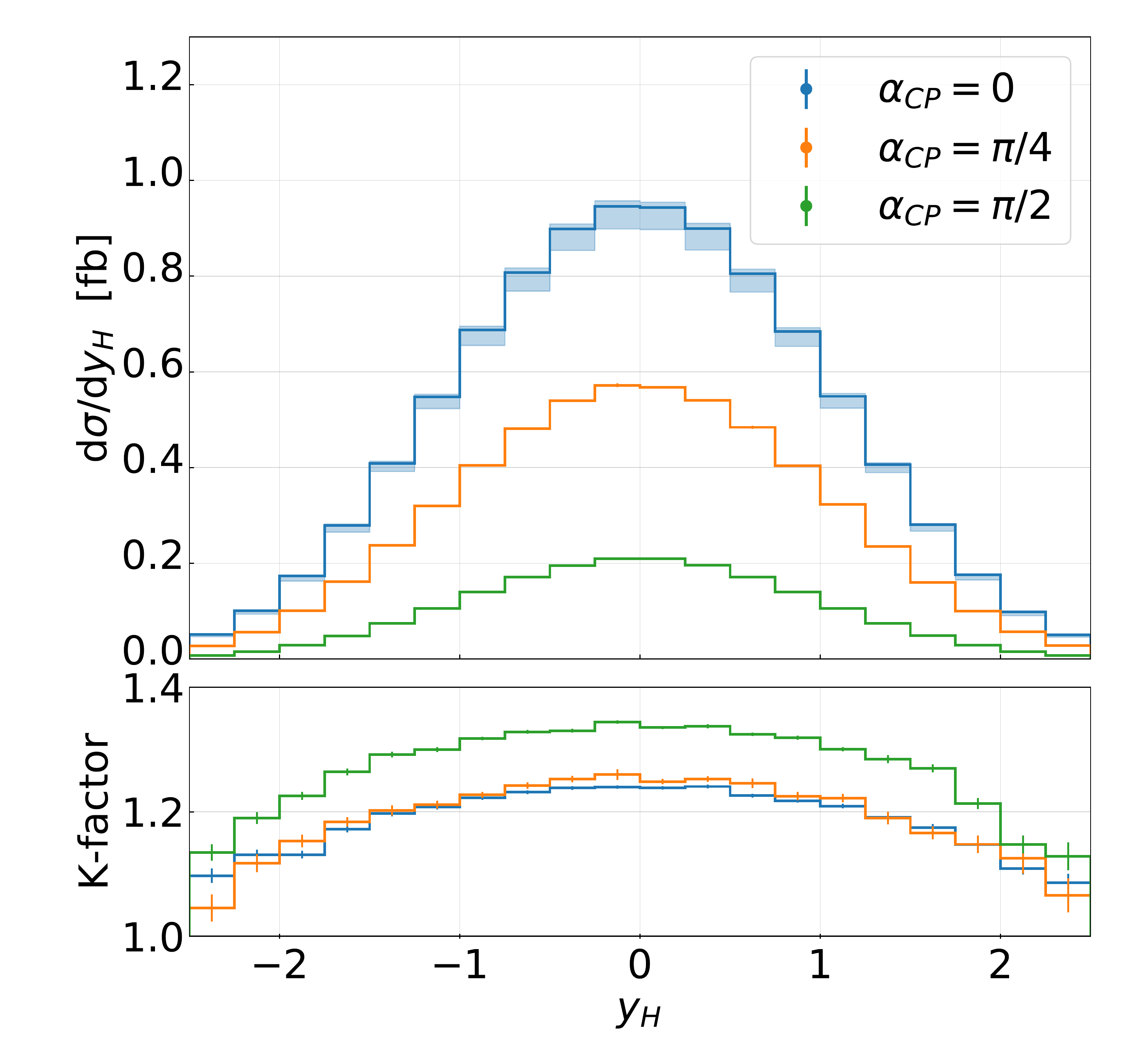}
	\end{center}
	\caption{\label{fig:kfac2b} \it
		Differential distributions at NLO in QCD for
                $\alpha_{CP}=0,{\pi}/{4}, {\pi}/{2}$ for the
                observables $y_{\mu^-}$ and $y_H$ for the $pp\to
                e^+\nu_e\, \mu^-\bar{\nu}_{\mu}\, b\bar{b} \, H$
                process at the LHC with $\sqrt{s}=13\textrm{
                  TeV}$. The lower panels show the differential
                $\mathcal{K}\textrm{-factor}$. Scale uncertainties are
                given for $\alpha_{CP}=0$ in the upper panel while
                Monte Carlo integration errors  are displayed in both
                panels.}
\end{figure}
%

In Figures \ref{fig:kfac2a} and \ref{fig:kfac2b} we show the
differential distributions and differential
$\mathcal{K}\textrm{-factors}$ for the dimensionless observables
$\cos\theta^*_{e^+\mu^-}$, the cosine of the opening angle between the
two charged leptons ($\cos\theta_{e^+\mu^-}$), the azimuthal angle
between the two charged leptons ($\Delta\phi_{e^+\mu^-}$), the
distance in the azimuthal angle rapidity plane between two charged
leptons ($\Delta R_{e^+\mu^-}$), the rapidity of the muon
($y_{\mu^-}$) and the rapidity of the Higgs boson ($y_H$). The
observable $\cos\theta^*_{e^+\mu^-}$ has been introduced in
Ref. \cite{Barr:2005dz} and is defined as
\begin{equation}
  \cos\theta^*_{e^+\mu^-}=\textrm{tanh}\left(\frac{y_{e^+}-y_{\mu^-}}{2}\right)\,.
\end{equation}
It should be distinguished from $\cos\theta_{e^+\mu^-}= \hat{p}_{e^+}
\cdot \,\hat{p}_{\mu^-}$ which we also use in the following. We note
that $\cos\theta^*_{e^+\mu^-}$ can be used to determine the ${\cal
CP}$ nature of spin-0 mediators in associated production of dark
matter and $t\bar{t}$ pairs, see
e.g. \cite{Haisch:2016gry,Hermann:2021xvs}. For this observable we
find that the NLO QCD corrections for the SM and ${\cal CP}$-mixed
Higgs boson again behave rather similarly. They amount to about $20\%$
for $\cos\theta^*_{e^+\mu^-}\approx 0$ and increase towards
$\cos\theta^*_{e^+\mu^-}\approx \pm 1$ up to $40\%$ for the SM and
$45\%$ for the ${\cal CP}$-mixed case. We find the same general
behaviour for the ${\cal CP}$-odd Higgs boson. However, in this case
the higher-order corrections can even reach $55 \%$. The differences
between LO and NLO distributions for $\cos\theta_{e^+\mu^-}$ are very
similar for all three ${\cal CP}$ configurations for small opening
angles between the two charged leptons and amount to $30\%-35\%$ for
these configurations. In the case of the SM and ${\cal CP}$-mixed
Higgs boson they decrease for increasing opening angle down to $15\%$
while they remain almost constant for the ${\cal CP}$-odd case. The
rather similar behaviour of the NLO QCD corrections in the ${\cal
CP}$-even and ${\cal CP}$-mixed Higgs boson cases is also present in
the $\Delta\phi_{e^+\mu^-}$ distributions. Here, the corrections
decrease with rising opening angle from about $40\%-45\%$ around
$\Delta\phi_{e^+\mu^-} \approx 0$ to $10\%$ for $\Delta\phi_{e^+\mu^-}
\approx \pi$. For the ${\cal CP}$-odd Higgs boson, higher-order
corrections lead to similar differences between the LO and NLO
distributions for small angles, but these effects decrease less
substantially and only down to $20\%$ for larger angles. Thus, shape
distortions are less pronounced for the pure pseudo-scalar Higgs boson
case.  For $\Delta R_{e^+\mu^-}$, the $\mathcal{O}(\alpha_s)$
corrections for the ${\cal CP}$-odd Higgs boson are rather flat up to
about $\Delta R_{e^+\mu^-}\approx 3$, whereas for $\Delta R_{e^+\mu^-}
> 3$ they increase from $30\%$ to $55\%$. On the other hand, we
observe significant shape distortions over the whole range for the SM
and ${\cal CP}$-mixed Higgs boson where the NLO QCD corrections vary
between $10 \%$ and $45\%-55\%$.  Finally, for both rapidity
distributions $y_{\mu^-}$ and $y_H$, the higher-order QCD effects for
the ${\cal CP}$-odd case exceed the ones for the SM and ${\cal
CP}$-mixed Higgs boson over the entire range by up to $20\%$ for
$y_{\mu^-}$ and $10\%$ for $y_H$. The SM and ${\cal CP}$-mixed cases
obtain higher-order corrections of about $20\%-25\%$ for $y_{\mu^-}$
while they vary between $10\%$ and $25\%$ for $y_H$.

Concluding, the NLO QCD corrections for the SM (${\cal CP}$-even) and
${\cal CP}$-mixed case are comparable for most of the observables we
have examined. A similar behaviour has already been found at the
integrated fiducial level for these two cases. Just like at the
integrated level, the $\mathcal{O}(\alpha_s)$ corrections for the
${\cal CP}$-odd Higgs are larger than for the other two cases. Indeed,
the harder Higgs-boson radiation for the pure pseudo-scalar case as
compared to the SM Higgs boson has an important impact on the size and
shape of the higher-order QCD corrections for dimensionful and
dimensionless observables. This is especially visible for observables
which obtain large NLO QCD corrections due to additional radiation.
%
\begin{figure}[t!]
	\begin{center}
		\includegraphics[width=0.49\textwidth]{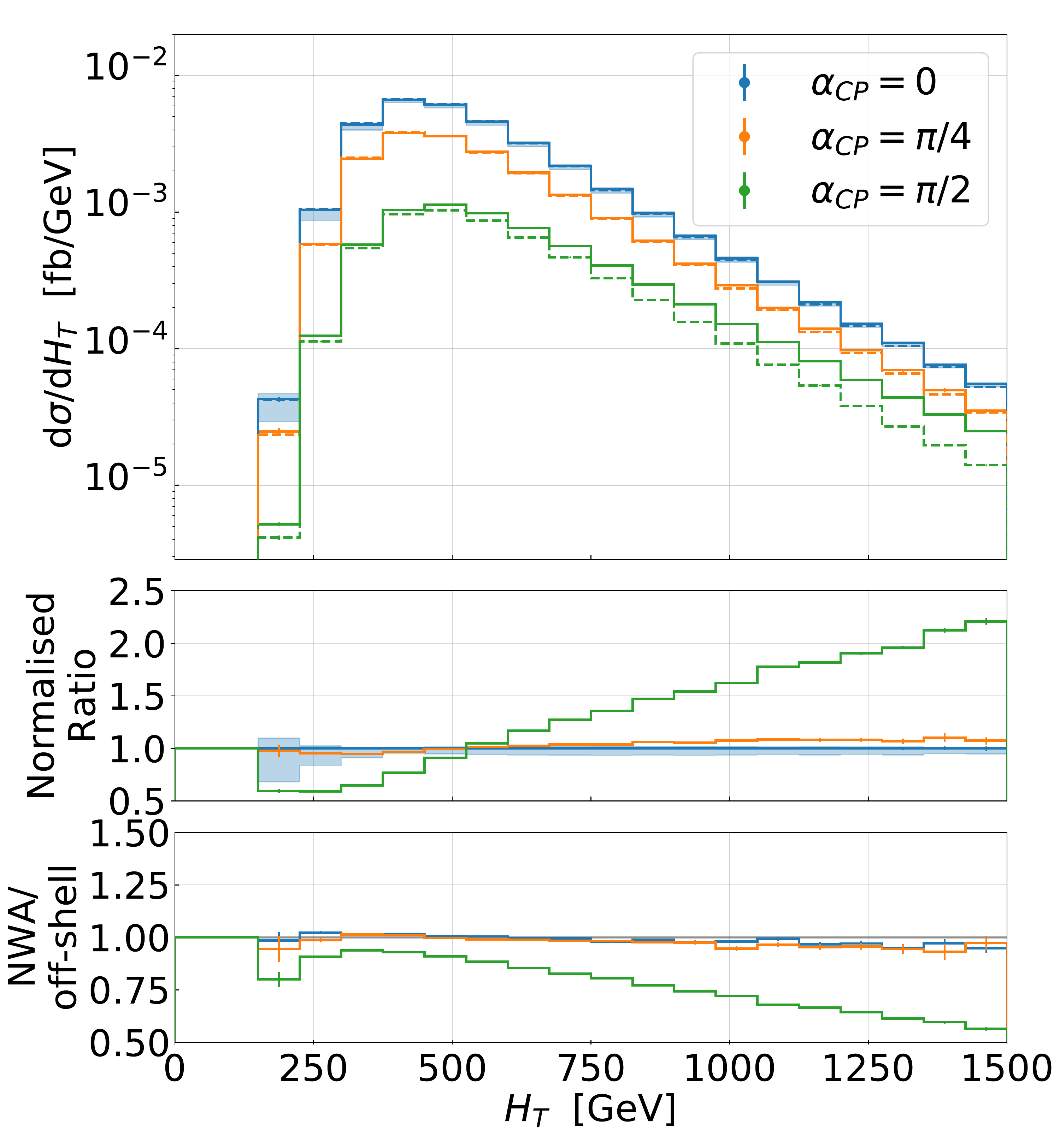}
		\includegraphics[width=0.49\textwidth]{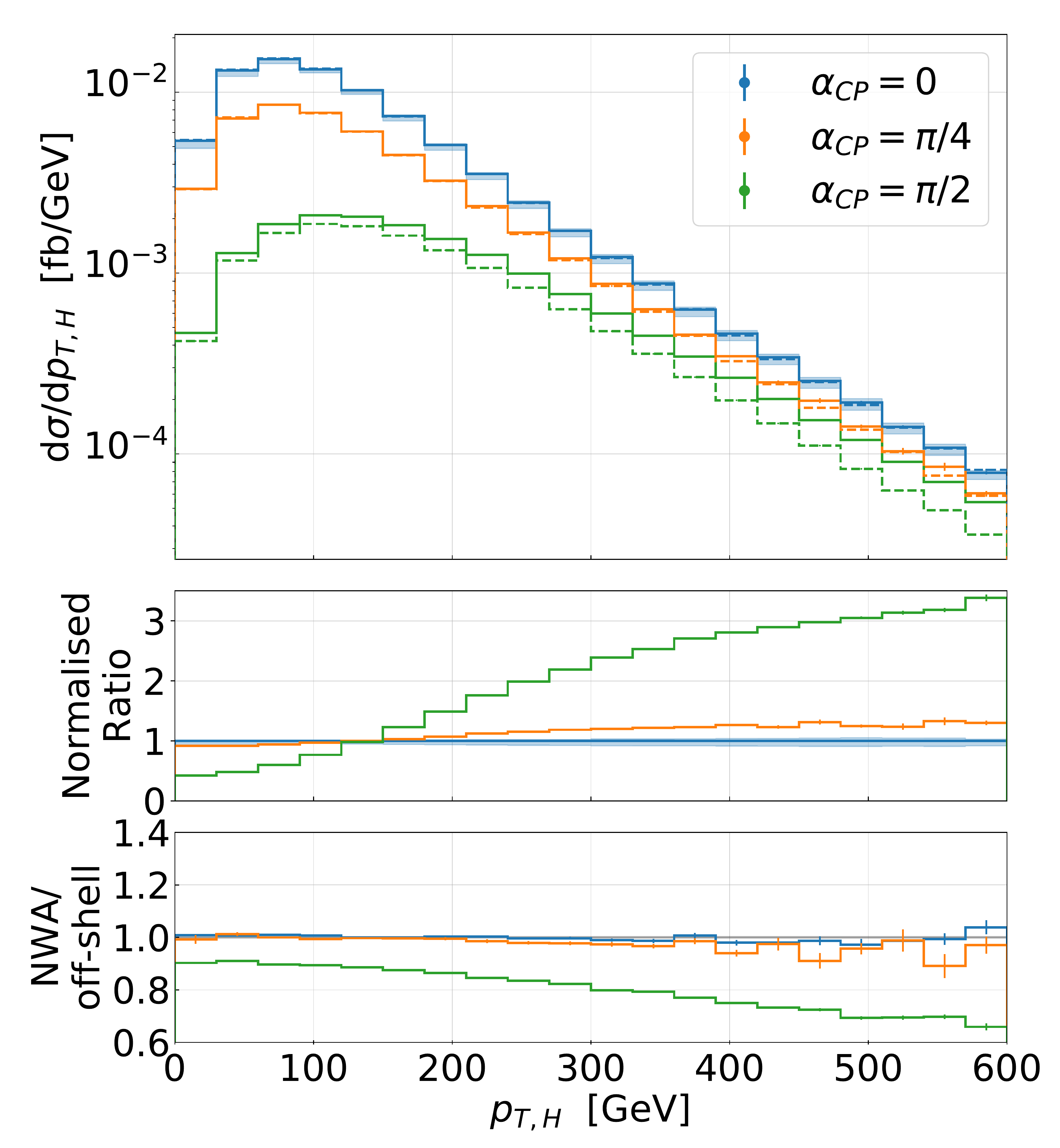}
		\includegraphics[width=0.49\textwidth]{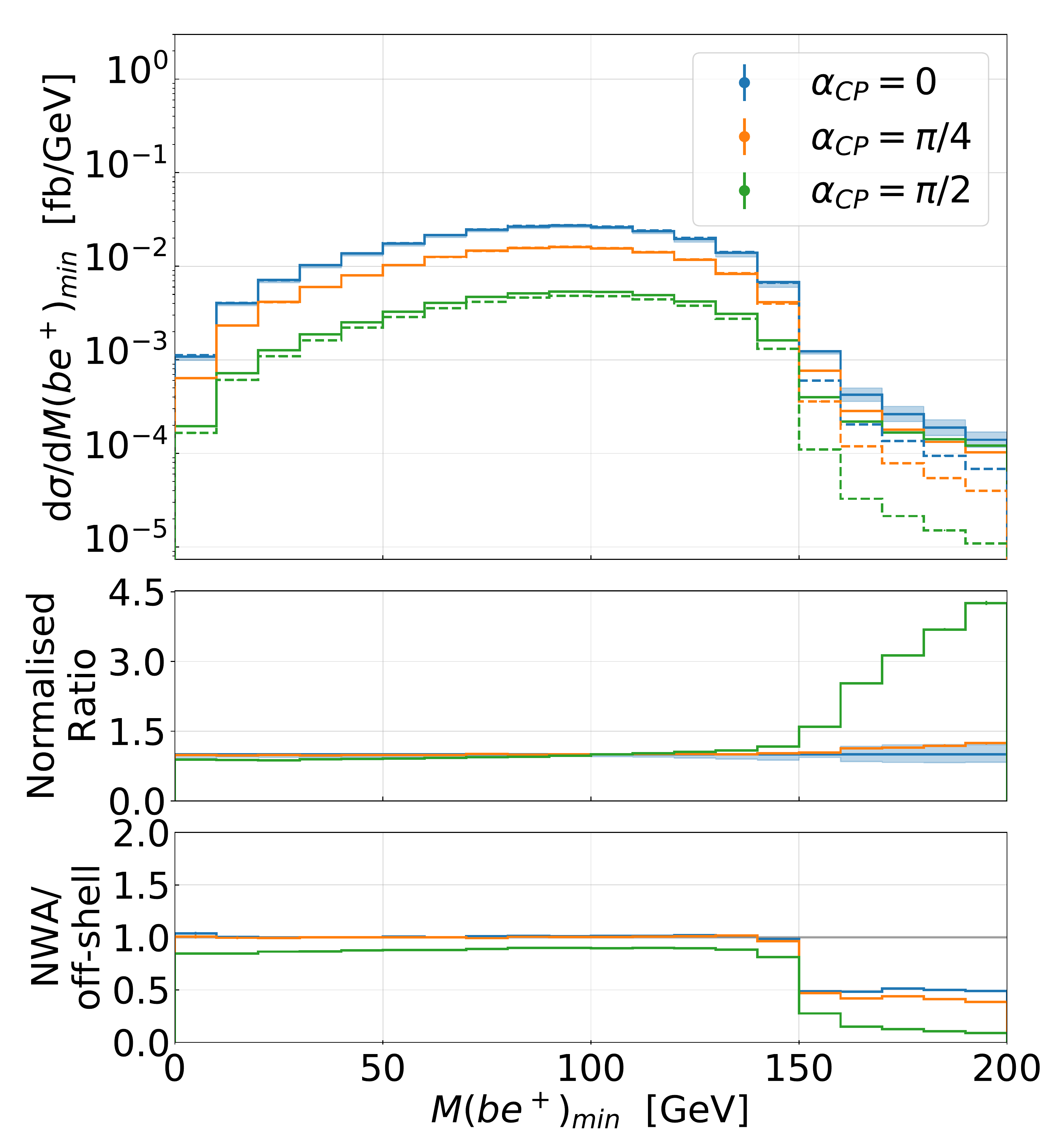}
		\includegraphics[width=0.49\textwidth]{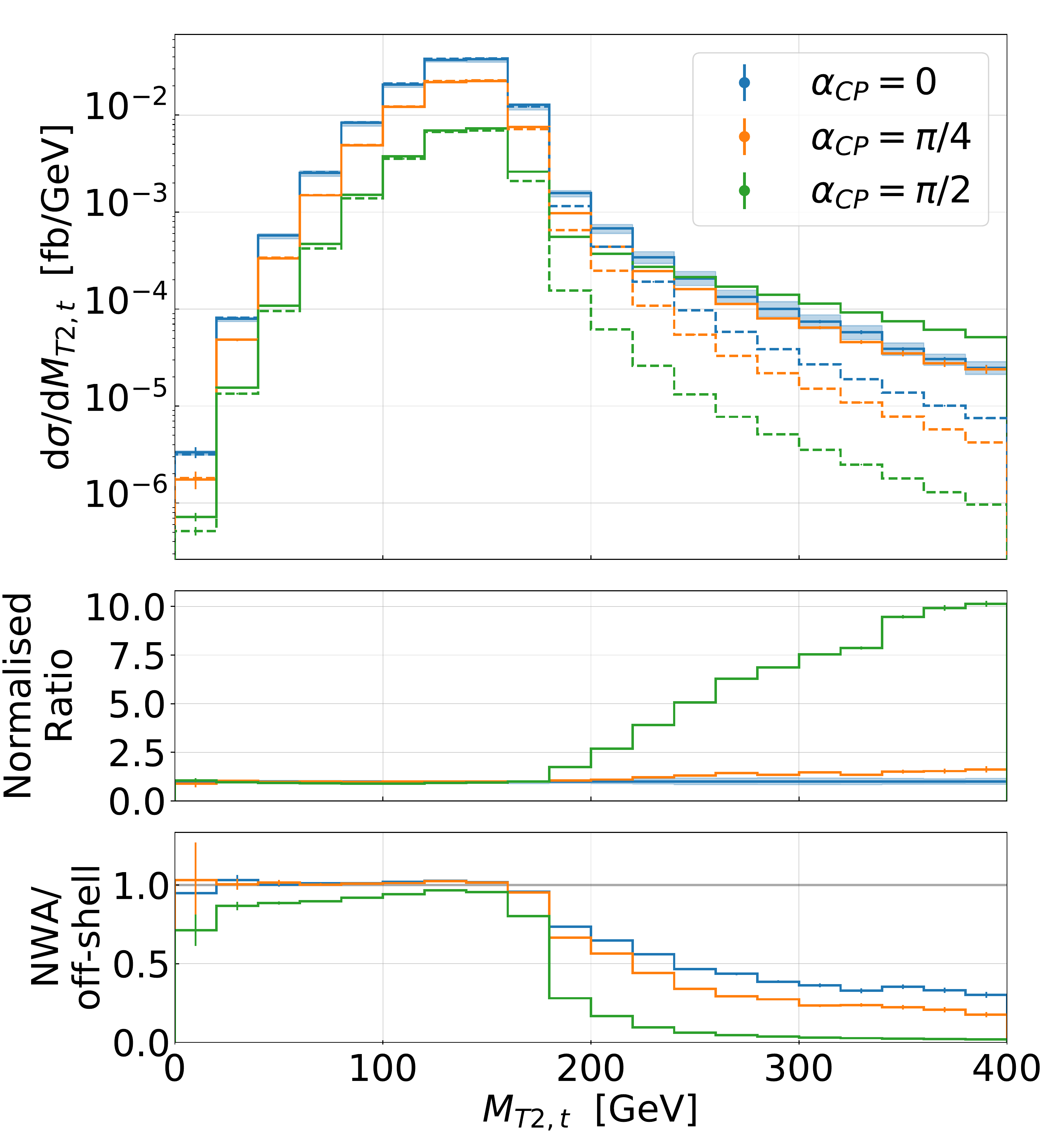}
	\end{center}
	\caption{\label{fig:cosa1} \it
		Differential distributions at NLO in QCD for
                $\alpha_{CP}=0,{\pi}/{4},{\pi}/{2}$ for the
                observables $H_T$, $p_{T,H}$, $M(be^+)_{min}$ and
                $M_{T2,t}$ for the $pp\to
                e^+\nu_e\,\mu^-\bar{\nu}_{\mu}\, b\bar{b}\,H$ process
                at the LHC with $\sqrt{s}=13\textrm{ TeV}$. The upper
                panels show full off-shell and NWA (dashed lines)
                results. The middle panels  provide the ratio to
                $\alpha_{CP}=0$ of the normalised differential
                distributions for the full off-shell case.  The ratio
                NWA/off-shell is given in the lower panels. Scale
                uncertainties are shown for $\alpha_{CP}=0$  in the
                upper and middle panels while Monte Carlo integration
                errors are  displayed in all panels.}
            \end{figure}
%
 
In the next step we discuss the shape distortions due to the different
Higgs boson ${\cal CP}$ configurations and the size of full off-shell
effects. The latter analysis is performed by comparing the full
off-shell calculation to the NWA predictions. In this way we can
estimate the impact of off-shell effects on dimensionful and
dimensionless observables for the three different ${\cal CP}$
states. We start this discussion with the four dimensionful
observables $H_T$, $p_{T,\,H}$, $M(be^+)_{min}$ and $M_{T2,\,t}$ shown
in Figure \ref{fig:cosa1}. In the upper panels we give the
differential distributions for the SM (${\cal CP}$-even), ${\cal
CP}$-mixed and ${\cal CP}$-odd Higgs boson in the full off-shell
calculation (solid lines) and in the NWA (dashed lines). The middle
panels provide the ratio to the $\alpha_{CP}=0$ case for the
normalised differential distributions in the full off-shell
approach. The ratio $\textrm{NWA}/\textrm{off-shell}$ for each
$\alpha_{CP}$ value is displayed in the lower panels. The scale
uncertainties are provided as reference for the SM case in the two
upper panels while Monte Carlo errors are reported in all three
panels. For $H_T$, as defined in Eq. (\ref{eq:ht}), we find that
compared to the SM Higgs boson, the spectra for the ${\cal CP}$-mixed
and ${\cal CP}$-odd case are harder towards the tails. Specifically,
we find an increase of only about $10\%$ for the ${\cal CP}$-mixed
case while for the ${\cal CP}$-odd case the increase is very
significant with more than $100\%$.  In both cases these effects are
larger than the corresponding scale uncertainties for the ${\cal
CP}$-even case which only amount to up to $6\%$. The full off-shell
effects for the SM and ${\cal CP}$-mixed Higgs boson cases are very
similar to each other and amount to about $5\%$ in the tails which is
comparable to the size to the scale uncertainties. On the other hand,
we obtain large off-shell effects of about $45\%$ for the ${\cal
CP}$-odd Higgs boson. The latter exceed the full off-shell effects at
the integrated level by more than a factor of three. Since these
off-shell effects rise towards the tails, they also have a significant
impact on the normalised ratios shown in the middle panels. Had we
taken the NWA predictions instead of the full off-shell ones, the
differences between the ${\cal CP}$ states would only be up to $60\%$
in the tails instead of about $120\%$. The shape differences in the
tails of $p_{T,\,H}$ are even larger than for $H_T$. In particular, we
observe differences with respect to the SM case at the level of $30\%$
for the ${\cal CP}$-mixed and effects larger than $200\%$ for the
${\cal CP}$-odd Higgs boson. Thus, these differences are significant
when compared to the scale uncertainties of about $10\%$ for the SM
Higgs boson case. Consequently, the $p_{T,\, H}$ observable is an
excellent example of a dimensionful differential cross-section
distribution that might be employed to decipher the ${\cal CP}$
quantum numbers of the Higgs boson and look for any hints of new
physics in the Higgs-top sector. In addition, the off-shell effects
are of the order of $5\%$ for the SM and ${\cal CP}$-mixed case and
thus would not have an impact on this comparison. For the ${\cal
CP}$-odd case, on the other hand, the full off-shell effects increase
up to $35\%$ in the tails of the $p_{T,\, H}$ distribution rendering
them indispensable in studies of the ${\cal CP}$ state of the Higgs
boson. If we had set $\kappa_{Ht\bar{t}}=\kappa_{At\bar{t}}=1$
instead, then the absolute differential distributions for $p_{T,\,H}$
would coincide in the tails for all three ${\cal CP}$ cases in the
NWA, see e.g. Refs. \cite{Frederix:2011zi,Demartin:2014fia}. Taking
into account the fragmentation functions in
Eq. (\ref{eq:fragmentation}), this behaviour becomes clear since the
soft enhancement for the scalar Higgs boson is absent in the tails of
this distribution. On the other hand, in the full off-shell
calculation we would not obtain such a behaviour due to the large
effects coming from single- and non-resonant top-quark contributions
as well as their interference effects with double-resonant ones.
For $M(be^+)_{min}$ the SM and ${\cal CP}$-mixed Higgs boson cases
lead to the exact same normalised distributions below the kinematical
edge around $M(be^+)_{min} \approx 153$ GeV.  Above this edge the
difference between the two scenarios is in the $20\%-25\%$
range. Nevertheless, it is still comparable to the scale uncertainties
of the ${\cal CP}$-even case, which are of the order of $20\%$ in this
phase-space region.  In the ${\cal CP}$-odd case the ratio of the
normalised differential distribution to the SM distribution is larger
than four. Even the absolute differential cross-section distribution
for the ${\cal CP}$-odd case is larger than the ${\cal CP}$-mixed one
in this phase-space region, even though the overall normalisation is
almost three times smaller. This behaviour is closely related to the
size of the full off-shell effects, which reach up to $90\%$ for the
${\cal CP}$-odd Higgs boson but only $50\%$ for the SM and $60\%$ for
the ${\cal CP}$-mixed Higgs boson. This is a result of a larger
relative contribution from single- and non-resonant diagrams for the
${\cal CP}$-odd Higgs boson which is partially caused by the
non-vanishing $HVV$ coupling.
A similar phenomenon can be observed in the stransverse mass
$M_{T2,t}$ \cite{Barr:2003rg,Lester:1999tx,Lester:2014yga} which is a
generalisation of the transverse mass in case of pairs of unstable
particles and is defined as
\begin{equation}
  M^2_{T2,t}=\textrm{min}_{\mathbf{p}^{\nu_1}_T+\mathbf{p}^{\nu_1}_T
    =\mathbf{p}_{T,miss}}\left[\textrm{max}\left\{
    M_T^2\left(\mathbf{p}_T^{(lb)_1},\mathbf{p}_T^{\nu_1}  \right),
    M_T^2\left(\mathbf{p}_T^{(lb)_2},\mathbf{p}_T^{\nu_2}  \right)
  \right\} \right],
\end{equation}
where the transverse mass of the lepton+$b$-jet system in presence of
a missing transverse momentum $\mathbf{p}_T^{\nu_i}$ is given by
\begin{equation}
M_T^2\left(\mathbf{p}_T^{(lb)_i},\mathbf{p}_T^{\nu_i}
\right)=M^2_{(lb)_i}+2\left(E_T^{(lb)_i}
  E_T^{\nu_i}-\mathbf{p}_T^{(lb)_i}\cdot\mathbf{p}_T^{\nu_i}  \right).
\end{equation}
We choose the lepton+$b$-jet pairs by minimising the value
$Q=M_{e^+b_i}+M_{\mu^-b_j}$ so that we do not take into account the
charge of the $b$-jets. Similarly to $M(be^+)_{min}$, the stransverse
mass has a kinematical edge around the top-quark mass ($m_t=172.5$
GeV). For $M_{T2,t} > m_t$, large suppression of double-resonant
top-quark contributions occurs. Thus, the absolute value of the
cross-section is significantly reduced above this edge and the full
off-shell effects, which are driven by the single-resonant top-quark
contributions, increase.  Following the same reasoning as before, this
reduction is suppressed for the ${\cal CP}$-odd case. When examining
the absolute differential cross-section distributions we can observe
that the ${\cal CP}$-odd case exceeds even the SM results in these
phase-space regions. The difference between the normalised $M_{T2,t}$
distributions can reach a factor of $10$. A less significant
enhancement occurs for the ${\cal CP}$-mixed Higgs boson so that the
SM and ${\cal CP}$-mixed cases are very similar towards the end of the
absolute $M_{T2,t}$ spectrum. This is due to the large shape
differences of about $50\%-60\%$ which are substantial compared to the
scale uncertainties. The latter are in the $15\%-18\%$ range for the
${\cal CP}$-even case. Compared to the $M(be^+)_{min}$ differential
cross-section distribution the full off-shell effects are even larger
here. Indeed, they are of the order of $70\%$, $80\%$ and $98\%$ for
the ${\cal CP}$-even, ${\cal CP}$-mixed and ${\cal CP}$-odd case,
respectively. Both $M(be^+)_{min}$ and $M_{T2,t}$ are very sensitive
to the various Higgs boson ${\cal CP}$ numbers. However, they are also
subject to large off-shell effects. Not only the finite top-quark
width effects should be properly taken into account for these
observables but also single-resonant top-quark contributions and
interference effects must be incorporated to adequately describe the
phase-space regions above these kinematical edges.
%
\begin{figure}[t!]
	\begin{center}
		\includegraphics[width=0.49\textwidth]{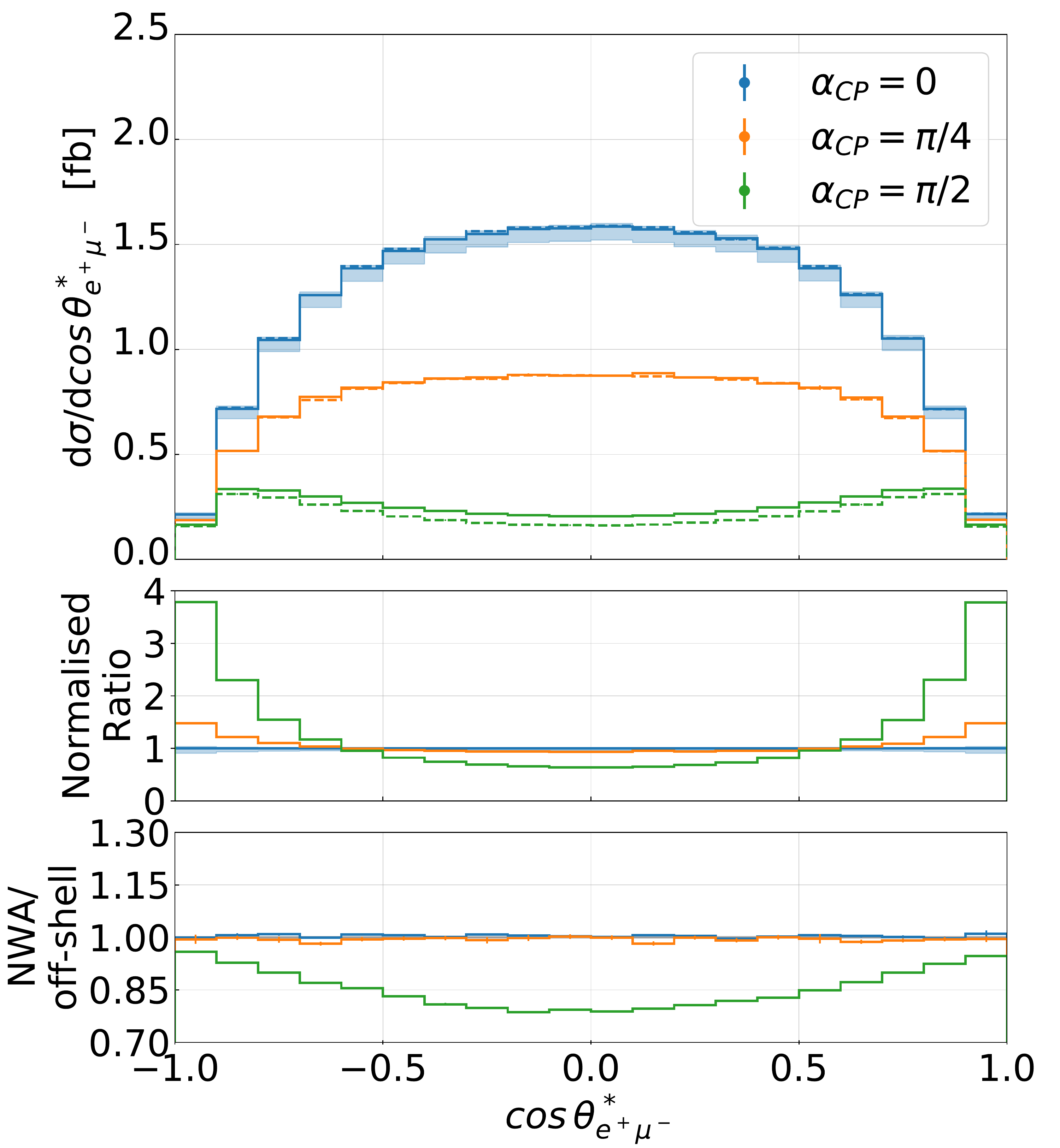}
		\includegraphics[width=0.49\textwidth]{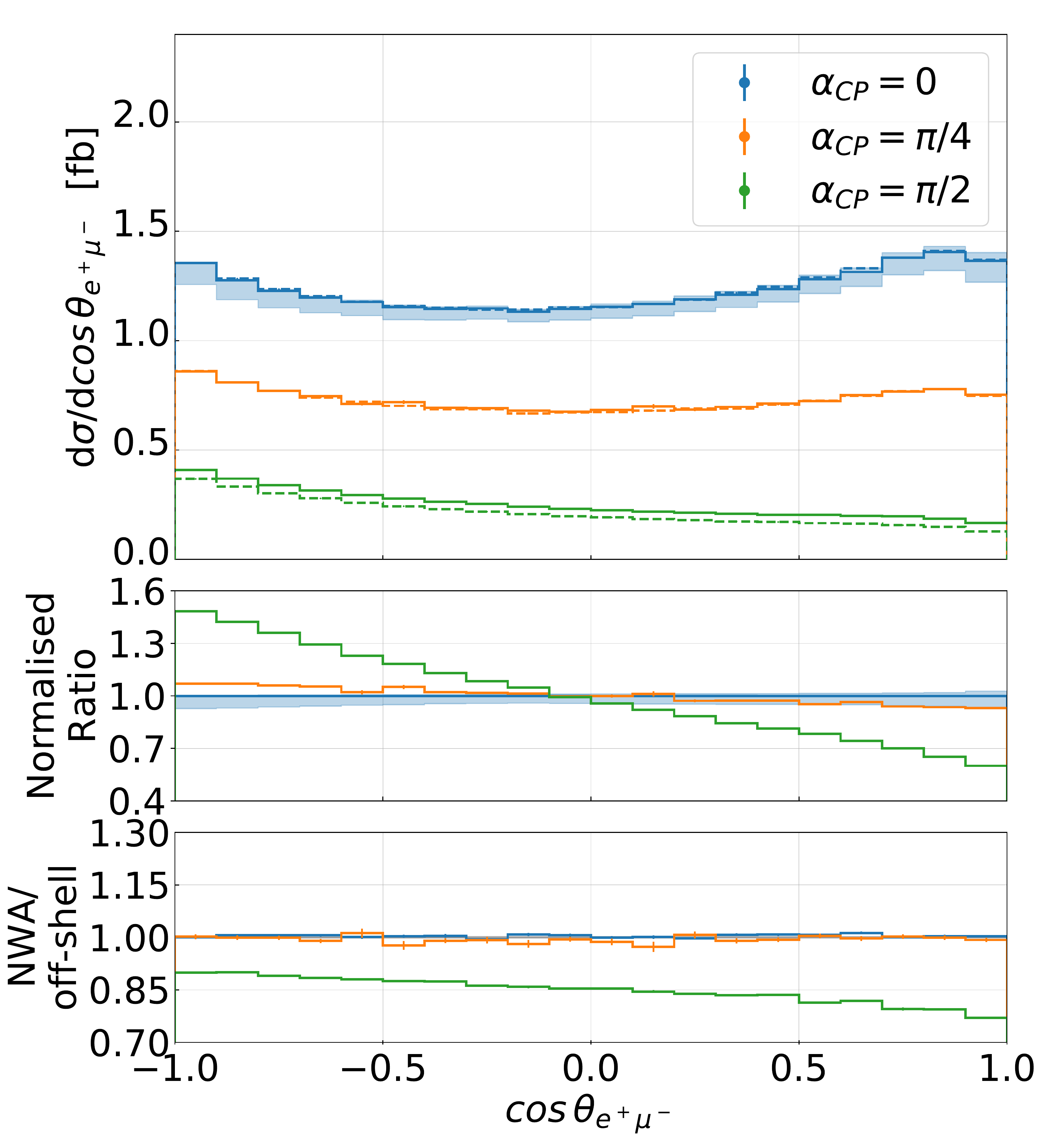}
		\includegraphics[width=0.49\textwidth]{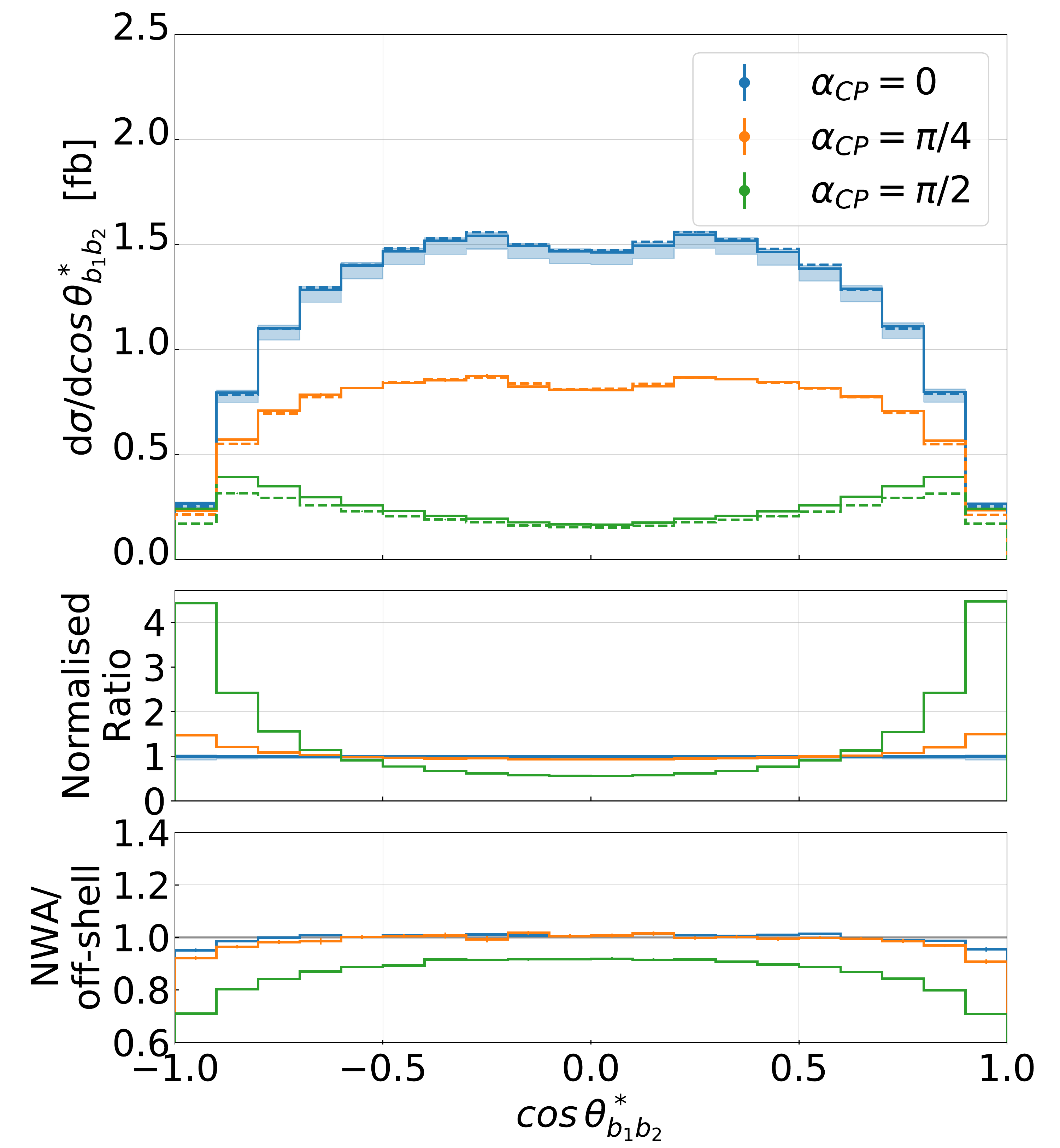}
		\includegraphics[width=0.49\textwidth]{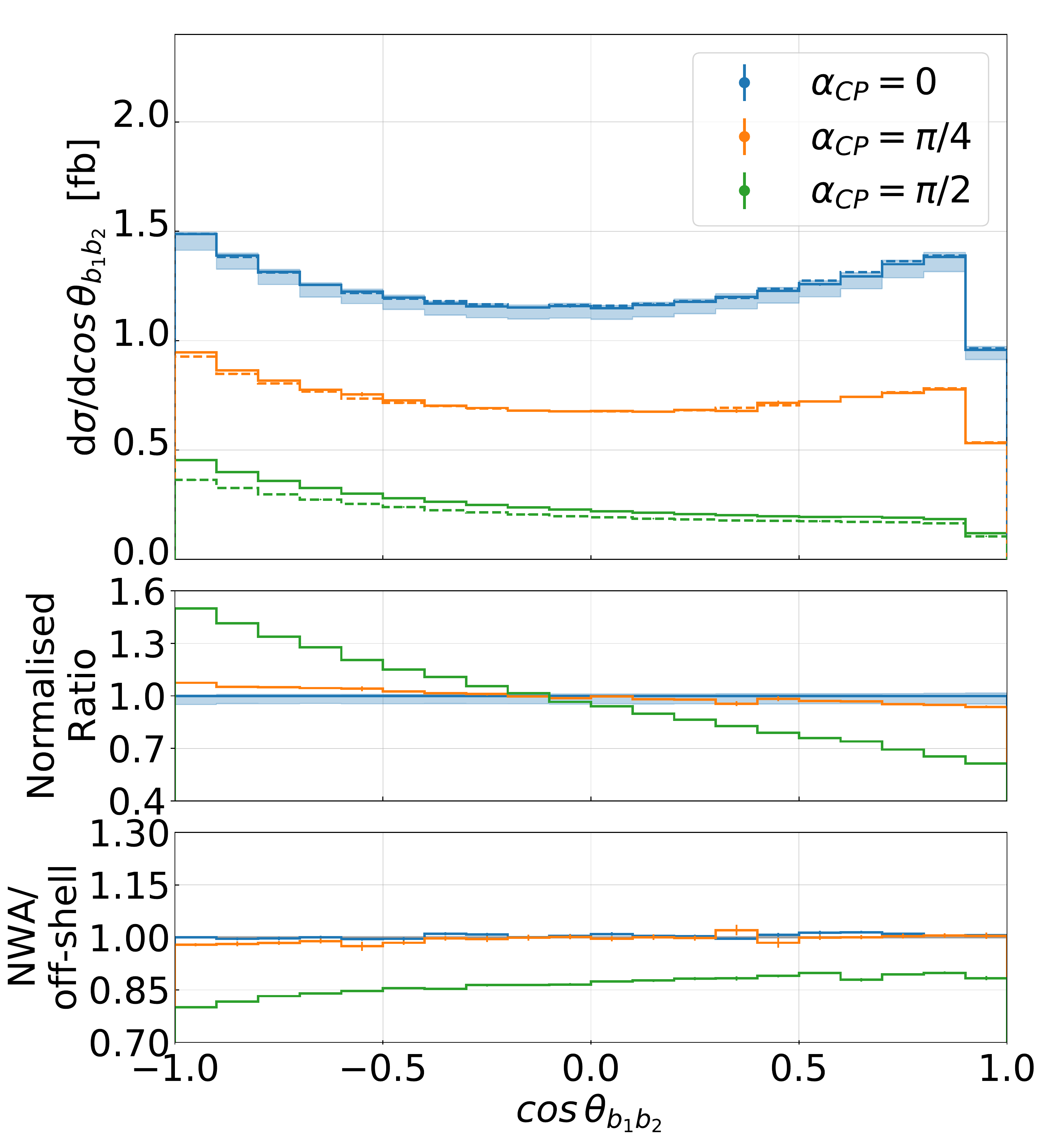}
	\end{center}
	\caption{\label{fig:cosa2} \it
		Differential distributions at NLO in QCD for
                $\alpha_{CP}=0,{\pi}/{4},{\pi}/{2}$ for the
                observables $\cos\theta^*_{e^+\mu^-}$,
                $\cos\theta_{e^+\mu^-}$, $\cos\theta^*_{b_1b_2}$ and
                $\cos\theta_{b_1b_2}$ for the $pp\to e^+\nu_e\,
                \mu^-\bar{\nu}_{\mu}\, b\bar{b}\,H$ process at the LHC
                with $\sqrt{s}=13\textrm{ TeV}$. The upper panels show
                full off-shell and NWA (dashed lines) results. The
                middle panels provide the ratio to $\alpha_{CP}=0$ of
                the normalised differential distributions for the full
                off-shell case. The ratio NWA/off-shell is given in
                the lower panels. Scale uncertainties are shown for
                $\alpha_{CP}=0$ in the upper and middle panels while
                Monte Carlo integration errors are  displayed in all
                panels.}
 \end{figure}
%

In the next step we discuss the shape differences between the ${\cal
CP}$-even, ${\cal CP}$-mixed and ${\cal CP}$-odd Higgs boson as well
as the respective size of off-shell effects in several angular
observables.  In Figure \ref{fig:cosa2} we present the differential
cross-section distributions for $\cos\theta^*_{e^+\mu^-}$,
$\cos\theta_{e^+\mu^-}$, $\cos\theta^*_{b_1b_2}$ and
$\cos\theta_{b_1b_2}$. In the case of $\cos\theta^*_{e^+\mu^-}$ we
find that the prediction for the central value of the scale is smaller
by about $7\%$ for the ${\cal CP}$-mixed and $35\%$ for ${\cal
CP}$-odd Higgs boson compared to the SM case. These differences
increase significantly towards $\cos\theta^*_{e^+\mu^-} \approx \pm 1$
to about $50\%$ for the ${\cal CP}$-mixed and more than $200\%$ for
the ${\cal CP}$-odd case. Thus, the two leptons tend to be more
separated in the rapidity plane in the case of the ${\cal CP}$-mixed
and ${\cal CP}$-odd Higgs boson.  Furthermore, the theoretical
uncertainties due to the scale dependence are not only very small but
also very similar for all the four observables. More specifically, we
have theoretical uncertainties at the level of $5\%$ in the central
regions for the SM Higgs boson case, whereas towards
$\cos\theta^*_{e^+\mu^-} \approx \pm 1$ they increase only slightly up
to $7\%-9\%$. This means that the shape differences outlined above are
much larger than the scale uncertainties. As a result, we can conclude
that the $\cos\theta^*_{e^+\mu^-}$ angular differential cross-section
distribution is a promising ${\cal CP}$-sensitive observable to be
measured at the LHC.  This is further underlined when examining the
full off-shell effects for the SM and ${\cal CP}$-mixed Higgs
boson. We can observe that these effects are less than $2\%$ in both
cases. Thus, they are even comparable in size to the full off-shell
top-quark effects at the integrated fiducial cross-section level.
However, the situation is rather different for the ${\cal CP}$-odd
Higgs boson for which the off-shell effects lead to a wrong
normalisation as well as additional shape distortions and we find that
the off-shell effects are reduced from the centre of the plot towards
the edges from $20\%$ to $5\%$.  Thus, even for angular distributions,
single- and non-resonant contributions can significantly affect the
shape, especially in the $\cos\theta^*_{e^+\mu^-}\approx 0$
phase-space region where theoretical uncertainties are four times
smaller than the size of the full off-shell effects. Consequently, the
latter should be included in theoretical predictions when comparing to
experimental LHC data.  For $\cos\theta_{e^+\mu^-}$ we can see the
largest differences between the SM and the ${\cal CP}$-mixed Higgs
boson around $\cos\theta_{e^+\mu^-} \approx \pm 1$. At about $7\%$,
they are comparable in size to the scale uncertainties in those
phase-space regions. The situation is again quite different for the
${\cal CP}$-odd Higgs boson.  In this case, differences of up to
$40\%- 50\%$ are obtained for $\cos\theta_{e^+\mu^-} \approx \pm
1$. The off-shell effects for the SM and ${\cal CP}$-mixed Higgs boson
cases are again quite small and rather similar in size while for the
${\cal CP}$-odd one they significantly affect the normalisation and
shape of the distribution. In particular, the off-shell effects reduce
towards large opening angles from $20\%$ to $10\%$.

The observables $\cos\theta^*_{b_1b_2}$ and $\cos\theta_{b_1b_2}$,
which are the hadronic counterparts of the two previously discussed
observables, are important for leptonic Higgs boson decays since in
those cases we have a larger lepton multiplicity which makes a precise
measurement of leptonic observables more challenging. All in all, the
shape differences between the three ${\cal CP}$ states are similar to
those of the leptonic observables.  This is especially true for
$\cos\theta_{b_1b_2}$ for which the normalised ratios are essentially
the same as for $\cos\theta_{e^+\mu^-}$.  For the ${\cal CP}$-odd
Higgs boson, on the other hand, both the $\cos\theta^*_{b_1b_2}\approx
0$ and $\cos\theta^*_{b_1b_2} \approx \pm 1$ phase-space regions
receive additional enhancements.  Compared to the corresponding scale
uncertainties the off-shell effects of the ${\cal CP}$-even and ${\cal
CP}$-mixed Higgs boson are again negligible for both observables but
sizable for the ${\cal CP}$-odd case. In the latter case, they amount
to $10\%-30\%$ for $\cos\theta^*_{b_1b_2}$ and $10\%-20\%$ for
$\cos\theta_{b_1b_2}$ which substantially exceed the corresponding
theoretical uncertainties and lead to significant shape distortions.

 In the following, we investigate the contribution of
double-, single- and non-resonant fiducial phase-space regions to the
integrated and differential cross-sections for the full off-shell
case. This should give us a qualitative picture of the importance of
the individual contributions, as well as their general distribution
for a given value of the ${\cal CP}$-mixing angle of the Higgs boson.
To identify these contributions, we use the method introduced in
Ref. \cite{Kauer:2001sp}. This approach was further modified and
employed in various studies, see
e.g. \cite{Liebler:2015ipp,Baskakov:2018huw,Bevilacqua:2019quz,
Bevilacqua:2020srb}. Specifically, at NLO in QCD we have the following
three resonance histories
\begin{equation}
\begin{array}{llll}
  (\mathrm{i}) \quad & t = W^+(\to e^+ \nu_e) \, b  &
                                                  \quad    \quad \quad
                                                      \mathrm{and}
                                                      \quad \quad
                                                      \quad
  & \bar{t}  =W^-(\to \mu^- \bar{\nu}_\mu) \,  \bar{b} \,,\\[0.2cm]
  (\mathrm{ii}) \quad &   t = W^+( \to e^+ \nu_e ) \,b j &
                                                             \quad
                                                                \quad
                                                                \quad
                                                                \mathrm{and}  \quad \quad \quad
  & \bar{t} = W^-(\to \mu^- \bar{\nu}_\mu)\,  \bar{b} \,,\\[0.2cm]
  (\mathrm{iii}) \quad &  t = W^+( \to e^+ \nu_e) \,b &
                                                 \quad       \quad
                                                        \quad
                                                        \mathrm{and}
                                                        \quad \quad
                                                        \quad
  & \bar{t} = W^-(\to \mu^- \bar{\nu}_\mu)\,  \bar{b} j\,,\\
\end{array}
\end{equation}
where the light jet (if resolved by the {\it anti-k${}_{T}$} jet
algorithm with $R = 0.4$) is added to the list only if it passes the
following cuts: $|y_j|<2.5$ and $p_{T,\, j} >15$ GeV.  In this way, we
closely mimic what is done on the experimental side for the light jet
to be observed in the ATLAS or CMS  detector. For each history
we compute the following quantity
\begin{equation}
{\cal Q} = |M_{t} - m_t| + |M_{\bar{t}} - m_t|\,,
\end{equation}  
where $m_t=172.5$ GeV and $M_{t}$ ($M_{\bar{t}}$) is the
invariant mass of the reconstructed top quark (anti-top quark).
For each phase-space point, one history is selected
by minimising the value of ${\cal Q}$. The boundaries of the fiducial
double-resonant (DR), single-resonant (SR) and non-resonant (NR)
regions are expressed in terms of $\Gamma_t^{\rm NLO}$ and the
parameter $n=15$. The value of the latter is rather
arbitrary. The DR phase-space region is defined according to
\begin{equation}
  |M_{t} -m_t|< n \, \Gamma_t^{\rm NLO} \,, \quad \quad \quad {\rm
    and}  \quad \quad \quad |M_{\bar{t}} -m_t|< n \, \Gamma_t^{\rm
    NLO} \,.
\end{equation}  
There are two  SR  regions  given by
\begin{equation}
|M_{t} -m_t|< n \, \Gamma_t^{\rm NLO} \,, \quad \quad \quad {\rm
  and}   \quad \quad \quad |M_{\bar{t}} -m_t|> n \, \Gamma_t^{\rm
  NLO}  \,,
\end{equation}
or
\begin{equation}
 |M_{t} -m_t|> n \, \Gamma_t^{\rm NLO} \,, \quad \quad \quad {\rm
   and}  \quad \quad \quad |M_{\bar{t}} -m_t|< n \, \Gamma_t^{\rm
   NLO}  \,.
\end{equation}
The NR region is determined via
\begin{equation}
  |M_{t} -m_t|> n \, \Gamma_t^{\rm NLO} \,, \quad \quad \quad {\rm
    and}  \quad \quad \quad |M_{\bar{t}} -m_t|> n \, \Gamma_t^{\rm
    NLO} \,.
\end{equation}
For the above outlined procedure, the contributions at the integrated
fiducial cross-section level for these three regions for the ${\cal
CP}$-even ($\alpha_{CP}=0$), ${\cal CP}$-mixed ($\alpha_{CP}=\pi/4$)
and ${\cal CP}$-odd ($\alpha_{CP}=\pi/2$) case are given in Table
\ref{table:PS_regions}. The integrated fiducial cross-sections for all
three cases are dominated by the DR contributions. For the SM and
${\cal CP}$-mixed Higgs boson, more than $90\%$ of $\sigma^{\rm
NLO}_{\rm Off-shell}$ comes from the DR contribution. For the ${\cal
CP}$-odd Higgs boson, however, the DR contribution is reduced to
$80\%$. A similar behaviour can be observed for the remaining fiducial
phase-space regions.  For both $\alpha_{CP}=0$ and $\alpha_{CP}=\pi/4$,
the SR region is of the order of $9\%$, while the NR regions of the
phase space are completely negligible.  Again, for $\alpha_{CP}=\pi/2$,
the SR and NR contributions are larger at the level of $19\%$ and
$1\%$, respectively.  Hence, in the case of the ${\cal CP}$-odd Higgs
boson, the SR regions of the phase space are more substantial, which
is reflected in the larger size of the full off-shell effects as we
have already seen in Table \ref{table:integrated}.
%
\begin{table*}[t!]
  \caption{\it NLO QCD integrated fiducial
cross-sections as calculated in the full off-shell approach together
with predictions for DR, SR and NR fiducial phase-space regions for
$\alpha_{CP} = 0, \pi/4$ and $\pi/2$.  Also shown are their respective
Monte Carlo integration errors. All values are given for the $pp \to
e^+ \nu_e \,\mu^- \bar{\nu}_\mu \,b \bar{b} \,H$ process at the LHC with
$\sqrt{s}= 13$ TeV.} 
    \label{table:PS_regions}
    
    \centering
    \renewcommand{\arraystretch}{1.5}
    \begin{tabular}{c@{\hskip 10mm}c@{\hskip 10mm}c@{\hskip
      10mm}c@{\hskip 10mm}c}
        \hline\noalign{\smallskip}
      $\alpha_{CP}$ & Off-shell   & Double-resonant & Single-resonant & Non-resonant \\
    &  [fb]  & [fb] & [fb] & [fb] \\
        \noalign{\smallskip}\midrule[0.5mm]\noalign{\smallskip} 
       $0$ (SM) & 2.466(2) & 2.246(2) & 0.2138(5)  & 0.00594(7)\\
       $\pi/4$               & 1.465(2)  & 1.325(2)& 0.136(1) & 0.0039(1)\\
      $\pi/2$ & 0.5018(3) & 0.4037(9) & 0.0948(4)&  0.00346(5)\\
        \noalign{\smallskip}\hline\noalign{\smallskip}
    \end{tabular}
  \end{table*}
\begin{figure}[t!]
	\begin{center}
		\includegraphics[width=0.49\textwidth]{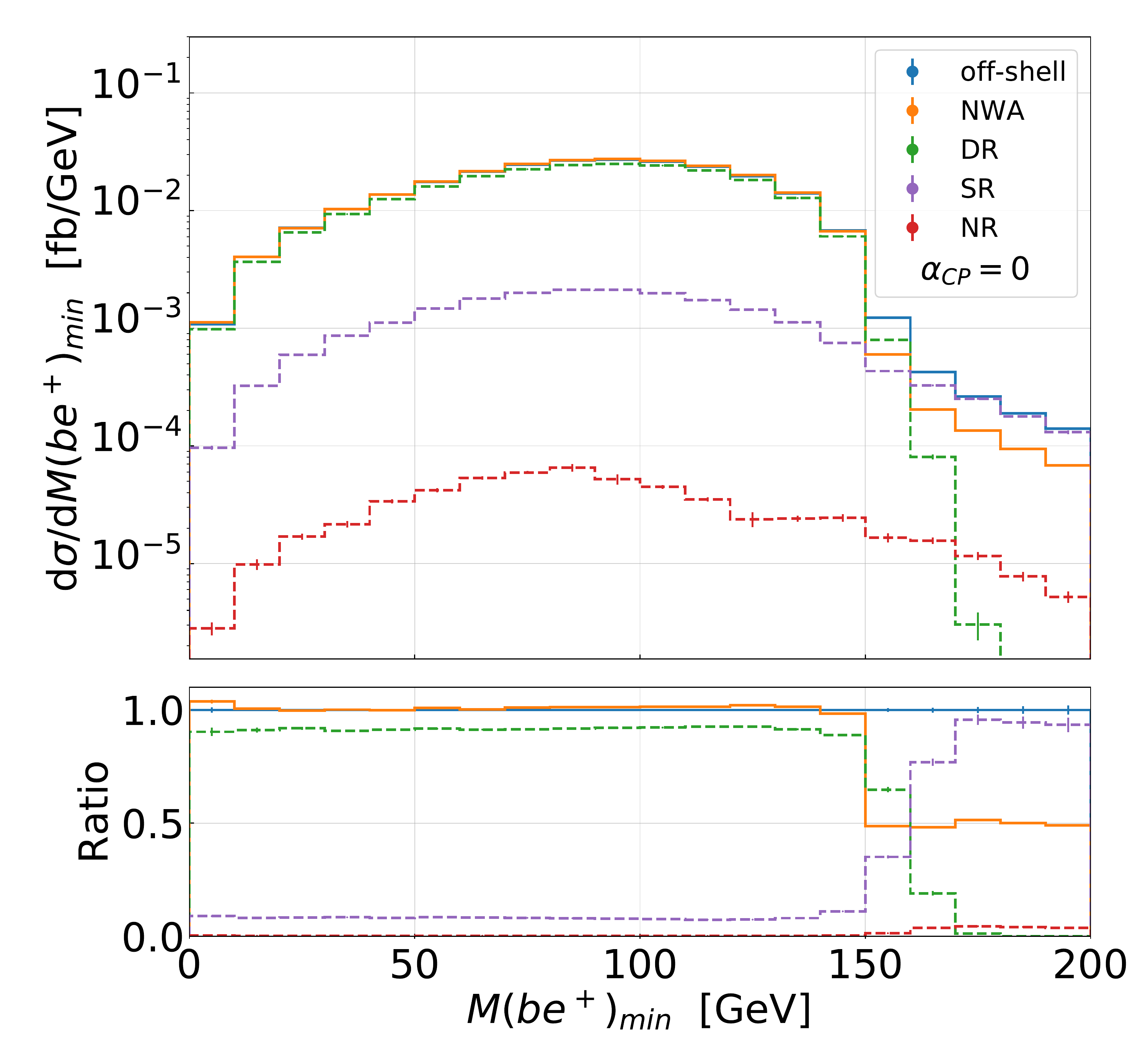}
		\includegraphics[width=0.49\textwidth]{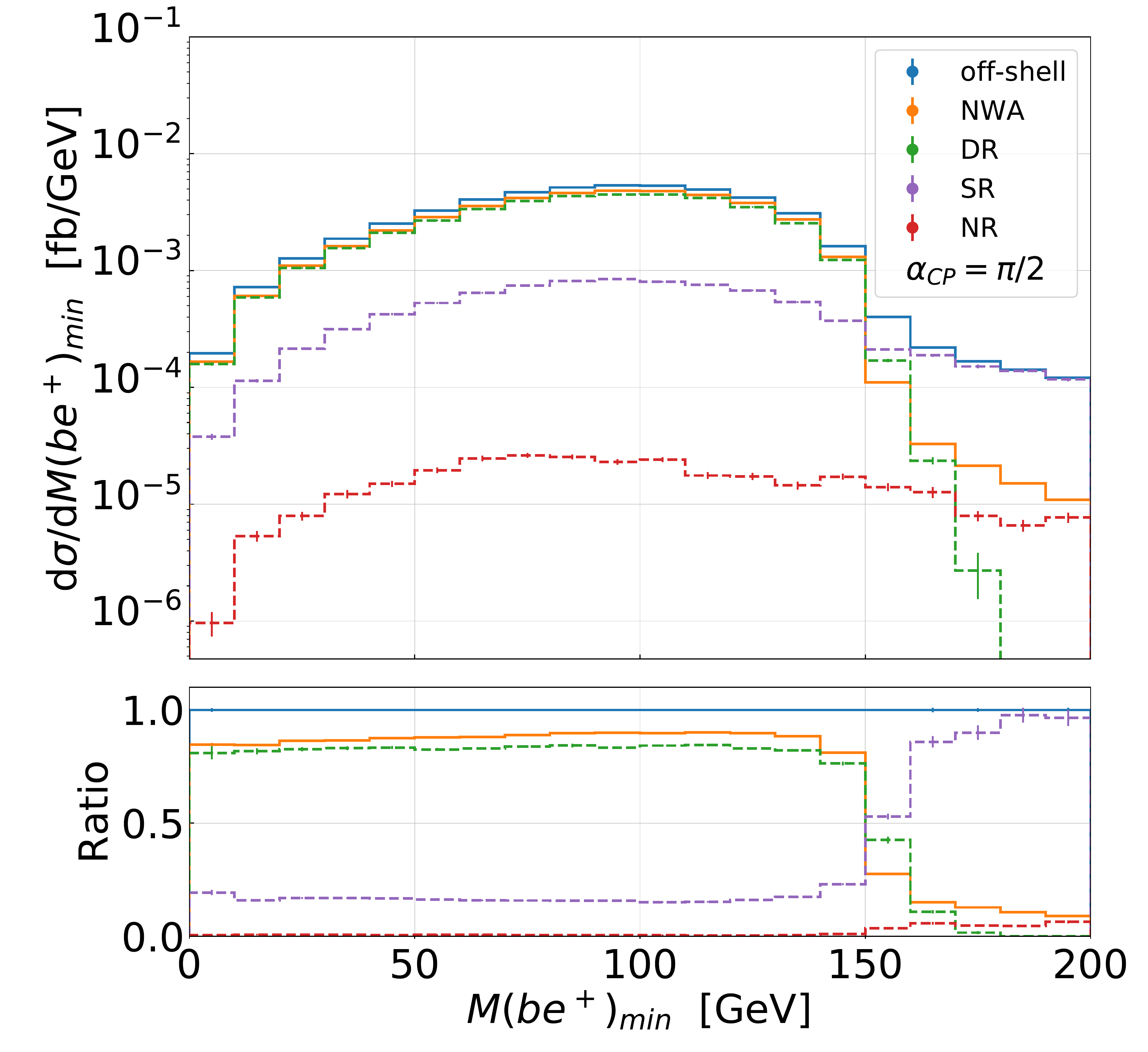}
		\includegraphics[width=0.49\textwidth]{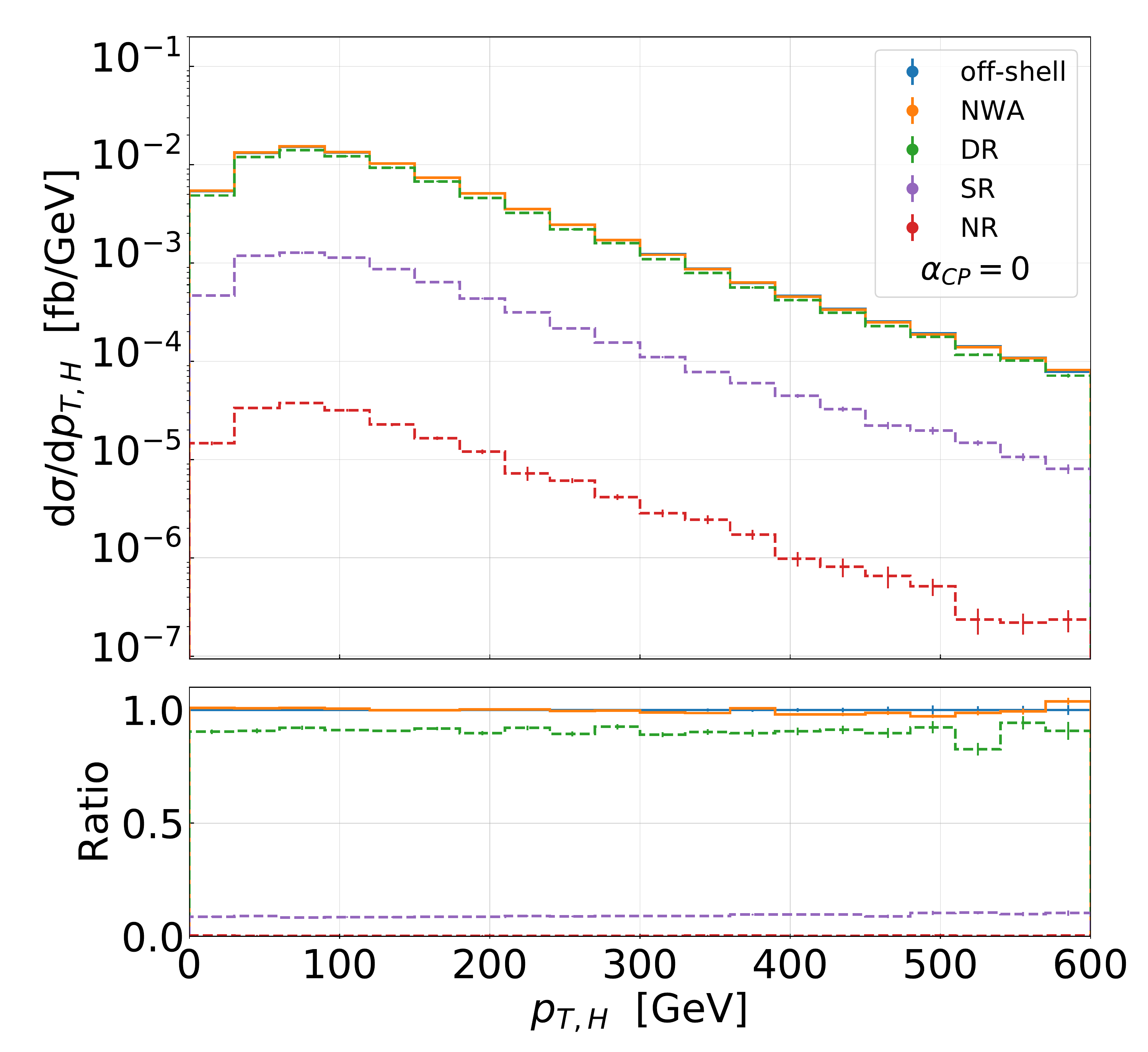}
		\includegraphics[width=0.49\textwidth]{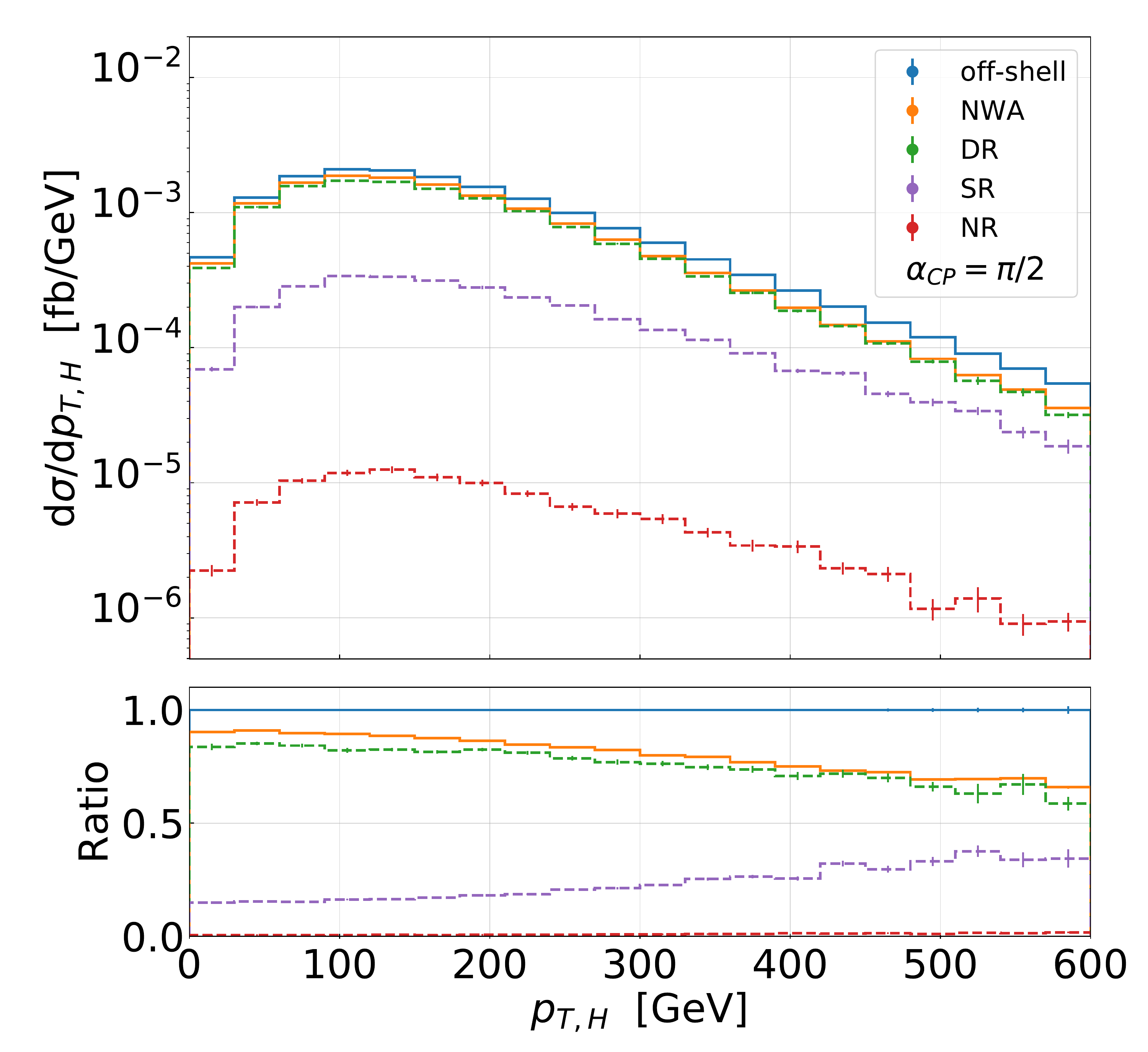}
              \end{center}
\caption{\label{fig:PS_regions} \it  Differential
distributions at NLO in QCD for $\alpha_{CP}=0$ and $\pi/2$ for the
observables $M(be^+)_{min}$ and $p_{T,\, H}$ for the $pp \to e^+ \nu_e
\,\mu^- \bar{\nu}_\mu \,b \bar{b} \,H$ process at the LHC with
$\sqrt{s}= 13$ TeV.  The upper panels show full off-shell and NWA
results as well as predictions for DR, SR and NR fiducial phase-space
regions. The ratios of  all contributions to the full off-shell
result are shown in the lower panels. Monte Carlo integration errors
are displayed in both panels.}
\end{figure}

We have confirmed these findings also at the differential level. In
Figure \ref{fig:PS_regions} we display the differential cross-section
distribution as a function of $M(be^+)_{min}$ and $p_{T,\,H}$ for
$\alpha_{CP}=0$ and $\pi/2$. The ${\cal CP}$-mixed case is omitted as
it follows the SM Higgs boson  very closely. For each plot, we show the
full off-shell and NWA predictions as well as the results for the DR,
SR and NR fiducial phase-space regions. The lower panels display the
ratio of all contributions to the full off-shell result.  For the
$M(be^+)_{min}$ distribution, we observe that in the case of the
${\cal CP}$-even Higgs boson, the SR part rapidly increases  in the
region that starts to be sensitive to off-shell effects, i.e. for
$M(be^+)_{min}\approx 153$ GeV, and even starts to dominate the full
off-shell result towards the end of the spectrum.  This effect is enhanced for
the ${\cal CP}$-odd Higgs boson. As a matter of fact, we have the
following distribution of DR, SR and NR phase-space regions for
$M(be^+)_{min}$ around $153$ GeV; DR $65\%$ ($42\%$), SR $35\%$
($53\%$) and NR $1\%$ ($4\%$) for $\alpha_{CP}=0$
($\alpha_{CP}=\pi/2$). Even for $M(be^+)_{min} \le 153$
GeV, the SR contribution for $\alpha_{CP}=\pi/2$ is twice as large as
for the $\alpha_{CP}=0$ case.  A similar effect can be observed for
$p_{T,\, H}$. Even though in the high $p_T$ tail of the differential
cross-section distribution the SR part does not dominate the full
off-shell result, its contribution increases substantially from $10\%$
for the ${\cal CP}$-even Higgs boson to $34\%$ for the ${\cal CP}$-odd
Higgs boson. Again, even at the beginning of the spectrum, the SR
contribution is of the order of $16\%$ for $\alpha_{CP}=\pi/2$, while
for the SM Higgs boson case we have a rather constant contribution of
the order of $10\%$ for the whole plotted range. Consequently, the DR
part  is reduced from around $90\%$ for the $\alpha_{CP}=0$ case  down
to  $84\%$ for the $\alpha_{CP}=\pi/2$ one.
%
\begin{figure}[t!]
	\begin{center}
		\includegraphics[width=0.49\textwidth]{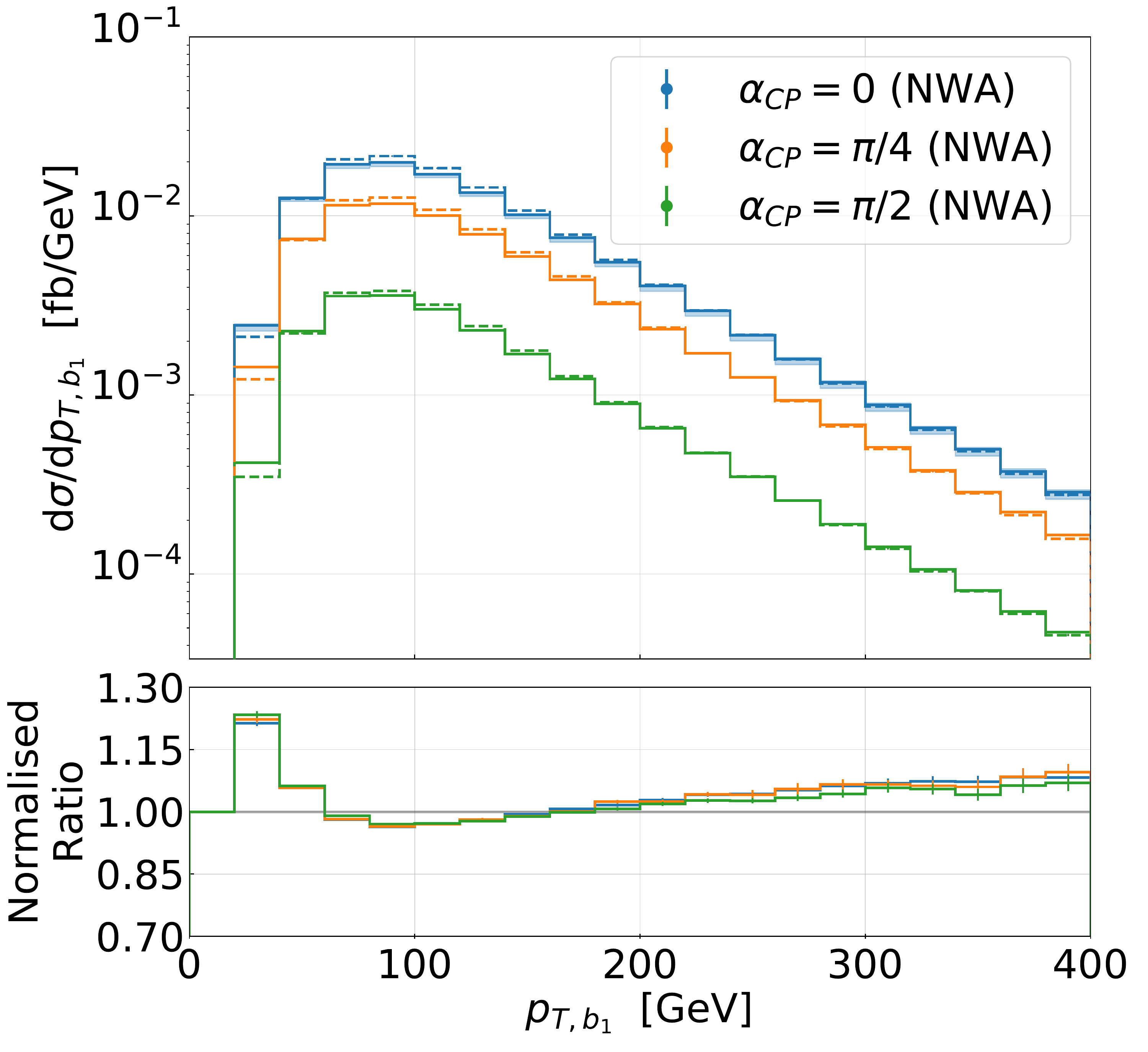}
		\includegraphics[width=0.49\textwidth]{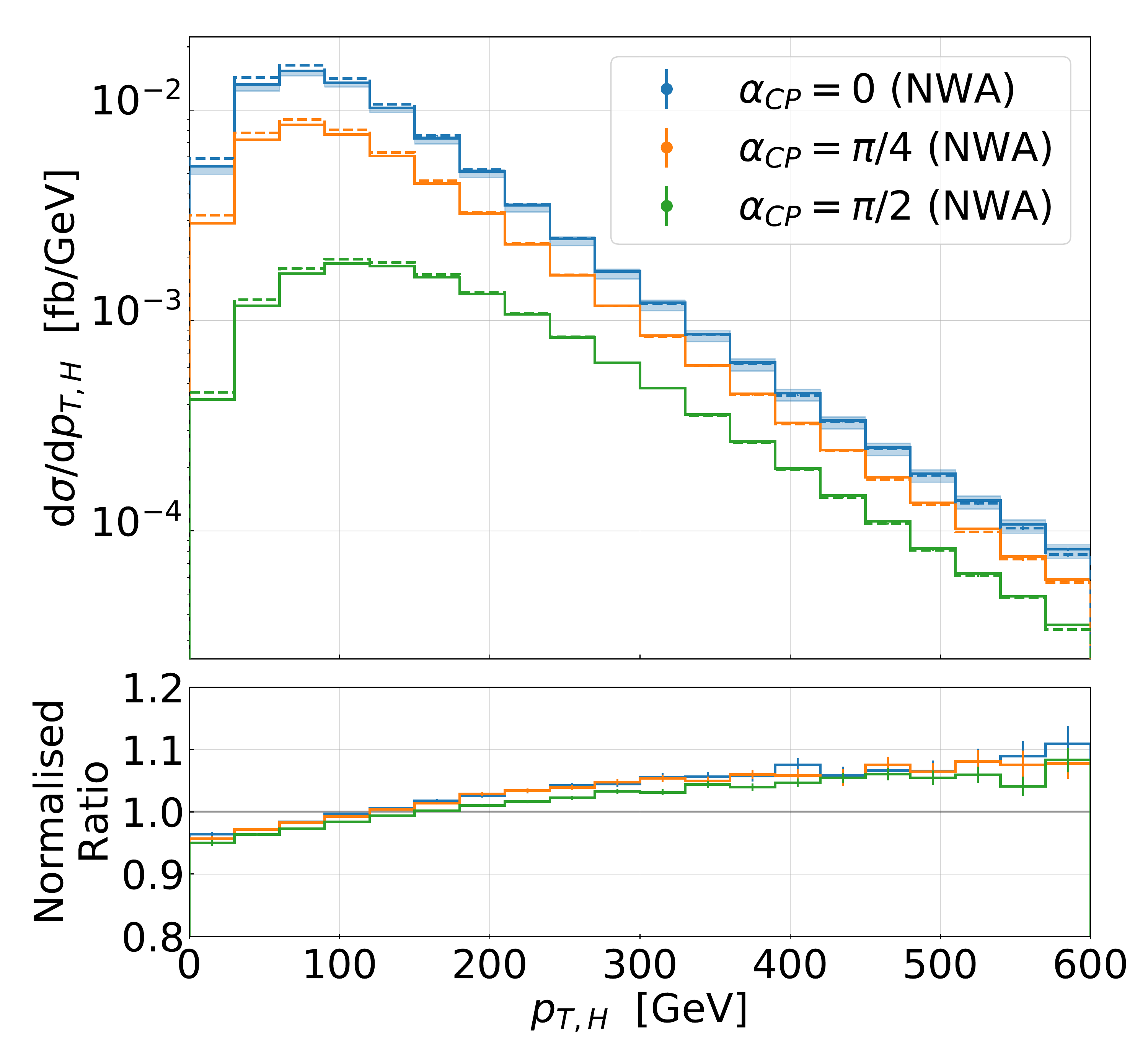}
		\includegraphics[width=0.49\textwidth]{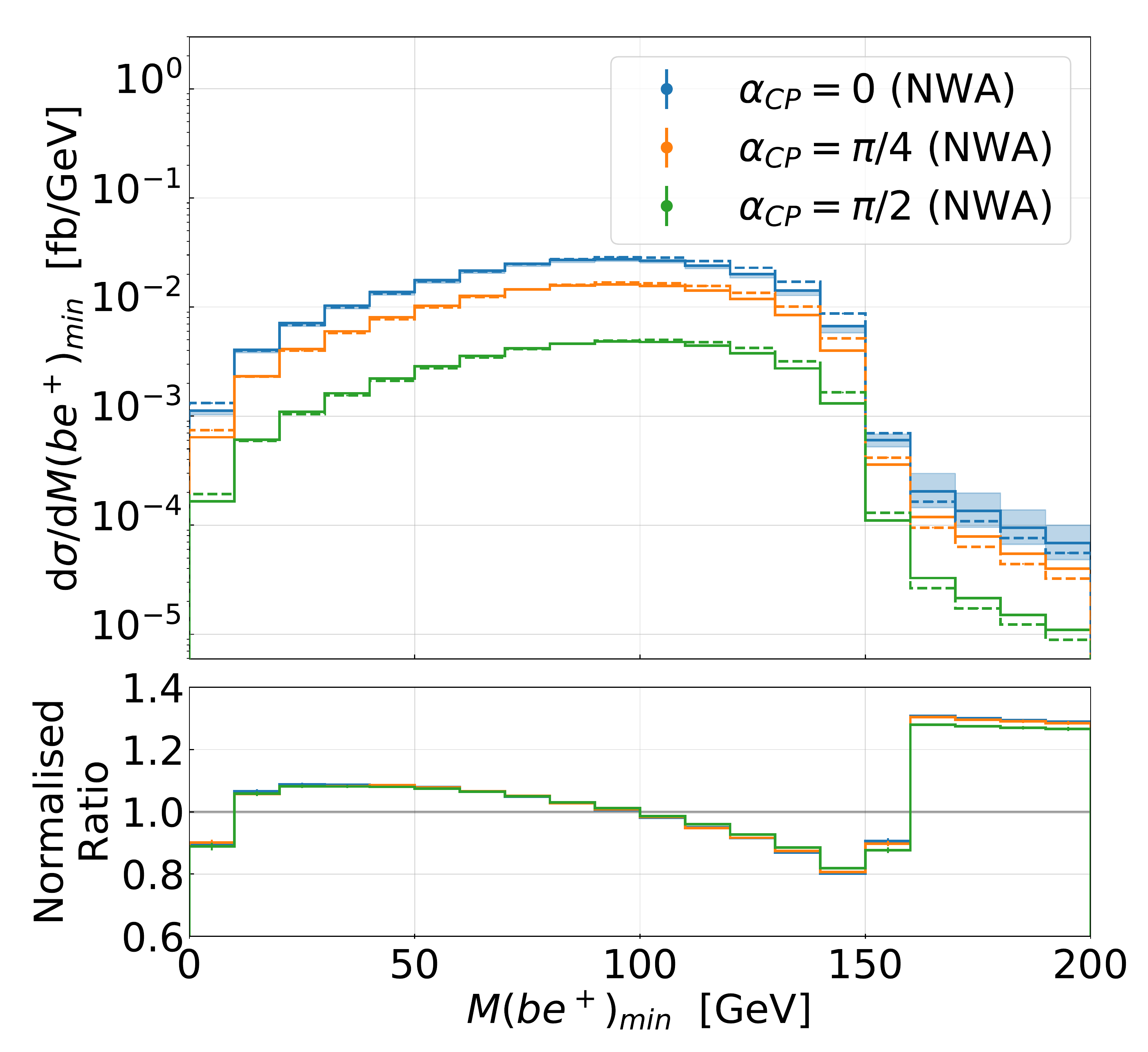}
		\includegraphics[width=0.49\textwidth]{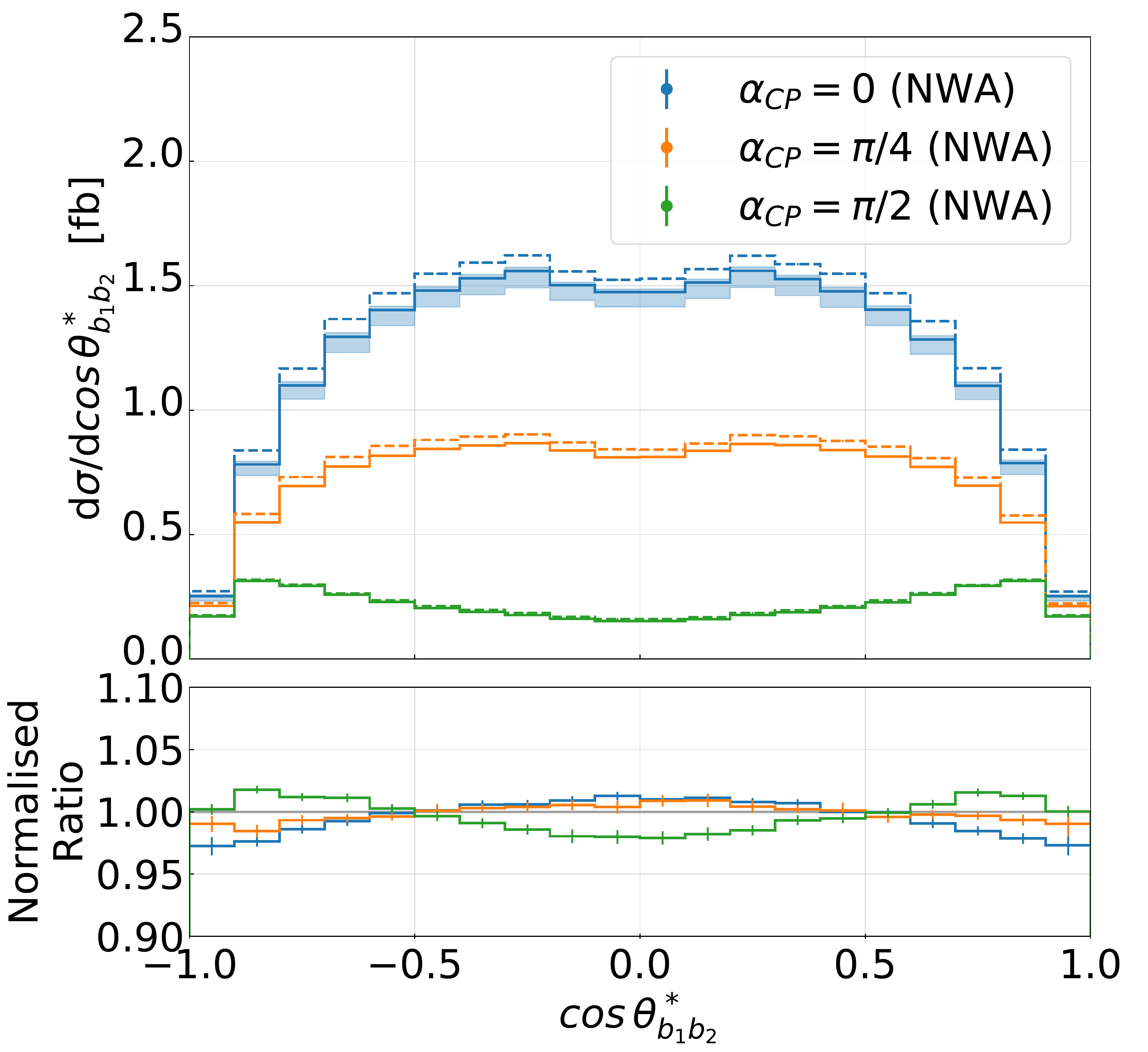}
	\end{center}
	\caption{\label{fig:cosa_lodec} \it
		Differential distributions at NLO in QCD for
                $\alpha_{CP}=0, {\pi}/{4}, {\pi}/{2}$ for the
                observables $p_{T,b_1}$, $p_{T,H}$, $M(be^+)_{min}$
                and $\cos\theta^*_{b_1b_2}$ for the $pp\to
                e^+\nu_e\,\mu^-\bar{\nu}_{\mu}\,b\bar{b}\,H$ process
                at the LHC with $\sqrt{s}=13\textrm{ TeV}$. The upper
                panels show NWA and $\textrm{NWA}_{\textrm{LOdec}}$
                (dashed lines) results. The lower panels provide the
                ratio NWA/$\textrm{NWA}_{\textrm{LOdec}}$ of the
                normalised distributions. Scale uncertainties are
                given for $\alpha_{CP}=0$ in the upper panel while
                Monte Carlo integration errors are displayed in both
                panels.}
\end{figure}
%

Finally, we investigate the effects of the inclusion of higher-order
QCD corrections to top-quark decays in an NLO calculation.  As an
example, we display the observables $p_{T,\,b_1}$, $p_{T,\,H}$,
$M(be^+)_{min}$ and $\cos\theta^*_{b_1b_2}$ in Figure
\ref{fig:cosa_lodec}.  In each case we plot the full NWA predictions
(solid lines) and the $\textrm{NWA}_{\textrm{LOdec}}$ ones (dashed
lines).  The lower panels display the ratio
NWA/$\textrm{NWA}_{\textrm{LOdec}}$ of the normalised differential
cross-section distributions.  For $p_{T,\, b_1}$ we find that the NLO
QCD corrections to the top-quark decays amount to about $20\%$ for
small transverse momenta and about $10\%$ for larger values of
$p_{T,\, b_1}$. They are very similar for all three ${\cal CP}$
configurations as the corresponding normalized ratios differ by less
than $2\%$ amongst themselves.  For $p_{T,H}$ the shape distortions
due to QCD corrections in the decay stage range from $-5\%$ to $10\%$
for all three ${\cal CP}$ configurations.  Thus again, the
higher-order effects do not depend on the ${\cal CP}$ nature of the
Higgs boson. Even for $M(be^+)_{min}$, for which we have found huge
differences between the three different ${\cal CP}$ states in the full
off-shell calculation, the inclusion of corrections in top-quark
decays is similar in all three cases.  Still, we find a large
enhancement of QCD corrections for $M(be^+)_{min} > 153$ GeV of the
order of $30\%$. Even for $M(be^+)_{min} < 153$ GeV the
$\mathcal{O}(\alpha_s)$ corrections to the top-quark decays lead to
substantial shape changes ranging from about $-20\%$ up to
$+10\%$. Therefore, in the vicinity of the kinematical edge, the
overall shape distortions reach up to $50\%$.  The dimensionless
observables are less affected by higher-order effects in the top-quark
decays. For angular distributions, like $\cos\theta^*_{b_1b_2}$, the
shape distortions in the normalised ratio do not exceed $4\%$ for all
three ${\cal CP}$ configurations. Furthermore, these shape distortions
do not differ by more than $4\%$ amongst the different ${\cal CP}$
states.

Thus, as expected and already discussed at the integrated fiducial
cross-section level, the NLO QCD corrections to top-quark decays are
similar in size and shape for all three considered ${\cal CP}$ states
of the Higgs boson in differential cross-section
distributions. Accordingly, on their own they do not help in
deciphering the ${\cal CP}$ nature of the Higgs boson.  Nevertheless,
higher-order corrections to top-quark decays cannot simply be
omitted. Not only do they range from a few percent up to $30\%$
depending on the phase-space region, introducing vast distortions in
the shape of various observables, but they also significantly impact
the final theoretical uncertainties. Therefore, they should not be
ignored in high-precision theoretical predictions for the $pp\to
e^+\nu_e\,\mu^-\bar{\nu}_{\mu}\,b\bar{b}\,H$ process.

In summary, when considering high precision theoretical predictions
for the associated production of the Higgs boson and top-quark pairs
various effects have to be taken into account. We have seen that NLO
QCD corrections to top-quark production and decays are indispensable
when studying the ${\cal CP}$ property of the Higgs
boson. Furthermore, the non-factorisable higher-order corrections,
which introduce the cross-talk between the production and decay stage,
play a crucial role. The latter cannot simply be approximated by some
overall factor. Indeed, they are vastly different for various
observables and their size depends on the examined phase-space region.

%
\section{Summary}
\label{sec:sum}
%

In this paper we have presented NLO QCD calculations for the $p p \to
e^+ \nu_e\, \mu^- \bar{\nu}_\mu\,b\bar{b}\,H$ process at the LHC with
$\sqrt{s}=13$ TeV for the ${\cal CP}$-even and ${\cal CP}$-odd state
of the Higgs boson as well as for possible mixing between these two
states.  We have placed particular emphasis on how the different
${\cal CP}$ states are affected by higher-order corrections and full
off-shell top-quark and $W$ gauge boson effects.  To this end, we have
compared integrated fiducial cross-sections and differential
distributions including full off-shell effects to those computed in
the narrow-width approximation for the ${\cal CP}$-even
($\alpha_{CP}=0$), ${\cal CP}$-odd ($\alpha_{CP}=\pi/2$) and ${\cal
CP}$-mixed ($\alpha_{CP}=\pi/4$) states.  For the full off-shell
calculations we have taken into account all double-, single- and
non-resonant diagrams as well as interference effects between those
and introduced Breit-Wigner distributions for the unstable top quarks
and gauge bosons through the complex-mass scheme to describe their
invariant mass distributions.

As a first step, we have demonstrated that even at the level of
integrated fiducial cross-sections, off-shell effects can
significantly exceed the typical size of $\mathcal{O} (\Gamma/
m)$. While these effects are below one percent for the ${\cal
CP}$-even and -mixed scenarios and thus still well within scale
uncertainties, they reach $13\% - 14\%$ for the pseudoscalar Higgs
boson.   This increase for the $\alpha_{CP}=\pi/2$ case
is driven by a substantially larger contribution from the
single-resonant top-quark region of the phase space. In addition, we
have investigated how the $\alpha_{CP}$-dependence of the integrated
cross-section changes between the NWA and the full off-shell
treatment.  Due to the absence of any $HVV$ couplings in the NWA, the
corresponding integrated cross-section is symmetric around
$\alpha_{CP} = \pi/2$.  However, this symmetry is broken if full
off-shell effects are included.  This is a result of interference
effects between diagrams in which the Higgs boson is radiated of a
gauge boson and those in which it is emitted from a top quark.  These
contributions are proportional to $\cos \left( \alpha_{CP} \right)$
instead of $\cos^2 \left( \alpha_{CP} \right)$ or $\sin^2 \left(
\alpha_{CP} \right)$ and thus not symmetric around $\alpha_{CP} =
\pi/2$.  Consequently, the integrated fiducial cross-section is about
$7 \%$ larger for $\alpha_{CP}=\pi$ than for the SM case.

Just like the off-shell effects, NLO corrections are also largest for
the ${\cal CP}$-odd scenario. We have shown that the
$\mathcal{K}-$factor is of the order of $1.21-1.23$ for the SM and
${\cal CP}$-mixed cases and $1.31$ for the ${\cal CP}$-odd one.
Furthermore, theoretical scale uncertainties are substantially reduced
when higher-order effects are incorporated. Specifically, they are at
the level of $31\%-34\%$ at LO and in the range of $5\%-6\%$ for the
NLO predictions. When only considering NLO QCD corrections to the
production stage, we have observed that the size of higher-order
effects increases together with the corresponding scale
uncertainties. Indeed, not only the $\mathcal{K}-$factor is now in the
range of $1.27-1.28$ for the ${\cal CP}$-even and ${\cal CP}$-mixed
Higgs boson and up to ${\cal K}=1.34$ for the ${\cal CP}$-odd case,
but also theoretical uncertainties due to the scale dependence
increase up to $11\%$. Therefore, they are almost twice as large as in
the full NWA or when full off-shell effects are included.
Independently of the modeling, the scale dependence is roughly the
same when comparing the different Higgs boson ${\cal CP}$ states.

Next to the overall normalisation, the NLO QCD corrections also change
the shape of several differential cross-section distributions.  In
contrast to their effects on the normalisation at the fiducial
cross-section level, the shape distortions are generally larger for
the SM case.  This is particularly apparent in the case of the
$p_{T,\,\text{miss}}$ and $M(be^+)_{min}$ distributions.  For these
observables the smaller shape distortions for the pseudoscalar Higgs
boson can mainly be ascribed to the generally harder Higgs boson
radiation and the larger relative size of single- and non-resonant
contributions.  In general, the $\mathcal{K}-$factors are larger in
the high-$p_T$ and -mass regions, irrespective of the Higgs bosons
${\cal CP}$ state.  Just like in the case of $t\bar{t}$ production and
for other $t\bar{t}$-associated production processes, NLO QCD
corrections are especially large for the $M(be^+)_{min}$ observable
above the kinematical edge around $153$ GeV. In this phase-space
region higher-order effects can even exceed $100 \%$ in the ${\cal
CP}$-even and ${\cal CP}$-mixed cases.

In the second part of our discussion on differential cross-section
distributions we have investigated the impact of full off-shell
effects on the distribution shapes.  Just as for the integrated
fiducial cross-section, the effects are fairly small for the SM and
${\cal CP}$-mixed cases apart from the high-$M(be^+)_\text{min}$ and
-$M_{T2,t}$ regions.  Both of these observables have kinematical edges
which are related to the masses of unstable intermediate
particles. For these two observables full off-shell effects can exceed
$50 \%$ above these edges.  For the pseudoscalar Higgs boson, on the
other hand, all observables are significantly affected by off-shell
effects.  Again, these effects are largest for $M(be^+)_\text{min}$
and $M_{T2,t}$ at up to $90 \%$. However, even in observables without
kinematic edges the differences can reach $45 \%$ in dimensionful and
$20 \%$ in dimensionless observables.  In general, the full off-shell
effects are larger in the high-$p_T$ and -mass phase-space regions and
for small angles between the decay products coming from two different
top quarks.

These changes in the behaviour of differential distributions also have
a significant impact on how well one can distinguish the three Higgs
boson ${\cal CP}$ states from each other at the differential fiducial
cross-section level.  We have shown that while the normalised
distributions for the ${\cal CP}$-even and ${\cal CP}$-mixed cases are
generally quite similar to each other, the ${\cal CP}$-odd case leads
to much more pronounced distribution tails in transverse momentum and
invariant mass distributions.  Again, this is mostly a result of the
harder radiation of a pseudoscalar Higgs boson compared to a scalar
one.  For $M_{T2,t}$ this effect is so large that the ${\cal CP}$-odd
distribution actually exceeds the one for the SM case above $240$ GeV
even though the integrated cross-section is about five times as large
for the latter.  Although not quite as drastic, the shape differences
in $H_T$, $p_{T,\,H}$ and $M(be^+)_{\text{min}}$ still result in
ratios of normalised distributions that reach a factor $2$ or more in
the tails of the respective distributions. Amongst dimensionless
cross-section distributions we have found that the $\cos \theta^*_{e^+
\mu^-}$ observable is the most promising candidate for distinguishing
the different ${\cal CP}$ states of the Higgs boson.

In addition, we have discussed the effects of omitting NLO QCD
corrections in top-quark decays at the differential fiducial
cross-section level by comparing the full NWA with the
$\text{NWA}_{\text{LOdec}}$ predictions. In contrast to the full NLO
QCD corrections, the shape distortions coming from NLO QCD corrections
to top-quark decays are largely independent of the ${\cal CP}$ state,
even above kinematic edges. Nevertheless, they should be taken into
account as they can reach up to $30\%$ depending on the considered
observable and phase-space region. Furthermore, they introduce vast
shape distortions and significantly impact the size of scale
uncertainties.

In summary, both NLO QCD corrections and full off-shell effects of the
top quark and $W$ gauge boson have a significant impact on the
normalisation and shape of various differential cross-section
distributions for the $p p \to e^+ \nu_e\, \mu^-
\bar{\nu}_\mu\,b\bar{b}\,H$ process at the LHC with $\sqrt{s}=13$ TeV
for all three Higgs boson ${\cal CP}$ states. However, the most
affected is the ${\cal CP}$-odd Higgs boson. For the ${\cal CP}$-even
and -mixed cases higher-order corrections play a crucial role but full
off-shell effects come into play mostly in the case of observables
with kinematic edges such as $M(be^+)_\text{min}$ and $M_{T2,t}$.  Due
to these kinematic edges, these observables are also the ones that are
most sensitive to the ${\cal CP}$ state of the Higgs boson.

Finally, let us mention again that the calculations presented in this
paper and the \textsc{Helac-NLO} framework constructed for this work
could easily be altered in order to obtain results for a general
beyond the SM Higgs-boson sector and are thus not limited to
predictions for an SM-like Higgs boson.

%
\acknowledgments{
The work  was supported by the Deutsche Forschungsgemeinschaft (DFG)
under grant 396021762 $-$ TRR 257: {\it P3H - Particle Physics
Phenomenology after the Higgs Discovery} and  by the DFG under grant
400140256 - GRK 2497: {\it The physics of the heaviest particles at the
LHC}.

Support by a grant of the Bundesministerium f\"ur Bildung
und Forschung (BMBF) is additionally acknowledged.

The authors gratefully acknowledge the computing time granted by the
Resource Allocation Board and provided on the supercomputer
CLAIX at RWTH Aachen University as part of the NHR infrastructure. The
calculations for this research were conducted with computing resources
under the projects {\tt rwth0414} and {\tt rwth0871}. }

\providecommand{\href}[2]{#2}\begingroup\raggedright\endgroup


\begin{thebibliography}{100}

\bibitem{ATLAS:2012yve}
{\scshape ATLAS} collaboration, \emph{{Observation of a new particle in the
  search for the Standard Model Higgs boson with the ATLAS detector at the
  LHC}}, \href{https://doi.org/10.1016/j.physletb.2012.08.020}{\emph{Phys.
  Lett. B} {\bfseries 716} (2012) 1}
  [\href{https://arxiv.org/abs/1207.7214}{{\ttfamily 1207.7214}}].

\bibitem{CMS:2012jmk}
{\scshape CMS} collaboration, \emph{{Observation of a New Boson at a Mass of
  125 GeV with the CMS Experiment at the LHC}},
  \href{https://doi.org/10.1016/j.physletb.2012.08.021}{\emph{Phys. Lett. B}
  {\bfseries 716} (2012) 30} [\href{https://arxiv.org/abs/1207.7235}{{\ttfamily
  1207.7235}}].

\bibitem{CMS:2014nkk}
{\scshape CMS} collaboration, \emph{{Constraints on the spin-parity and
  anomalous HVV couplings of the Higgs boson in proton collisions at 7 and 8
  TeV}}, \href{https://doi.org/10.1103/PhysRevD.92.012004}{\emph{Phys. Rev. D}
  {\bfseries 92} (2015) 012004}
  [\href{https://arxiv.org/abs/1411.3441}{{\ttfamily 1411.3441}}].

\bibitem{ATLAS:2015zhl}
{\scshape ATLAS} collaboration, \emph{{Study of the spin and parity of the
  Higgs boson in diboson decays with the ATLAS detector}},
  \href{https://doi.org/10.1140/epjc/s10052-015-3685-1}{\emph{Eur. Phys. J. C}
  {\bfseries 75} (2015) 476}
  [\href{https://arxiv.org/abs/1506.05669}{{\ttfamily 1506.05669}}].

\bibitem{ATLAS:2016ifi}
{\scshape ATLAS} collaboration, \emph{{Test of CP Invariance in vector-boson
  fusion production of the Higgs boson using the Optimal Observable method in
  the ditau decay channel with the ATLAS detector}},
  \href{https://doi.org/10.1140/epjc/s10052-016-4499-5}{\emph{Eur. Phys. J. C}
  {\bfseries 76} (2016) 658}
  [\href{https://arxiv.org/abs/1602.04516}{{\ttfamily 1602.04516}}].

\bibitem{CMS:2017len}
{\scshape CMS} collaboration, \emph{{Constraints on anomalous Higgs boson
  couplings using production and decay information in the four-lepton final
  state}}, \href{https://doi.org/10.1016/j.physletb.2017.10.021}{\emph{Phys.
  Lett. B} {\bfseries 775} (2017) 1}
  [\href{https://arxiv.org/abs/1707.00541}{{\ttfamily 1707.00541}}].

\bibitem{CMS:2019ekd}
{\scshape CMS} collaboration, \emph{{Measurements of the Higgs boson width and
  anomalous $HVV$ couplings from on-shell and off-shell production in the
  four-lepton final state}},
  \href{https://doi.org/10.1103/PhysRevD.99.112003}{\emph{Phys. Rev. D}
  {\bfseries 99} (2019) 112003}
  [\href{https://arxiv.org/abs/1901.00174}{{\ttfamily 1901.00174}}].

\bibitem{CMS:2019jdw}
{\scshape CMS} collaboration, \emph{{Constraints on anomalous $HVV$ couplings
  from the production of Higgs bosons decaying to $\tau$ lepton pairs}},
  \href{https://doi.org/10.1103/PhysRevD.100.112002}{\emph{Phys. Rev. D}
  {\bfseries 100} (2019) 112002}
  [\href{https://arxiv.org/abs/1903.06973}{{\ttfamily 1903.06973}}].

\bibitem{ATLAS:2020evk}
{\scshape ATLAS} collaboration, \emph{{Test of CP invariance in vector-boson
  fusion production of the Higgs boson in the $H\to \tau^+\tau^-$ channel in
  $pp$ collisions at $\sqrt{s}=13$ TeV with the ATLAS detector}},
  \href{https://doi.org/10.1016/j.physletb.2020.135426}{\emph{Phys. Lett. B}
  {\bfseries 805} (2020) 135426}
  [\href{https://arxiv.org/abs/2002.05315}{{\ttfamily 2002.05315}}].

\bibitem{CMS:2021nnc}
{\scshape CMS} collaboration, \emph{{Constraints on anomalous Higgs boson
  couplings to vector bosons and fermions in its production and decay using the
  four-lepton final state}},
  \href{https://doi.org/10.1103/PhysRevD.104.052004}{\emph{Phys. Rev. D}
  {\bfseries 104} (2021) 052004}
  [\href{https://arxiv.org/abs/2104.12152}{{\ttfamily 2104.12152}}].

\bibitem{ATLAS:2013dos}
{\scshape ATLAS} collaboration, \emph{{Measurements of Higgs boson production
heis  and couplings in diboson final states with the ATLAS detector at the LHC}},
  \href{https://doi.org/10.1016/j.physletb.2014.05.011}{\emph{Phys. Lett. B}
  {\bfseries 726} (2013) 88} [\href{https://arxiv.org/abs/1307.1427}{{\ttfamily
  1307.1427}}].

\bibitem{CMS:2014fzn}
{\scshape CMS} collaboration, \emph{{Precise determination of the mass of the
  Higgs boson and tests of compatibility of its couplings with the standard
  model predictions using proton collisions at 7 and 8 $\,\text {TeV}$}},
  \href{https://doi.org/10.1140/epjc/s10052-015-3351-7}{\emph{Eur. Phys. J. C}
  {\bfseries 75} (2015) 212} [\href{https://arxiv.org/abs/1412.8662}{{\ttfamily
  1412.8662}}].

\bibitem{ATLAS:2016neq}
{\scshape ATLAS, CMS} collaboration, \emph{{Measurements of the Higgs boson
  production and decay rates and constraints on its couplings from a combined
  ATLAS and CMS analysis of the LHC pp collision data at $ \sqrt{s}=7 $ and 8
  TeV}}, \href{https://doi.org/10.1007/JHEP08(2016)045}{\emph{JHEP} {\bfseries
  08} (2016) 045} [\href{https://arxiv.org/abs/1606.02266}{{\ttfamily
  1606.02266}}].

\bibitem{Sakharov_1991}
A.~D. Sakharov, \emph{Violation of cp invariance, c asymmetry, and baryon
  asymmetry of the universe},
  \href{https://doi.org/10.1070/pu1991v034n05abeh002497}{\emph{Soviet Physics
  Uspekhi} {\bfseries 34} (1991) 392}.

\bibitem{Kuzmin:1985mm}
V.~A. Kuzmin, V.~A. Rubakov and M.~E. Shaposhnikov, \emph{{On the Anomalous
  Electroweak Baryon Number Nonconservation in the Early Universe}},
  \href{https://doi.org/10.1016/0370-2693(85)91028-7}{\emph{Phys. Lett. B}
  {\bfseries 155} (1985) 36}.

\bibitem{Gunion:1989we}
J.~F. Gunion, H.~E. Haber, G.~L. Kane and S.~Dawson, \emph{{The Higgs Hunter's
  Guide}}, vol.~80. 2000.

\bibitem{Gunion:2002zf}
J.~F. Gunion and H.~E. Haber, \emph{{The CP conserving two Higgs doublet model:
  The Approach to the decoupling limit}},
  \href{https://doi.org/10.1103/PhysRevD.67.075019}{\emph{Phys. Rev. D}
  {\bfseries 67} (2003) 075019}
  [\href{https://arxiv.org/abs/hep-ph/0207010}{{\ttfamily hep-ph/0207010}}].

\bibitem{Branco:2011iw}
G.~C. Branco, P.~M. Ferreira, L.~Lavoura, M.~N. Rebelo, M.~Sher and J.~P.
  Silva, \emph{{Theory and phenomenology of two-Higgs-doublet models}},
  \href{https://doi.org/10.1016/j.physrep.2012.02.002}{\emph{Phys. Rept.}
  {\bfseries 516} (2012) 1} [\href{https://arxiv.org/abs/1106.0034}{{\ttfamily
  1106.0034}}].

\bibitem{Grzadkowski:2014ada}
B.~Grzadkowski, O.~M. Ogreid and P.~Osland, \emph{{Measuring CP violation in
  Two-Higgs-Doublet models in light of the LHC Higgs data}},
  \href{https://doi.org/10.1007/JHEP11(2014)084}{\emph{JHEP} {\bfseries 11}
  (2014) 084} [\href{https://arxiv.org/abs/1409.7265}{{\ttfamily 1409.7265}}].

\bibitem{Ellis:2013yxa}
J.~Ellis, D.~S. Hwang, K.~Sakurai and M.~Takeuchi, \emph{{Disentangling
  Higgs-Top Couplings in Associated Production}},
  \href{https://doi.org/10.1007/JHEP04(2014)004}{\emph{JHEP} {\bfseries 04}
  (2014) 004} [\href{https://arxiv.org/abs/1312.5736}{{\ttfamily 1312.5736}}].

\bibitem{Boudjema:2015nda}
F.~Boudjema, R.~M. Godbole, D.~Guadagnoli and K.~A. Mohan, \emph{{Lab-frame
  observables for probing the top-Higgs interaction}},
  \href{https://doi.org/10.1103/PhysRevD.92.015019}{\emph{Phys. Rev. D}
  {\bfseries 92} (2015) 015019}
  [\href{https://arxiv.org/abs/1501.03157}{{\ttfamily 1501.03157}}].

\bibitem{Buckley:2015vsa}
M.~R. Buckley and D.~Goncalves, \emph{{Boosting the Direct CP Measurement of
  the Higgs-Top Coupling}},
  \href{https://doi.org/10.1103/PhysRevLett.116.091801}{\emph{Phys. Rev. Lett.}
  {\bfseries 116} (2016) 091801}
  [\href{https://arxiv.org/abs/1507.07926}{{\ttfamily 1507.07926}}].

\bibitem{AmorDosSantos:2017ayi}
S.~Amor Dos~Santos et~al., \emph{{Probing the CP nature of the Higgs coupling
  in $t{\bar t}h$ events at the LHC}},
  \href{https://doi.org/10.1103/PhysRevD.96.013004}{\emph{Phys. Rev. D}
  {\bfseries 96} (2017) 013004}
  [\href{https://arxiv.org/abs/1704.03565}{{\ttfamily 1704.03565}}].

\bibitem{CMS:2017dib}
{\scshape CMS} collaboration, \emph{{Measurements of properties of the Higgs
  boson decaying into the four-lepton final state in pp collisions at $
  \sqrt{s}=13 $ TeV}},
  \href{https://doi.org/10.1007/JHEP11(2017)047}{\emph{JHEP} {\bfseries 11}
  (2017) 047} [\href{https://arxiv.org/abs/1706.09936}{{\ttfamily
  1706.09936}}].

\bibitem{CMS:2018piu}
{\scshape CMS} collaboration, \emph{{Measurements of Higgs boson properties in
  the diphoton decay channel in proton-proton collisions at $\sqrt{s} =$ 13
  TeV}}, \href{https://doi.org/10.1007/JHEP11(2018)185}{\emph{JHEP} {\bfseries
  11} (2018) 185} [\href{https://arxiv.org/abs/1804.02716}{{\ttfamily
  1804.02716}}].

\bibitem{ATLAS:2018doi}
{\scshape ATLAS} collaboration, \emph{{Combined measurements of Higgs boson
  production and decay using up to 80 fb$^{-1}$ of proton--proton collision
  data at $\sqrt{s}=$ 13 TeV collected with the ATLAS experiment, {\tt
  [ATLAS-CONF-2018-031]}}}, .

\bibitem{CMS:2018uag}
{\scshape CMS} collaboration, \emph{{Combined measurements of Higgs boson
  couplings in proton\textendash{}proton collisions at $\sqrt{s}=13\,\text
  {Te}\text {V} $}},
  \href{https://doi.org/10.1140/epjc/s10052-019-6909-y}{\emph{Eur. Phys. J. C}
  {\bfseries 79} (2019) 421}
  [\href{https://arxiv.org/abs/1809.10733}{{\ttfamily 1809.10733}}].

\bibitem{ATLAS:2018mme}
{\scshape ATLAS} collaboration, \emph{{Observation of Higgs boson production in
  association with a top quark pair at the LHC with the ATLAS detector}},
  \href{https://doi.org/10.1016/j.physletb.2018.07.035}{\emph{Phys. Lett. B}
  {\bfseries 784} (2018) 173}
  [\href{https://arxiv.org/abs/1806.00425}{{\ttfamily 1806.00425}}].

\bibitem{CMS:2018uxb}
{\scshape CMS} collaboration, \emph{{Observation of $t\overline{t}H$
  production}},
  \href{https://doi.org/10.1103/PhysRevLett.120.231801}{\emph{Phys. Rev. Lett.}
  {\bfseries 120} (2018) 231801}
  [\href{https://arxiv.org/abs/1804.02610}{{\ttfamily 1804.02610}}].

\bibitem{ATLAS:2020ior}
{\scshape ATLAS} collaboration, \emph{{$CP$ Properties of Higgs Boson
  Interactions with Top Quarks in the $t\bar{t}H$ and $tH$ Processes Using $H
  \rightarrow \gamma\gamma$ with the ATLAS Detector}},
  \href{https://doi.org/10.1103/PhysRevLett.125.061802}{\emph{Phys. Rev. Lett.}
  {\bfseries 125} (2020) 061802}
  [\href{https://arxiv.org/abs/2004.04545}{{\ttfamily 2004.04545}}].

\bibitem{CMS:2020cga}
{\scshape CMS} collaboration, \emph{{Measurements of ${t\bar{t}}H$ Production
  and the CP Structure of the Yukawa Interaction between the Higgs Boson and
  Top Quark in the Diphoton Decay Channel}},
  \href{https://doi.org/10.1103/PhysRevLett.125.061801}{\emph{Phys. Rev. Lett.}
  {\bfseries 125} (2020) 061801}
  [\href{https://arxiv.org/abs/2003.10866}{{\ttfamily 2003.10866}}].

\bibitem{CMS:2022def}
{\scshape CMS} collaboration, \emph{{Search for CP violation in $t\bar{t}H$ and
  $tH$ production in multilepton channels at $\sqrt{s} = 13~\mathrm{TeV}$, {\tt
  [CMS-PAS-HIG-21-006]}}}, .

\bibitem{CMS-PAS-HIG-20-006}
{\scshape CMS} collaboration, \emph{{Analysis of the CP structure of the Yukawa
  coupling between the Higgs boson and $\tau$ leptons in proton-proton
  collisions at $\sqrt{s}=13~\mathrm{TeV}$, {\tt [CMS-PAS-HIG-20-006]}}},
  tech. rep., CERN, Geneva, 2020.

\bibitem{Artoisenet:2013puc}
P.~Artoisenet et~al., \emph{{A framework for Higgs characterisation}},
  \href{https://doi.org/10.1007/JHEP11(2013)043}{\emph{JHEP} {\bfseries 11}
  (2013) 043} [\href{https://arxiv.org/abs/1306.6464}{{\ttfamily 1306.6464}}].

\bibitem{Hou:2018uvr}
W.-S. Hou, M.~Kohda and T.~Modak, \emph{{Probing for extra top Yukawa couplings
  in light of $t\bar th(125)$ observation}},
  \href{https://doi.org/10.1103/PhysRevD.98.075007}{\emph{Phys. Rev. D}
  {\bfseries 98} (2018) 075007}
  [\href{https://arxiv.org/abs/1806.06018}{{\ttfamily 1806.06018}}].

\bibitem{Bahl:2020wee}
H.~Bahl, P.~Bechtle, S.~Heinemeyer, J.~Katzy, T.~Klingl, K.~Peters et~al.,
  \emph{{Indirect $\mathcal{CP}$ probes of the Higgs-top-quark interaction:
  current LHC constraints and future opportunities}},
  \href{https://doi.org/10.1007/JHEP11(2020)127}{\emph{JHEP} {\bfseries 11}
  (2020) 127} [\href{https://arxiv.org/abs/2007.08542}{{\ttfamily
  2007.08542}}].

\bibitem{Azevedo:2020fdl}
D.~Azevedo, R.~Capucha, E.~Gouveia, A.~Onofre and R.~Santos, \emph{{Light Higgs
  searches in $ t\overline{t}\phi $ production at the LHC}},
  \href{https://doi.org/10.1007/JHEP04(2021)077}{\emph{JHEP} {\bfseries 04}
  (2021) 077} [\href{https://arxiv.org/abs/2012.10730}{{\ttfamily
  2012.10730}}].

\bibitem{Bortolato:2020zcg}
B.~Bortolato, J.~F. Kamenik, N.~Ko\v{s}nik and A.~Smolkovi\v{c},
  \emph{{Optimized probes of $CP$ -odd effects in the $t \bar t h$ process at
  hadron colliders}},
  \href{https://doi.org/10.1016/j.nuclphysb.2021.115328}{\emph{Nucl. Phys. B}
  {\bfseries 964} (2021) 115328}
  [\href{https://arxiv.org/abs/2006.13110}{{\ttfamily 2006.13110}}].

\bibitem{Martini:2021uey}
T.~Martini, R.-Q. Pan, M.~Schulze and M.~Xiao, \emph{{Probing the CP structure
  of the top quark Yukawa coupling: Loop sensitivity versus on-shell
  sensitivity}}, \href{https://doi.org/10.1103/PhysRevD.104.055045}{\emph{Phys.
  Rev. D} {\bfseries 104} (2021) 055045}
  [\href{https://arxiv.org/abs/2104.04277}{{\ttfamily 2104.04277}}].

\bibitem{Bahl:2021dnc}
H.~Bahl and S.~Brass, \emph{{Constraining $ \mathcal{CP} $-violation in the
  Higgs-top-quark interaction using machine-learning-based inference}},
  \href{https://doi.org/10.1007/JHEP03(2022)017}{\emph{JHEP} {\bfseries 03}
  (2022) 017} [\href{https://arxiv.org/abs/2110.10177}{{\ttfamily
  2110.10177}}].

\bibitem{Frederix:2011zi}
R.~Frederix, S.~Frixione, V.~Hirschi, F.~Maltoni, R.~Pittau and P.~Torrielli,
  \emph{{Scalar and pseudoscalar Higgs production in association with a
  top\textendash{}antitop pair}},
  \href{https://doi.org/10.1016/j.physletb.2011.06.012}{\emph{Phys. Lett. B}
  {\bfseries 701} (2011) 427}
  [\href{https://arxiv.org/abs/1104.5613}{{\ttfamily 1104.5613}}].

\bibitem{Garzelli:2011vp}
M.~V. Garzelli, A.~Kardos, C.~G. Papadopoulos and Z.~Trocsanyi, \emph{{Standard
  Model Higgs boson production in association with a top anti-top pair at NLO
  with parton showering}},
  \href{https://doi.org/10.1209/0295-5075/96/11001}{\emph{EPL} {\bfseries 96}
  (2011) 11001} [\href{https://arxiv.org/abs/1108.0387}{{\ttfamily
  1108.0387}}].

\bibitem{Artoisenet:2012st}
P.~Artoisenet, R.~Frederix, O.~Mattelaer and R.~Rietkerk, \emph{{Automatic
  spin-entangled decays of heavy resonances in Monte Carlo simulations}},
  \href{https://doi.org/10.1007/JHEP03(2013)015}{\emph{JHEP} {\bfseries 03}
  (2013) 015} [\href{https://arxiv.org/abs/1212.3460}{{\ttfamily 1212.3460}}].

\bibitem{Biswas:2014hwa}
S.~Biswas, R.~Frederix, E.~Gabrielli and B.~Mele, \emph{{Enhancing the
  $t\bar{t}H$ signal through top-quark spin polarization effects at the LHC}},
  \href{https://doi.org/10.1007/JHEP07(2014)020}{\emph{JHEP} {\bfseries 07}
  (2014) 020} [\href{https://arxiv.org/abs/1403.1790}{{\ttfamily 1403.1790}}].

\bibitem{Demartin:2014fia}
F.~Demartin, F.~Maltoni, K.~Mawatari, B.~Page and M.~Zaro, \emph{{Higgs
  characterisation at NLO in QCD: CP properties of the top-quark Yukawa
  interaction}},
  \href{https://doi.org/10.1140/epjc/s10052-014-3065-2}{\emph{Eur. Phys. J. C}
  {\bfseries 74} (2014) 3065}
  [\href{https://arxiv.org/abs/1407.5089}{{\ttfamily 1407.5089}}].

\bibitem{Hartanto:2015uka}
H.~B. Hartanto, B.~Jager, L.~Reina and D.~Wackeroth, \emph{{Higgs boson
  production in association with top quarks in the POWHEG BOX}},
  \href{https://doi.org/10.1103/PhysRevD.91.094003}{\emph{Phys. Rev. D}
  {\bfseries 91} (2015) 094003}
  [\href{https://arxiv.org/abs/1501.04498}{{\ttfamily 1501.04498}}].

\bibitem{Demartin:2015uha}
F.~Demartin, F.~Maltoni, K.~Mawatari and M.~Zaro, \emph{{Higgs production in
  association with a single top quark at the LHC}},
  \href{https://doi.org/10.1140/epjc/s10052-015-3475-9}{\emph{Eur. Phys. J. C}
  {\bfseries 75} (2015) 267}
  [\href{https://arxiv.org/abs/1504.00611}{{\ttfamily 1504.00611}}].

\bibitem{Demartin:2016axk}
F.~Demartin, B.~Maier, F.~Maltoni, K.~Mawatari and M.~Zaro, \emph{{tWH
  associated production at the LHC}},
  \href{https://doi.org/10.1140/epjc/s10052-017-4601-7}{\emph{Eur. Phys. J. C}
  {\bfseries 77} (2017) 34} [\href{https://arxiv.org/abs/1607.05862}{{\ttfamily
  1607.05862}}].

\bibitem{Frixione:2008yi}
S.~Frixione, E.~Laenen, P.~Motylinski, B.~R. Webber and C.~D. White,
  \emph{{Single-top hadroproduction in association with a W boson}},
  \href{https://doi.org/10.1088/1126-6708/2008/07/029}{\emph{JHEP} {\bfseries
  07} (2008) 029} [\href{https://arxiv.org/abs/0805.3067}{{\ttfamily
  0805.3067}}].

\bibitem{Jezo:2016ujg}
T.~Jezo, J.~M.~Lindert, P.~Nason, C.~Oleari and S.~Pozzorini,
\emph{An NLO+PS generator for $t\bar{t}$ and $Wt$ production and decay
  including non-resonant and interference effects},
\href{https://doi.org/10.1140/epjc/s10052-016-4538-2}{\emph{Eur. Phys. J. C}
  {\bfseries 76} (2016) no.12, 691}
  [\href{https://arxiv.org/abs/1607.04538}{\ttfamily 1607.04538}].

\bibitem{Denner:2015yca}
A.~Denner and R.~Feger, \emph{{NLO QCD corrections to off-shell top-antitop
  production with leptonic decays in association with a Higgs boson at the
  LHC}}, \href{https://doi.org/10.1007/JHEP11(2015)209}{\emph{JHEP} {\bfseries
  11} (2015) 209} [\href{https://arxiv.org/abs/1506.07448}{{\ttfamily
  1506.07448}}].

\bibitem{Stremmer:2021bnk}
D.~Stremmer and M.~Worek, \emph{{Production and decay of the Higgs boson in
  association with top quarks}},
  \href{https://doi.org/10.1007/JHEP02(2022)196}{\emph{JHEP} {\bfseries 02}
  (2022) 196} [\href{https://arxiv.org/abs/2111.01427}{{\ttfamily
  2111.01427}}].

\bibitem{Denner:2016wet}
A.~Denner, J.-N. Lang, M.~Pellen and S.~Uccirati, \emph{{Higgs production in
  association with off-shell top-antitop pairs at NLO EW and QCD at the LHC}},
  \href{https://doi.org/10.1007/JHEP02(2017)053}{\emph{JHEP} {\bfseries 02}
  (2017) 053} [\href{https://arxiv.org/abs/1612.07138}{{\ttfamily
  1612.07138}}].

\bibitem{Harlander:2020cyh}
R.~V. Harlander, S.~Y. Klein and M.~Lipp, \emph{{FeynGame}},
  \href{https://doi.org/10.1016/j.cpc.2020.107465}{\emph{Comput. Phys. Commun.}
  {\bfseries 256} (2020) 107465}
  [\href{https://arxiv.org/abs/2003.00896}{{\ttfamily 2003.00896}}].

\bibitem{Denner:1999gp}
A.~Denner, S.~Dittmaier, M.~Roth and D.~Wackeroth, \emph{{Predictions for all
  processes $e^+ e^- \to$ 4 fermions $+$ $\gamma$}},
  \href{https://doi.org/10.1016/S0550-3213(99)00437-X}{\emph{Nucl. Phys. B}
  {\bfseries 560} (1999) 33}
  [\href{https://arxiv.org/abs/hep-ph/9904472}{{\ttfamily hep-ph/9904472}}].

\bibitem{Denner:2005fg}
A.~Denner, S.~Dittmaier, M.~Roth and L.~H. Wieders, \emph{{Electroweak
  corrections to charged-current $e^+ e^-\to 4$ fermion processes: Technical
  details and further results}},
  \href{https://doi.org/10.1016/j.nuclphysb.2011.09.001}{\emph{Nucl. Phys. B}
  {\bfseries 724} (2005) 247}
  [\href{https://arxiv.org/abs/hep-ph/0505042}{{\ttfamily hep-ph/0505042}}].

\bibitem{Bevilacqua:2010qb}
G.~Bevilacqua, M.~Czakon, A.~van Hameren, C.~G. Papadopoulos and M.~Worek,
  \emph{{Complete off-shell effects in top quark pair hadroproduction with
  leptonic decay at next-to-leading order}},
  \href{https://doi.org/10.1007/JHEP02(2011)083}{\emph{JHEP} {\bfseries 02}
  (2011) 083} [\href{https://arxiv.org/abs/1012.4230}{{\ttfamily 1012.4230}}].

\bibitem{Denner:2012yc}
A.~Denner, S.~Dittmaier, S.~Kallweit and S.~Pozzorini, \emph{{NLO QCD
  corrections to off-shell top-antitop production with leptonic decays at
  hadron colliders}},
  \href{https://doi.org/10.1007/JHEP10(2012)110}{\emph{JHEP} {\bfseries 10}
  (2012) 110} [\href{https://arxiv.org/abs/1207.5018}{{\ttfamily 1207.5018}}].

\bibitem{Bevilacqua:2011xh}
G.~Bevilacqua, M.~Czakon, M.~V. Garzelli, A.~van Hameren, A.~Kardos, C.~G.
  Papadopoulos et~al., \emph{{HELAC-NLO}},
  \href{https://doi.org/10.1016/j.cpc.2012.10.033}{\emph{Comput. Phys. Commun.}
  {\bfseries 184} (2013) 986}
  [\href{https://arxiv.org/abs/1110.1499}{{\ttfamily 1110.1499}}].

\bibitem{Bevilacqua:2019quz}
G.~Bevilacqua, H.~B. Hartanto, M.~Kraus, T.~Weber and M.~Worek,
  \emph{{Off-shell vs on-shell modelling of top quarks in photon associated
  production}}, \href{https://doi.org/10.1007/JHEP03(2020)154}{\emph{JHEP}
  {\bfseries 03} (2020) 154}
  [\href{https://arxiv.org/abs/1912.09999}{{\ttfamily 1912.09999}}].

\bibitem{Czakon:2009ss}
M.~Czakon, C.~G. Papadopoulos and M.~Worek, \emph{{Polarizing the Dipoles}},
  \href{https://doi.org/10.1088/1126-6708/2009/08/085}{\emph{JHEP} {\bfseries
  08} (2009) 085} [\href{https://arxiv.org/abs/0905.0883}{{\ttfamily
  0905.0883}}].

\bibitem{vanHameren:2009dr}
A.~van Hameren, C.~G. Papadopoulos and R.~Pittau, \emph{{Automated one-loop
  calculations: A Proof of concept}},
  \href{https://doi.org/10.1088/1126-6708/2009/09/106}{\emph{JHEP} {\bfseries
  09} (2009) 106} [\href{https://arxiv.org/abs/0903.4665}{{\ttfamily
  0903.4665}}].

\bibitem{Draggiotis:1998gr}
P.~Draggiotis, R.~H.~P. Kleiss and C.~G. Papadopoulos, \emph{{On the
  computation of multigluon amplitudes}},
  \href{https://doi.org/10.1016/S0370-2693(98)01015-6}{\emph{Phys. Lett. B}
  {\bfseries 439} (1998) 157}
  [\href{https://arxiv.org/abs/hep-ph/9807207}{{\ttfamily hep-ph/9807207}}].

\bibitem{Draggiotis:2002hm}
P.~D. Draggiotis, R.~H.~P. Kleiss and C.~G. Papadopoulos, \emph{{Multijet
  production in hadron collisions}},
  \href{https://doi.org/10.1007/s10052-002-0955-5}{\emph{Eur. Phys. J. C}
  {\bfseries 24} (2002) 447}
  [\href{https://arxiv.org/abs/hep-ph/0202201}{{\ttfamily hep-ph/0202201}}].

\bibitem{Papadopoulos:2005ky}
C.~G. Papadopoulos and M.~Worek, \emph{{Multi-parton cross sections at hadron
  colliders}}, \href{https://doi.org/10.1140/epjc/s10052-007-0246-2}{\emph{Eur.
  Phys. J. C} {\bfseries 50} (2007) 843}
  [\href{https://arxiv.org/abs/hep-ph/0512150}{{\ttfamily hep-ph/0512150}}].

\bibitem{vanHameren:2007pt}
A.~van Hameren, \emph{{PARNI for importance sampling and density estimation}},
  {\emph{Acta Phys. Polon. B} {\bfseries 40} (2009) 259}
  [\href{https://arxiv.org/abs/0710.2448}{{\ttfamily 0710.2448}}].

\bibitem{vanHameren:2010gg}
A.~van Hameren, \emph{{Kaleu: A General-Purpose Parton-Level Phase Space
  Generator}},  \href{https://arxiv.org/abs/1003.4953}{{\ttfamily 1003.4953}}.

\bibitem{Bevilacqua:2018woc}
G.~Bevilacqua, H.~B. Hartanto, M.~Kraus, T.~Weber and M.~Worek, \emph{{Hard
  Photons in Hadroproduction of Top Quarks with Realistic Final States}},
  \href{https://doi.org/10.1007/JHEP10(2018)158}{\emph{JHEP} {\bfseries 10}
  (2018) 158} [\href{https://arxiv.org/abs/1803.09916}{{\ttfamily
  1803.09916}}].

\bibitem{Bevilacqua:2019cvp}
G.~Bevilacqua, H.~B. Hartanto, M.~Kraus, T.~Weber and M.~Worek, \emph{{Towards
  constraining Dark Matter at the LHC: Higher order QCD predictions for
  $t\bar{t}+Z(Z\to \nu_\ell \bar{\nu}_\ell)$}},
  \href{https://doi.org/10.1007/JHEP11(2019)001}{\emph{JHEP} {\bfseries 11}
  (2019) 001} [\href{https://arxiv.org/abs/1907.09359}{{\ttfamily
  1907.09359}}].

\bibitem{Bevilacqua:2020pzy}
G.~Bevilacqua, H.-Y. Bi, H.~B. Hartanto, M.~Kraus and M.~Worek, \emph{{The
  simplest of them all: $t\bar{t} W^\pm$ at NLO accuracy in QCD}},
  \href{https://doi.org/10.1007/JHEP08(2020)043}{\emph{JHEP} {\bfseries 08}
  (2020) 043} [\href{https://arxiv.org/abs/2005.09427}{{\ttfamily
  2005.09427}}].

\bibitem{Ossola:2007ax}
G.~Ossola, C.~G. Papadopoulos and R.~Pittau, \emph{{CutTools: A Program
  implementing the OPP reduction method to compute one-loop amplitudes}},
  \href{https://doi.org/10.1088/1126-6708/2008/03/042}{\emph{JHEP} {\bfseries
  03} (2008) 042} [\href{https://arxiv.org/abs/0711.3596}{{\ttfamily
  0711.3596}}].

\bibitem{Ossola:2006us}
G.~Ossola, C.~G. Papadopoulos and R.~Pittau, \emph{{Reducing full one-loop
  amplitudes to scalar integrals at the integrand level}},
  \href{https://doi.org/10.1016/j.nuclphysb.2006.11.012}{\emph{Nucl. Phys. B}
  {\bfseries 763} (2007) 147}
  [\href{https://arxiv.org/abs/hep-ph/0609007}{{\ttfamily hep-ph/0609007}}].

\bibitem{vanHameren:2010cp}
A.~van Hameren, \emph{{OneLOop: For the evaluation of one-loop scalar
  functions}}, \href{https://doi.org/10.1016/j.cpc.2011.06.011}{\emph{Comput.
  Phys. Commun.} {\bfseries 182} (2011) 2427}
  [\href{https://arxiv.org/abs/1007.4716}{{\ttfamily 1007.4716}}].

\bibitem{Catani:1996vz}
S.~Catani and M.~H. Seymour, \emph{{A General algorithm for calculating jet
  cross-sections in NLO QCD}},
  \href{https://doi.org/10.1016/S0550-3213(96)00589-5}{\emph{Nucl. Phys. B}
  {\bfseries 485} (1997) 291}
  [\href{https://arxiv.org/abs/hep-ph/9605323}{{\ttfamily hep-ph/9605323}}].

\bibitem{Catani:2002hc}
S.~Catani, S.~Dittmaier, M.~H. Seymour and Z.~Trocsanyi, \emph{{The Dipole
  formalism for next-to-leading order QCD calculations with massive partons}},
  \href{https://doi.org/10.1016/S0550-3213(02)00098-6}{\emph{Nucl. Phys. B}
  {\bfseries 627} (2002) 189}
  [\href{https://arxiv.org/abs/hep-ph/0201036}{{\ttfamily hep-ph/0201036}}].

\bibitem{Bevilacqua:2013iha}
G.~Bevilacqua, M.~Czakon, M.~Kubocz and M.~Worek, \emph{{Complete Nagy-Soper
  subtraction for next-to-leading order calculations in QCD}},
  \href{https://doi.org/10.1007/JHEP10(2013)204}{\emph{JHEP} {\bfseries 10}
  (2013) 204} [\href{https://arxiv.org/abs/1308.5605}{{\ttfamily 1308.5605}}].

\bibitem{Campbell:2004ch}
J.~M. Campbell, R.~K. Ellis and F.~Tramontano, \emph{{Single top production and
  decay at next-to-leading order}},
  \href{https://doi.org/10.1103/PhysRevD.70.094012}{\emph{Phys. Rev. D}
  {\bfseries 70} (2004) 094012}
  [\href{https://arxiv.org/abs/hep-ph/0408158}{{\ttfamily hep-ph/0408158}}].

\bibitem{Bevilacqua:2016jfk}
G.~Bevilacqua, H.~B. Hartanto, M.~Kraus and M.~Worek, \emph{{Off-shell Top
  Quarks with One Jet at the LHC: A comprehensive analysis at NLO QCD}},
  \href{https://doi.org/10.1007/JHEP11(2016)098}{\emph{JHEP} {\bfseries 11}
  (2016) 098} [\href{https://arxiv.org/abs/1609.01659}{{\ttfamily
  1609.01659}}].

\bibitem{Alwall:2006yp}
J.~Alwall et~al., \emph{{A Standard format for Les Houches event files}},
  \href{https://doi.org/10.1016/j.cpc.2006.11.010}{\emph{Comput. Phys. Commun.}
  {\bfseries 176} (2007) 300}
  [\href{https://arxiv.org/abs/hep-ph/0609017}{{\ttfamily hep-ph/0609017}}].

\bibitem{Antcheva:2009zz}
I.~Antcheva et~al., \emph{{ROOT: A C++ framework for petabyte data storage,
  statistical analysis and visualization}},
  \href{https://doi.org/10.1016/j.cpc.2009.08.005}{\emph{Comput. Phys. Commun.}
  {\bfseries 180} (2009) 2499}
  [\href{https://arxiv.org/abs/1508.07749}{{\ttfamily 1508.07749}}].

\bibitem{Alwall:2014hca}
J.~Alwall, R.~Frederix, S.~Frixione, V.~Hirschi, F.~Maltoni, O.~Mattelaer
  et~al., \emph{{The automated computation of tree-level and next-to-leading
  order differential cross sections, and their matching to parton shower
  simulations}}, \href{https://doi.org/10.1007/JHEP07(2014)079}{\emph{JHEP}
  {\bfseries 07} (2014) 079} [\href{https://arxiv.org/abs/1405.0301}{{\ttfamily
  1405.0301}}].

\bibitem{Hirschi:2011pa}
V.~Hirschi, R.~Frederix, S.~Frixione, M.~V. Garzelli, F.~Maltoni and R.~Pittau,
  \emph{{Automation of one-loop QCD corrections}},
  \href{https://doi.org/10.1007/JHEP05(2011)044}{\emph{JHEP} {\bfseries 05}
  (2011) 044} [\href{https://arxiv.org/abs/1103.0621}{{\ttfamily 1103.0621}}].

\bibitem{Kleiss:1985gy}
R.~Kleiss, W.~J. Stirling and S.~D. Ellis, \emph{{A New Monte Carlo Treatment
  of Multiparticle Phase Space at High-energies}},
  \href{https://doi.org/10.1016/0010-4655(86)90119-0}{\emph{Comput. Phys.
  Commun.} {\bfseries 40} (1986) 359}.

\bibitem{Nagy:1998bb}
Z.~Nagy and Z.~Trocsanyi, \emph{{Next-to-leading order calculation of four jet
  observables in electron positron annihilation}},
  \href{https://doi.org/10.1103/PhysRevD.62.099902}{\emph{Phys. Rev. D}
  {\bfseries 59} (1999) 014020}
  [\href{https://arxiv.org/abs/hep-ph/9806317}{{\ttfamily hep-ph/9806317}}].

\bibitem{Nagy:2003tz}
Z.~Nagy, \emph{{Next-to-leading order calculation of three jet observables in
  hadron hadron collision}},
  \href{https://doi.org/10.1103/PhysRevD.68.094002}{\emph{Phys. Rev. D}
  {\bfseries 68} (2003) 094002}
  [\href{https://arxiv.org/abs/hep-ph/0307268}{{\ttfamily hep-ph/0307268}}].

\bibitem{Bevilacqua:2009zn}
G.~Bevilacqua, M.~Czakon, C.~G. Papadopoulos, R.~Pittau and M.~Worek,
  \emph{{Assault on the NLO Wishlist: $pp \to t \bar{t} b \bar{b}$}},
  \href{https://doi.org/10.1088/1126-6708/2009/09/109}{\emph{JHEP} {\bfseries
  09} (2009) 109} [\href{https://arxiv.org/abs/0907.4723}{{\ttfamily
  0907.4723}}].

\bibitem{Czakon:2015cla}
M.~Czakon, H.~B. Hartanto, M.~Kraus and M.~Worek, \emph{{Matching the
  Nagy-Soper parton shower at next-to-leading order}},
  \href{https://doi.org/10.1007/JHEP06(2015)033}{\emph{JHEP} {\bfseries 06}
  (2015) 033} [\href{https://arxiv.org/abs/1502.00925}{{\ttfamily
  1502.00925}}].

\bibitem{Buckley:2014ana}
A.~Buckley, J.~Ferrando, S.~Lloyd, K.~Nordstr\"om, B.~Page, M.~R\"ufenacht
  et~al., \emph{{LHAPDF6: parton density access in the LHC precision era}},
  \href{https://doi.org/10.1140/epjc/s10052-015-3318-8}{\emph{Eur. Phys. J. C}
  {\bfseries 75} (2015) 132} [\href{https://arxiv.org/abs/1412.7420}{{\ttfamily
  1412.7420}}].

\bibitem{Ball:2017nwa}
{\scshape NNPDF} collaboration, \emph{{Parton distributions from high-precision
  collider data}},
  \href{https://doi.org/10.1140/epjc/s10052-017-5199-5}{\emph{Eur. Phys. J. C}
  {\bfseries 77} (2017) 663}
  [\href{https://arxiv.org/abs/1706.00428}{{\ttfamily 1706.00428}}].

\bibitem{Jezabek:1988iv}
M.~Jezabek and J.~H. Kuhn, \emph{{QCD Corrections to Semileptonic Decays of
  Heavy Quarks}},
  \href{https://doi.org/10.1016/0550-3213(89)90108-9}{\emph{Nucl. Phys. B}
  {\bfseries 314} (1989) 1}.

\bibitem{Cacciari:2008gp}
M.~Cacciari, G.~P. Salam and G.~Soyez, \emph{{The anti-$k_t$ jet clustering
  algorithm}}, \href{https://doi.org/10.1088/1126-6708/2008/04/063}{\emph{JHEP}
  {\bfseries 04} (2008) 063} [\href{https://arxiv.org/abs/0802.1189}{{\ttfamily
  0802.1189}}].

\bibitem{ATLAS-CONF-2021-014}
{\scshape ATLAS} collaboration, \emph{{Measurements of gluon fusion and
  vector-boson-fusion production of the Higgs boson in $H\rightarrow W W^*
  \rightarrow e\nu \mu\nu$ decays using $pp$ collisions at $\sqrt{s}=13$ TeV
  with the ATLAS detector, \tt[ATLAS-CONF-2021-014]}},  tech. rep., CERN,
  Geneva, Mar, 2021.

\bibitem{CMS-PAS-HIG-18-029}
{\scshape CMS} collaboration, \emph{{Measurements of Higgs boson production via
  gluon fusion and vector boson fusion in the diphoton decay channel at
  $\sqrt{s} = 13$ TeV, \tt[CMS-PAS-HIG-18-029]}},  tech. rep., CERN, Geneva,
  2019.

\bibitem{Aoki:2009ha}
M.~Aoki, S.~Kanemura, K.~Tsumura and K.~Yagyu, \emph{{Models of Yukawa
  interaction in the two Higgs doublet model, and their collider
  phenomenology}},
  \href{https://doi.org/10.1103/PhysRevD.80.015017}{\emph{Phys. Rev. D}
  {\bfseries 80} (2009) 015017}
  [\href{https://arxiv.org/abs/0902.4665}{{\ttfamily 0902.4665}}].

\bibitem{Plehn:2001nj}
T.~Plehn, D.~L.~Rainwater and D.~Zeppenfeld,
\emph{Determining the Structure of Higgs Couplings at the LHC}, 
\href{https://journals.aps.org/prl/abstract/10.1103/PhysRevLett.88.051801}{
  \emph{Phys. Rev.  Lett.} {\bfseries 88}  (2002), 051801}
[\href{https://arxiv.org/abs/hep-ph/0105325}{\ttfamily hep-ph/0105325}].
  
\bibitem{Haisch:2016gry}
U.~Haisch, P.~Pani and G.~Polesello, \emph{{Determining the CP nature of spin-0
  mediators in associated production of dark matter and $ t\overline{t} $
  pairs}}, \href{https://doi.org/10.1007/JHEP02(2017)131}{\emph{JHEP}
  {\bfseries 02} (2017) 131}
  [\href{https://arxiv.org/abs/1611.09841}{{\ttfamily 1611.09841}}].

\bibitem{Dittmaier:2000tc}
S.~Dittmaier, M.~Kramer, Y.~Liao, M.~Spira and P.~M. Zerwas, \emph{{Higgs
  radiation off quarks in supersymmetric theories at $e^+ e^-$ colliders}},
  \href{https://doi.org/10.1016/S0370-2693(00)00278-1}{\emph{Phys. Lett. B}
  {\bfseries 478} (2000) 247}
  [\href{https://arxiv.org/abs/hep-ph/0002035}{{\ttfamily hep-ph/0002035}}].

\bibitem{Dawson:1997im}
S.~Dawson and L.~Reina, \emph{{QCD corrections to associated Higgs boson
  production}}, \href{https://doi.org/10.1103/PhysRevD.57.5851}{\emph{Phys.
  Rev. D} {\bfseries 57} (1998) 5851}
  [\href{https://arxiv.org/abs/hep-ph/9712400}{{\ttfamily hep-ph/9712400}}].

\bibitem{Fadin:1993kt}
V.~S. Fadin, V.~A. Khoze and A.~D. Martin, \emph{{How suppressed are the
  radiative interference effects in heavy instable particle production?}},
  \href{https://doi.org/10.1016/0370-2693(94)90837-0}{\emph{Phys. Lett. B}
  {\bfseries 320} (1994) 141}
  [\href{https://arxiv.org/abs/hep-ph/9309234}{{\ttfamily hep-ph/9309234}}].

\bibitem{Hermann:2021xvs}
J.~Hermann and M.~Worek, \emph{{The impact of top-quark modelling on the
  exclusion limits in ${t\bar{t}}+{DM}$ searches at the LHC}},
  \href{https://doi.org/10.1140/epjc/s10052-021-09831-0}{\emph{Eur. Phys. J. C}
  {\bfseries 81} (2021) 1029}
  [\href{https://arxiv.org/abs/2108.01089}{{\ttfamily 2108.01089}}].

\bibitem{Bevilacqua:2022nrm}
G.~Bevilacqua, H.~B. Hartanto, M.~Kraus, J.~Nasufi and M.~Worek, \emph{{NLO QCD
  corrections to full off-shell production of $t\bar{t}Z$ including leptonic
  decays}},
\href{https://doi.org/10.1007/JHEP08(2022)060}{\emph{JHEP} {\bfseries 08}
  (2022) 060} [\href{https://arxiv.org/abs/2203.15688}{{\ttfamily 2203.15688}}].

\bibitem{Czakon:2020qbd}
M.~Czakon, A.~Mitov and R.~Poncelet, \emph{{NNLO QCD corrections to leptonic
  observables in top-quark pair production and decay}},
  \href{https://doi.org/10.1007/JHEP05(2021)212}{\emph{JHEP} {\bfseries 05}
  (2021) 212} [\href{https://arxiv.org/abs/2008.11133}{{\ttfamily
  2008.11133}}].

\bibitem{Barr:2005dz}
A.~J. Barr, \emph{{Measuring slepton spin at the LHC}},
  \href{https://doi.org/10.1088/1126-6708/2006/02/042}{\emph{JHEP} {\bfseries
  02} (2006) 042} [\href{https://arxiv.org/abs/hep-ph/0511115}{{\ttfamily
  hep-ph/0511115}}].

\bibitem{Barr:2003rg}
A.~Barr, C.~Lester and P.~Stephens, \emph{{m(T2): The Truth behind the
  glamour}}, \href{https://doi.org/10.1088/0954-3899/29/10/304}{\emph{J. Phys.
  G} {\bfseries 29} (2003) 2343}
  [\href{https://arxiv.org/abs/hep-ph/0304226}{{\ttfamily hep-ph/0304226}}].

\bibitem{Lester:1999tx}
C.~G. Lester and D.~J. Summers, \emph{{Measuring masses of semiinvisibly
  decaying particles pair produced at hadron colliders}},
  \href{https://doi.org/10.1016/S0370-2693(99)00945-4}{\emph{Phys. Lett. B}
  {\bfseries 463} (1999) 99}
  [\href{https://arxiv.org/abs/hep-ph/9906349}{{\ttfamily hep-ph/9906349}}].

\bibitem{Lester:2014yga}
C.~G. Lester and B.~Nachman, \emph{{Bisection-based asymmetric M$_{T2}$
  computation: a higher precision calculator than existing symmetric methods}},
  \href{https://doi.org/10.1007/JHEP03(2015)100}{\emph{JHEP} {\bfseries 03}
  (2015) 100} [\href{https://arxiv.org/abs/1411.4312}{{\ttfamily 1411.4312}}].

\bibitem{Kauer:2001sp}
N.~Kauer and D.~Zeppenfeld,
\emph{{Finite width effects in top quark production at hadron
  colliders}}, 
\href{https://journals.aps.org/prd/abstract/10.1103/PhysRevD.65.014021}{
\emph{Phys. Rev.}  {\bfseries D 65} (2002) 014021}
[\href{https://arxiv.org/abs/hep-ph/0107181}{\ttfamily hep-ph/0107181}].

\bibitem{Liebler:2015ipp}
S.~Liebler, G.~Moortgat-Pick and A.~S.~Papanastasiou,
\emph{Probing the top-quark width through ratios of resonance
  contributions of $e^+e^-\rightarrow W^+W^-b\bar{b}$},
\href{https://link.springer.com/article/10.1007/JHEP03(2016)099}{
  \emph{JHEP} {\bfseries 03} (2016) 099}
[\href{https://arxiv.org/abs/1511.02350}{\ttfamily 1511.02350}].

\bibitem{Baskakov:2018huw}
A.~Baskakov, E.~Boos and L.~Dudko,
\emph{Model independent top quark width measurement using a
  combination of resonant and nonresonant cross sections},
\href{https://journals.aps.org/prd/abstract/10.1103/PhysRevD.98.116011}{
  \emph{Phys. Rev.}  {\bfseries  D 98} (2018) no.11, 116011}
[\href{https://arxiv.org/abs/1807.11193}{\ttfamily 1807.11193}].

\bibitem{Bevilacqua:2020srb}
G.~Bevilacqua, H.~Y.~Bi, H.~B.~Hartanto, M.~Kraus, J.~Nasufi and M.~Worek,
\emph{NLO QCD corrections to off-shell ${t{\bar{t}}W^\pm }$ production
  at the LHC: correlations and asymmetries},
\href{https://link.springer.com/article/10.1140/epjc/s10052-021-09478-x}{
  \emph{Eur. Phys. J.}  {\bfseries C 81} (2021) no.7, 675}
[\href{https://arxiv.org/abs/2012.01363}{\ttfamily 2012.01363}].

\end{thebibliography}
\end{document}